\documentclass{book_study2}
\usepackage{graphicx}
\usepackage{longtable}
\usepackage{supertabular}
\usepackage{amsmath}
\usepackage{color}
\usepackage[figuresright]{rotfloat}[1995/03/30]
\usepackage{shortvrb}
\usepackage{afterpage}
\usepackage{xr}
\usepackage[hang]{caption}
\usepackage{chappg}
\usepackage[dvips,colorlinks=true,citecolor=blue]{hyperref}
\renewcommand{\chaptermark}[1]{%
\markboth{\MakeUppercase{\chaptername\ \thechapter.\ #1}}{}}

\makeatletter
\newcommand\ps@juanpreface{%
  \renewcommand\@oddhead{}%
  \renewcommand\@evenhead{}%
  \renewcommand\@oddfoot{\hfil\thepage\hfil}%
  \renewcommand\@evenfoot{\hfil\thepage\hfil}%
  }
\newcommand\ps@juan{%
  \renewcommand\@oddhead{\hfil\rightmark}%
  \renewcommand\@evenhead{\rightmark\hfil}%
  \renewcommand\@oddfoot{\hfil\thepage\hfil}%
  \renewcommand\@evenfoot{\hfil\thepage\hfil}%
  }
\makeatother
\newcommand{\beq}{\begin{equation}}
\newcommand{\eeq}{\end{equation}}
\newcommand{\beqs}{\begin{eqnarray}}
\newcommand{\eeqs}{\end{eqnarray}}
\newcommand{\lsim}{\mathrel{\raisebox{-.6ex}{$\stackrel{\textstyle<}{\sim}$} }}
\newcommand{\gsim}{\mathrel{\raisebox{-.6ex}{$\stackrel{\textstyle>}{\sim}$} }}
\newcommand{\rr}{\color{red}}
\def\bi{\begin{itemize}}
\def\ei{\end{itemize}}
\def\bt{\vskip.1in \begin{tabular}}
\def\et{\end{tabular} \vskip.1in}

\def \nufact {$\nu$-Factory}
\newcommand\as{\alpha_S}

\def\bit{\begin{itemize}}
\def\eit{\end{itemize}}

\def\what{\widehat}
\def\fbi{~{\rm fb}^{-1}}
\def\call{{\cal L}}
\def\lyear{L_{\rm year}}
\def\sig{\sigma}
\def\gamhsmtot{\Gamma_{\hsm}^{\rm tot}}
\def\br{{\rm BF}}
\def\gamhtot{\Gamma_{\h}^{\rm tot}}

\def\hl{h^0}
\def\ha{A^0}
\def\hh{H^0}

\def\mha{m_{\ha}}
\def\mhh{m_{\hh}}

\def\h{h}
\def\mh{m_{\h}}
\def\epem{e^+e^-}
\def\mupmum{\mu^+\mu^-}
\def\anti{\overline}
\def\tanb{\tan\beta}
\def\what{\widehat}
\def\rts{\sqrt s}
\def\hsm{h_{SM}}
\def\mhsm{m_{\hsm}}

\def\gev{~{\rm GeV}}

\def\srts{\sigma_{\!\!\!\sqrt s}^{\vphantom y}}
\newcommand{\alt}{\mathrel{\raisebox{-.6ex}{$\stackrel{\textstyle<}{\sim}$}}}
\newcommand{\agt}{\mathrel{\raisebox{-.6ex}{$\stackrel{\textstyle>}{\sim}$}}}
\def\lsim{\alt}
\def\gsim{\agt}
\def\beq{\begin{equation}}
\def\eeq{\end{equation}}
\def\bea{\begin{eqnarray}}
\def\eea{\end{eqnarray}}
\begin{document}
\title{
\textbf{The Program in Muon and Neutrino Physics: 
Super Beams, Cold Muon Beams, Neutrino Factory and the Muon Collider}
}
\author{\textbf{Editor:}~Rajendran~Raja$^{1}$\\
$^{1}$Fermi National Accelerator Laboratory, Batavia, IL 60510, USA\\
\textbf{Members of the Executive Board of the Muon Collaboration}
\\
D.~Cline,$^{2}$
J.~Gallardo,$^{3}$
S.~Geer,$^{1}$
D.~Kaplan,$^{4}$\\
K.~McDonald,$^{5}$
R.~Palmer,$^{3}$
A.~Sessler,$^{6}$\thanks{Co-Editor}
A.N.~Skrinsky,$^{7}$\\
D.~Summers,$^{8}$
M.~Tigner,$^{9}$
A.~Tollestrup,$^{1}$
J.~Wurtele,$^{6}$
M.~Zisman$^{6}$\thanks{Co-Editor}    \\
\\
$^{2}$University of California-Los Angeles, LA, CA 90095\\
$^{3}$Brookhaven National Laboratory, Upton, NY 11973\\
$^{4}$Illinois Institute of Technology, Chicago, IL 60616\\
$^{5}$Princeton University, Princeton, NJ 08544\\
$^{6}$Lawrence Berkeley National Laboratory, Berkeley, CA 94720\\
$^{7}$Budker Institute of Nuclear Physics, 630090 Novosibirsk, Russia\\
$^{8}$University of Mississippi, Oxford, MS 38677\\
$^{9}$Cornell University, Ithaca, NY 14853}
\pagenumbering{roman}
\thispagestyle{juanpreface}
\pagestyle{juanpreface}
\maketitle

\newpage

\newpage
\tableofcontents
\listoffigures    
\listoftables   
\thispagestyle{juan}
\pagestyle{juan}
\pagenumbering{arabic}
\renewcommand{\thepage}{\thechapter~-~\arabic{page}}
%

\chapter{Executive Summary}

\label{summary}
 Recent results from the SNO collaboration~\cite{snolatest}
coupled with data from the SuperK collaboration~\cite{superk} have provided
convincing evidence that neutrinos oscillate and that they very likely do so
among the three known neutrino species. Experiments currently under way or
planned in the near future will shed further light on the nature of these
mixings among neutrino species and the magnitudes of the mass differences
between them. Neutrino oscillations and the implied non-zero masses and
mixings represent the first experimental evidence of effects beyond the
Standard Model, and as such are worthy of our utmost attention.

This document points the way towards establishing an ongoing program of
research in accelerator and experimental physics that can be implemented in
an incremental fashion. At each step, one opens up new physics vistas,
leading eventually to a Neutrino Factory and a Muon Collider. One of the
first steps toward a Neutrino Factory is a proton driver that can be used to
provide intense beams of conventional neutrinos in addition to providing the
intense source of low energy muons from pion decay that must be cooled to be
accelerated and stored. While the proton driver is being constructed, we
will simultaneously engage in R\&D on collecting and cooling muons. A source
of intense cold muons can be immediately used to do physics on such items as
measuring the electric and magnetic dipole moments of the muon to higher
precision, muonium-antimuonium oscillations, rare muon decays and so on.
Once we develop the capability of cooling and accelerating muons, the
storage ring for such muons will be the first Neutrino Factory. Its precise
energy and its distance from the long-baseline experiment will be chosen
using the knowledge of neutrino oscillation parameters gleaned from the
present generation of solar and accelerator experiments (Homestake,
Kamiokande, SuperKamiokande, SAGE, GALLEX, K2K, SNO), the next generation
experiments (miniBOONE, MINOS, CNGS, KamLAND, Borexino), and the
high-intensity conventional beam experiments that would already have taken
place.

A Neutrino Factory provides both $\nu _{\mu }$ and $\anti\nu _{e}$ beams of
equal intensity for stored $\mu ^{-}$ beams and their charge conjugate beams
for stored $\mu ^{+}$ beams. Beams from a Neutrino Factory are intense. In
addition, they have smaller divergence than conventional neutrino beams of
comparable energy. These properties permit the study of non-oscillation
physics at near detectors and the measurement of structure functions and
associated parameters in non-oscillation physics to unprecedented accuracy.
They also permit long-baseline experiments that can determine oscillation
parameters to unprecedented accuracy. Depending on the value of the
parameter $\sin ^{2}2\theta _{13}$ in the three-neutrino oscillation
formalism, one can expect to measure the oscillation $\nu _{e}\rightarrow
\nu _{\mu }$. By comparing the rates for this channel with its
charge-conjugate channel $\anti\nu _{e}\rightarrow \anti\nu _{\mu }$ , one
can determine the sign of the leading mass difference in neutrinos, $\delta
m_{32}^{2}$, by making use of their passage through matter in a
long-baseline experiment. Such experiments can also shed light on the CP
violating phase, $\delta $, in the lepton mixing matrix and enable us to
study CP violation in the lepton sector. It is known that CP violation in
the quark sector is insufficient to explain the baryon asymmetry of the
Universe. Perhaps the lepton sector CP violation plays a crucial role in
creating this asymmetry during the initial phases of the Big Bang.

While the Neutrino Factory is being constructed, R\&D can be performed to
make the Muon Collider a reality. This would require orders of magnitude
more cooling. Muon Colliders, if realized, provide a tool to explore
Higgs-like objects by direct $s$-channel fusion, much as LEP explored the $Z$
. They also provide a means to reach higher energies (3--4~TeV in the center
of mass) using compact collider rings.

These concepts and ideas have aroused significant interest throughout the
world scientific community. In the U.S., a formal collaboration of some 140
scientists, the Neutrino Factory and Muon Collider Collaboration (MC)~\cite
{EPP:collaboration}, has undertaken the study of designing a Neutrino
Factory, along with R\&D activities in support of a Muon Collider design.

\section{Feasibility Studies}

In the fall of 1999, Fermilab, with help from the MC, undertook a
Feasibility Study (``Study-I'') of an entry-level Neutrino Factory~\cite
{INTRO:ref1}. One aim of Study-I was to assess the extent to which the
Fermilab accelerator complex could be made to evolve into a Neutrino
Factory. Study-I showed that such an evolution was clearly possible. The
performance reached in Study-I, characterized in terms of the number of muon
decays aimed at a detector located 3000 km away from the muon storage ring,
was $N$ = 2 $\times $ 10$^{19}$ decays per ``Snowmass year'' (10$^{7}$ s)
per MW of protons on target.

Simultaneously, Fermilab launched a study of the physics that might be
addressed by such a facility~\cite{INTRO:ref9} and, more recently, initiated
a study to compare the physics reach of a Neutrino Factory with that of
conventional neutrino beams~\cite{superbeams} powered by a high intensity
proton driver (referred to as ``superbeams''). It was determined that a
steady and diverse physics program will result from following the
evolutionary path from a superbeam to a full-fledged Neutrino Factory.

After the completion of Study-I, BNL organized a follow-on study
(``Study-II'') on a high-performance Neutrino Factory sited at BNL, also in
collaboration with the MC. An important goal of Study-II was to evaluate
whether BNL was a suitable site for a Neutrino Factory. Study-II has
recently answered that question affirmatively. A second goal of Study-II was
to examine various site-independent means of enhancing the performance of a
Neutrino Factory. Based on the improvements in Study-II, the number of muons
delivered to the storage ring per Snowmass year from a 1-MW proton driver
would be:

\begin{eqnarray*}
\mu /\text{year} &=&10^{14}\text{ ppp}\times 2.5\text{ Hz}\times 10^{7}\text{
s/year}\times 0.17\text{ }\mu /\text{p}\times 0.81 \\
&=&3.4\times 10^{20}
\end{eqnarray*}

\noindent where the last factor (0.81) is the estimated efficiency of the
acceleration system. For the case of an upgraded 4 MW proton driver, the
muon production would increase to 1.4 $\times $ $10^{21}$ $\mu $ /year.
(R\&D to develop a target capable of handling this beam power would be
needed.)

The number of muons decaying in the production straight section per Snowmass
year would be 35\% of this number, or 1.2 $\times $ 10$^{20}$ decays for a 1
MW proton driver (4.8 $\times $ 10$^{20}$ decays for a 4 MW proton driver).
Though these neutrinos are potentially available for experiments, in the
current storage ring design the angular divergence at both ends of the
production straight section is higher than desirable for the physics
program. This can be improved in a straightforward manner and we are
confident that storage ring designs allowing 30--40\% of useful muon decays
are feasible.

Both Study-I and -II are site specific in that each has a few site-dependent
aspects; otherwise, they are generic. In particular, Study-II uses BNL
site-specific proton driver specifications corresponding to an upgrade of
the 24-GeV AGS complex and a BNL-specific layout of the storage ring, which
is housed in an above-ground berm to avoid penetrating the local water
table. Study-I uses a new Fermilab booster to achieve its beam intensities
and an underground storage ring. The primary substantive difference between
the two studies is that Study-II is aimed at a lower muon energy (20 GeV),
but higher intensity (for physics reach). Taking the two Feasibility Studies
together, we conclude that a high-performance Neutrino Factory could easily
be sited at either BNL or Fermilab.

It is worthwhile noting that a $\mu ^{+}$ storage ring with an average
neutrino energy of 15~GeV and $2\times 10^{20}$ useful muon decays will
yield (in the absence of oscillations) $\approx $30,000 charged-current
events in the $\nu _{e}$ channel per kiloton-year in a detector located
732~km away. In comparison, a 1.6~MW superbeam~\cite{superbeams} from the
Fermilab Main Injector with an average neutrino energy of 15~GeV will yield $
\approx $13,000 $\nu _{\mu }$ charged-current events per kiloton-year.
However, a superbeam has a significant $\nu _{e}$ contamination, which will
be the major background in $\nu _{\mu }\rightarrow \nu _{e}$ appearance
searches. It is much easier to detect the oscillation $\nu _{e}\rightarrow
\nu _{\mu }$ from muon storage rings than the oscillation $\nu _{\mu
}\rightarrow \nu _{e}$ from conventional neutrino beams, since the electron
final state from conventional beams has significant background contribution
from $\pi ^{0}$'s produced in the events.

\section{Neutrino Factory Description}

The muons we use result from decays of pions produced when an intense proton
beam bombards a high-power production target. The target and downstream
transport channel are surrounded by superconducting solenoids to contain the
pions and muons, which are produced with a larger spread of transverse and
longitudinal momenta than can be conveniently transported through an
acceleration system. To prepare a beam suitable for subsequent acceleration,
we first perform a ``phase rotation,'' during which the initial large energy
spread and small time spread are interchanged using induction linacs. Next,
to reduce the transverse momentum spread, the resulting long bunch, with an
average momentum of about 250 MeV/$c$, is bunched into a 201.25-MHz bunch
train and sent through an ionization cooling channel consisting of LH$_{2}$
energy absorbers interspersed with rf cavities to replenish the energy lost
in the absorbers. The resulting beam is then accelerated to its final energy
using a superconducting linac to make the beam relativistic, followed by one
or more recirculating linear accelerators (RLAs). Finally, the muons are
stored in a racetrack-shaped ring with one long straight section aimed at a
detector located at a distance of roughly 3000 km.

A list of the main ingredients of a Neutrino Factory is given below. Details
of the design described here are based on the specific scenario of sending a
neutrino beam from Brookhaven to a detector in Carlsbad, New Mexico. More
generally, however, the design exemplifies a Neutrino Factory for which the
two Feasibility Studies demonstrated technical feasibility (provided the
challenging component specifications are met), established a cost baseline,
and established the expected range of physics performance.

\begin{itemize}
\item  \textbf{Proton Driver:} Provides 1--4 MW of protons on target from an
upgraded AGS; a new booster at Fermilab would perform equivalently.

\item  \textbf{Target and Capture:} A high-power target immersed in a 20-T
superconducting solenoidal field to capture pions produced in proton-nucleus
interactions.

\item  \textbf{Decay and Phase Rotation:} Three induction linacs, with
internal superconducting solenoidal focusing to contain the muons from pion
decays, that provide nearly non-distorting phase rotation; a
``mini-cooling'' absorber section is included after the first induction
linac to reduce the beam emittance and lower the beam energy to match the
cooling channel acceptance.

\item  \textbf{Bunching and Cooling:} A solenoidal focusing channel, with
high-gradient rf cavities and liquid-hydrogen absorbers, that bunches the
250~MeV/$c$ muons into 201.25-MHz rf buckets and cools their transverse
normalized emittance from 12 mm$\cdot $rad to 2.7 mm$\cdot $rad.

\item  \textbf{Acceleration:} A superconducting linac with solenoidal
focusing to raise the muon beam energy to 2.48 GeV, followed by a four-pass
superconducting RLA to provide a 20 GeV muon beam; a second RLA could
optionally be added to reach 50 GeV, if the physics requires this.

\item  \textbf{Storage Ring:} A compact racetrack-shaped superconducting
storage ring in which $\approx $35\% of the stored muons decay toward a
detector located about 3000 km from the ring.
\end{itemize}

\section{Detector}

The Neutrino Factory plus its long-baseline detector will have a physics
program that is a logical continuation of current and near-future neutrino
oscillation experiments in the U.S., Japan and Europe. Moreover, detector
facilities located in experimental areas near the neutrino source will have
access to integrated neutrino intensities $10^{4}$--$10^{5}$ times larger
than previously available ($10^{20}$ neutrinos per year compared with $
10^{15}$--$10^{16}$).

Specifications for the long-baseline Neutrino Factory detector are rather
typical for an accelerator-based neutrino experiment. However, because of
the need to maintain a high neutrino rate at these long distances ($\approx $
3000 km), the detectors considered here are 3--10 times more massive than
those in current neutrino experiments.

Several detector options are possible for the far detector:

\begin{itemize}
\item  A 50 kton steel--scintillator--proportional-drift-tube (PDT)
detector. The PDT detector would resemble MINOS. A detector with dimensions $
8~\text{m}\times 8~\text{m}\times 150$~m would record up to $4\times 10^{4}$ 
$\nu _{\mu }$ events per year.

\item  A large water-Cherenkov detector, similar to SuperKamiokande but with
either a magnetized water volume or toroids separating smaller water tanks.
This could be the UNO detector~\cite{DET:uno}, currently proposed to study
both proton decay and cosmic neutrinos. UNO would be a 650-kton
water-Cherenkov detector segmented into a minimum of three tanks. It would
have an active fiducial mass of 440~kton and would record up to $3\,\times
\,10^{5}$ $\nu _{\mu }$ events per year from the Neutrino Factory beam.

\item  A massive liquid-argon magnetized detector~\cite{landd} that would
attempt to detect proton decay, detect solar and supernova neutrinos, and
also serve as a Neutrino Factory detector.
\end{itemize}

For the near detector, a compact liquid-argon TPC (similar to the ICARUS
detector~\cite{ICARUS}) could be used. It would be cylindrically shaped with
a radius of 0.5 m and a length of 1~m, would have an active volume of $10^{3}
$ kg, and would provide a neutrino event rate \textsl{O}(10~Hz). The TPC
could be combined with a downstream magnetic spectrometer for muon and
hadron momentum measurements. At these neutrino intensities, it is even
possible to envision an experiment with a relatively thin Pb target (1~$
L_{rad}$~), followed by a standard fixed-target spectrometer containing
tracking chambers, time-of-flight and calorimetry, with an event rate 
\textsl{O}(1~Hz).

\section{R\&D Program\label{RDprog}}

Successful construction of a muon storage ring to provide a copious source
of neutrinos requires many novel approaches to be developed and
demonstrated. To construct a high-luminosity Muon Collider is an even
greater extrapolation of the present state of accelerator design. Thus,
reaching the full facility performance in either case requires an extensive
R\&D program.

Each of the major systems has significant issues that must be addressed by
R\&D activities, including a mix of theoretical, simulation, modeling, and
experimental studies, as appropriate. Component specifications need to be
verified. For example, the cooling channel assumes a normal conducting rf
(NCRF) cavity gradient of 17 MV/m at 201.25 MHz, and the acceleration
section demands similar performance from superconducting rf (SCRF) cavities
at this frequency. In both cases, the requirements are beyond the
performance reached to date for cavities in this frequency range. The
ability of the induction linac units to coexist with their internal SC
solenoids must be verified, and the ability of the target to withstand a
proton beam power of up to 4 MW must be tested. Finally, some sort of
cooling demonstration experiment should be undertaken to validate the
implementation of the cooling channel.

To make progress on the R\&D program in a timely way, the required support
level is about \$15M per year. At present, the MC is getting only about \$8M
per year, so R\&D progress is less rapid than it could be.

\section{Cost Estimate}

As part of the Study, we have specified each system in sufficient detail to
obtain a ``top-down'' cost estimate for it. Clearly this estimate is not the
complete and detailed cost estimate that would come from preparing a full
Conceptual Design Report (CDR). However, there is considerable experience in
designing and building accelerators with similar components, so we have a
substantial knowledge base from which costs can be derived. With this
caveat, we find that the cost of such a facility is about \$1.9$\,$B in FY01
dollars. This value represents only direct costs, not including overhead or
contingency allowances.

It should be noted that the current design has erred on the side of
feasibility rather than costs. Thus, we do not yet have a fully
cost-optimized design, nor one that has been reviewed from the standpoint of
``value engineering.'' \ In that sense, there is hope that a detailed design
study will \textit{reduce} the costs compared with what we indicate here.

\section{Staging Scenario}

If desired by the particle physics community, a fast-track plan leading
directly to a Neutrino Factory could be executed. This would be done by
beginning now to create the required Proton Driver (see Stage 1 below),
using well-understood technology, while working in parallel on the R\&D
needed to complete a CDR for the Neutrino Factory facility. We estimate
that, with adequate R\&D support (see Section \ref{RDprog}), we could
complete a CDR in 2006 and be ready for construction in 2007. On the other
hand, the Neutrino Factory offers the distinct advantage that it can be
built in stages. This could satisfy both programmatic and cost constraints
by allowing an ongoing physics program while reducing the annual
construction funding needs. Depending on the results of our technical
studies and the results of ongoing searches for the Higgs boson, it is hoped
that the Neutrino Factory is really the penultimate stage, to be followed
later by a Muon Collider (e.g., a Higgs Factory). Below we list possible
stages for the evolution of a muon beam facility and give an indication of
incremental costs. These cost increments represent only machine-related
items and do not include detector costs.

\begin{description}
\item  \textbf{Stage 1:} \$250--330M (1 MW) or \$330--410M (4 MW)

\item  \qquad We envision a Proton Driver and a Target Facility. The Driver
could have a 1 MW beam level or be designed from the outset to reach 4 MW.
The Target Facility is built initially to accommodate a 4 MW beam. \ A 1 MW
beam would provide about $1.2\times 10^{14}$ $\mu $/s ($1.2\times 10^{21}$ $
\mu $/year) and a 4 MW beam about $5\times 10^{14}$ $\mu $/s ($5\times
10^{21}$ $\mu $/year) into a solenoid channel. Costs for this stage depend
on site-specific choices, e.g., beam energy. This stage could be
accomplished within the next 4--5 years if the particle physics community
considers it a high priority.

\item  \textbf{Stage 2:} \$660--840M

\item  \qquad We envision a muon beam that has been phase rotated and
transversely cooled. This provides a muon beam with a central momentum of
about 200 MeV/$c$, a transverse (normalized) emittance of 2.7 mm-rad and an
rms energy spread of about 4.5\%. The intensity of the beam would be about $
4\times 10^{13}$ $\mu $/s ($4\times 10^{20}$ $\mu $/year) at 1 MW, or $
1.7\times 10^{14}$ $\mu $/s ($1.7\times 10^{21}$ $\mu $/year) at 4 MW. The 
\textit{incremental} cost of this option is \$840M, based on taking the
cooling channel length adopted in Study-II. If more intensity were needed,
and if less cooling could be tolerated, the length of the cooling channel
could be reduced. \ Accepting twice the transverse emittance would reduce
the incremental cost by about \$180M. At this stage, physics with intense
cold muon beams can start and continue to the stage when the muons are
accelerated.

\item  \textbf{Stage 3:} \$220--250M

\item  \qquad We envision using the pre-acceleration Linac to raise the beam
energy to roughly 2.5 GeV. \ The incremental cost of this option is about
\$220M. At this juncture, it may be appropriate to consider a small storage
ring, comparable to the $g-2$ ring at BNL, to be used, perhaps, for the next
round of muon $g-2$ experiments. No cost estimate has been made for this
ring, but it would be expected to cost roughly \$30M.

\item  \textbf{Stage 4:} \$550M (20 GeV) or \$1250--1350M (50 GeV)

\item  \qquad We envision having a complete Neutrino Factory. For a 20 GeV
beam energy, the incremental cost of this stage, which includes the RLA and
the storage ring, is \$550M. If it were necessary to provide a 50 GeV muon
beam for physics reasons, an additional RLA and a larger storage ring would
be needed. \ The incremental cost would then increase by \$700--800M.

\item  \textbf{Stage 5}

\item  \qquad We envision an entry-level Muon Collider to operate as a Higgs
Factory. No cost estimate has yet been prepared for this stage, so we
mention here only the obvious ``cost drivers''---the additional cooling and
the additional acceleration. Future work will define the system requirements
better and permit a cost estimate of the same type provided for Studies-I
and -II.
\end{description}

\section{Muon Collider}

As is clear from the above discussion, a Neutrino Factory facility can be
viewed as a first critical step on the path toward an eventual high-energy
Muon Collider. Such a collider offers the potential of bringing the energy
frontier in particle physics within reach of a moderate sized machine. The
very fortuitous situation of having an intermediate step along this path
that offers a powerful and exciting physics program in its own right
presents an ideal opportunity, and it is hoped that the particle physics
community will have the resources to take advantage of it.

To reach the feasibility study stage, we must find robust technical
solutions to longitudinal emittance cooling, issues related to the high
bunch charges, techniques for cooling to the required final emittances, and
the design of a very low $\beta $* collider ring. We are confident that
solutions exist along the lines we have been investigating. We in the MC are
eager to advance to the stage of building a Muon Collider on the earliest
possible time scale. However, for that to happen there is an urgent need to
increase support for our R\&D so that we can address the vital issues.
Unless and until we obtain such support, it is hard to predict how long it
will take to solve the longitudinal emittance cooling and other
collider-specific problems.
\section{International Activities}

Work on Neutrino Factory R\&D is being carried out both in Europe and in
Japan. Communication between these groups and the MC is good. In addition to
having members of the MC Executive Board from these regions, there are
annual NUFACT workshops held to disseminate information. These meetings,
which rotate through the three regions, have been held in Lyon (1999), in
Monterey (2000), and in Tsukuba (2001); the next meeting will be held in
London.

Activities in Europe are centered at CERN but involve many European
universities and labs. Their concept for a Neutrino Factory is analogous to
that of the MC, but the implementation details differ. The European Proton
Driver is based on a 2.2-GeV superconducting proton linac that makes use of
the LEP rf cavity infrastructure. Phase rotation and cooling are based on rf
cavities operating at 44 and 88 MHz, along with appropriate LH$_{2}$
absorbers. R\&D on the rf cavities is in progress. CERN has mounted the HARP
experiment to measure particle yields in the energy regime of interest to
them (about 2 GeV). The CERN group is participating actively in the E951
Targetry experiment at BNL, and has provided some of the mercury-jet
apparatus that was tested successfully. European groups are also heavily
involved in the MUSCAT experiment at TRIUMF, where they play a lead role.

Activities in Japan have concentrated on the development of Fixed-Field
Alternating Gradient (FFAG) accelerators. These have very large transverse
and longitudinal acceptance, and thus have the potential of giving a
Neutrino Factory that does not require cooling. They are pursuing this
scheme. A proof-of-principle FFAG giving 500-keV protons has already been
built and tested, and plans exist for a 150 MeV version.  A 50-GeV 1-MW
Proton Driver is approved for construction in Japan, with a six-year
schedule. A collaboration with the MC on LH$_{2}$ absorber design is under
way, using U.S.-Japan funds.

On a global note, the three regions are in the process of developing a joint
proposal for an international Cooling Demonstration Experiment that could
begin in 2004. A Steering Committee has been set up for this purpose, with
representatives from all three regions.

\section{Conclusions}

In summary, the Muon Collaboration is developing the knowledge and ability
to create, manipulate, and accelerate muon beams. Our R\&D program will
position the HEP community such that, when it requires a Neutrino Factory or
a Muon Collider, we shall be in a position to provide it. A staged plan for
the deployment of a Neutrino Factory has been developed that provides an
active neutrino and muon physics program at each stage.
The requisite R\&D program, diversified over laboratories and
universities and having international participation, is currently
supported at the \$8M level, but requires of the order of \$15M per year
to make progress in a timely way.

\chapter{Introduction}
\label{intro}
\section{History}

The concept of a Muon Collider was first proposed by
Budker~\cite{PREFACE:budker} and by Skrinsky~\cite {PREFACE:skrinsky}
in the 60s and early 70s. However, there was little substance to the
concept until the idea of ionization cooling was developed by Skrinsky
and Parkhomchuk~\cite{INTRO:ref3}. The ionization cooling approach was
expanded by Neuffer~\cite{INTRO:ref4} and then by Palmer~\cite
{PREFACE:palmer}, whose work led to the formation of the Neutrino
Factory and Muon Collider Collaboration (MC)~\cite{EPP:collaboration} in
1995. \footnote{A good summary of the Muon Collider concept can be found in the
Status Report of 1999~\cite{INTRO:ref5}; an earlier
document~\cite{INTRO:ref6}, prepared for Snowmass-1996, is also useful
reading. MC Notes prepared by the MC are available
on the web~\cite{INTRO:ref11}}

The concept of a neutrino source based on a pion storage ring was
originally considered by Koshkarev~\cite{INTRO:ref7}. However, the
intensity of the muons created within the ring from pion decay was too
low to provide a useful neutrino source. The Muon Collider concept
provided a way to produce a very intense muon source.  The physics
potential of neutrino beams produced by muon storage rings was
investigated by Geer in 1997 at a Fermilab workshop~\cite{rajageer,geer}
where it became evident that the neutrino beams produced by muon
storage rings needed for the muon collider were exciting on their own
merit.  The neutrino factory concept quickly captured the imagination
of the particle physics community, driven in large part by the
exciting atmospheric neutrino deficit results from the
SuperKamiokande experiment.

  As a result, the MC realized that a Neutrino
Factory could be an important first step toward a Muon
Collider and  the physics that could be addressed by a
Neutrino Factory was interesting in its own right. With this in mind,
the MC has shifted its primary emphasis toward the issues  relevant
to a Neutrino Factory.  There is also considerable international
activity on Neutrino Factories, with international conferences held at
Lyon in 1999, Monterey in 2000~\cite {INTRO:ref13}, Tsukuba in 
2001~\cite{INTRO:ref14}, and another planned for London in 2002.

In the fall of 1999, Fermilab undertook a Feasibility Study
(``Study-I'') of an entry-level Neutrino
Factory~\cite{INTRO:ref1}. One of the aims of Study-I was to
determine to what extent the Fermilab accelerator complex could be made to 
evolve into a  Neutrino Factory. Study-I answered this question affirmatively.
Simultaneously Fermilab launched a study
of the physics that might be addressed by such a 
facility~\cite{INTRO:ref9}. More recently, Fermilab initiated a 
study to compare the
physics reach of a Neutrino Factory with that of conventional neutrino
beams~\cite{superbeams} powered by a high intensity proton driver, 
which are referred to as ``superbeams''. 
The aim  is to compare the physics reach of superbeams  with that
of a realistic Neutrino factory. Suffice it
to say, it was determined that a steady and diverse stream of  physics 
will result along this evolutionary path.

More recently, BNL organized a follow-on study (``Study-II'') on a
high-performance Neutrino Factory sited at BNL. Study-II was recently completed. 
Clearly, an important goal of Study-II was to evaluate
whether BNL was a suitable site for a Neutrino Factory. Based on the
work contained in Study-II, that question was answered affirmatively.

\begin{figure}[tbh!]
\centerline{\includegraphics[width=4.0in]{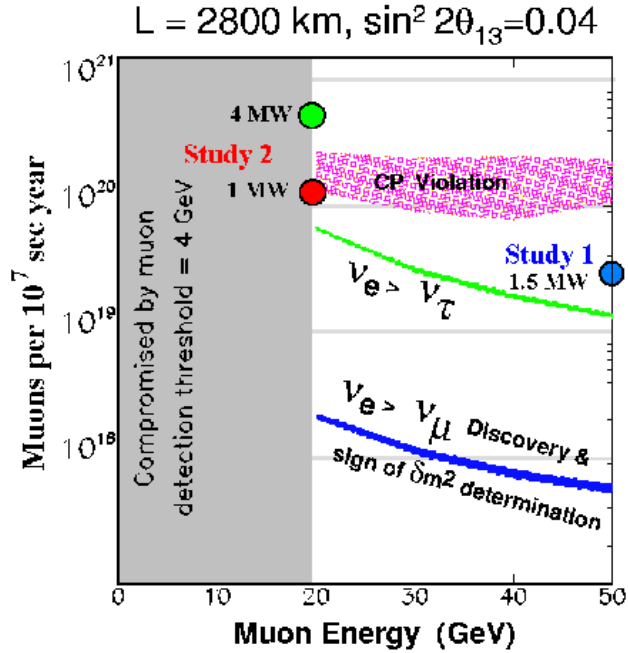}}
\caption[Muon decays in  a straight section \textit{vs.} muon energy]{
Muon decays in a straight section per $10^{7}\,$s \textit{vs.} muon
energy, with fluxes required for different physics searches assuming a
50~kT detector. Simulated performance of the two studies is
indicated.\label{studycomp}}
\end{figure}

Studies I and II are site specific in that in each study there are a
few site-dependent parts; otherwise, they are quite generic. In
particular, Study-II uses BNL site-specific proton driver
specifications and a BNL-specific layout of the storage ring,
especially the pointing angle of the straight sections. Study-I uses
an upgraded Fermilab booster to achieve its beam intensities. The
primary substantive difference between the two studies is that
Study-II is aimed at a lower muon energy (20 GeV), but higher
intensity (for physics reach). Figure \ref{studycomp} shows a
comparison of the performance of the neutrino factory designs in Study
I and Study II~\cite{INTRO:ref9}. Both Study-I and
Study-II were carried out jointly with the 
MC~\cite{EPP:collaboration}, which has over 140 members from many
institutions in the U.S. and abroad.

 Complementing the Feasibility Studies, the MC carries on an 
experimental and theoretical R\&D
program, including work on targetry, cooling, rf hardware (both normal
conducting and superconducting), high-field solenoids, LH$_{2}$
absorber design, theory, simulations, parameter studies, and emittance
exchange~\cite{INTRO:ref12}.

\section{General Scheme and Expected Performance}

\begin{figure}[tbh!]
\centerline{\includegraphics[width=4.0in,angle=-90]{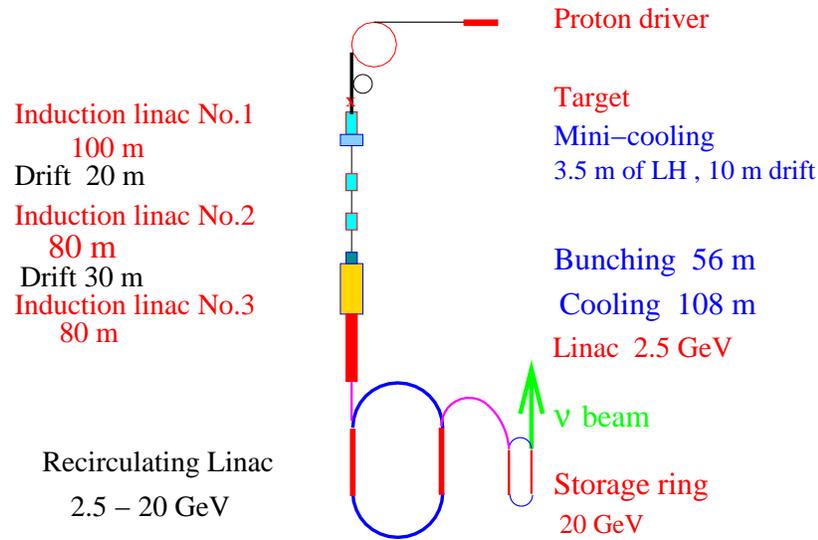}}
\caption[Schematic of the neutrino factory-Study II version]
{Schematic of the neutrino factory-Study II
version.\label{nufact-scheme-bnl}}
\end{figure}

Our present understanding of the design of a Neutrino Factory
and results for its simulated performance are
summarized here. Specific details can be found in the Study-II 
report~\cite{EPP:studyii}. 
A schematic layout is shown in
Fig.\ref{nufact-scheme-bnl}.

\subsection{Neutrino Factory Systems}

In overview, the muons result from decays of pions produced when an
intense proton beam bombards a high-power production target. The
target and downstream transport channel are surrounded by
superconducting solenoids to contain the pions and muons, which are
produced with a larger spread of transverse and longitudinal momenta
than can be conveniently injected into an acceleration system. 
To produce a beam suitable
for subsequent acceleration, the energy spread is reduced by ``phase
rotation'' where the initial large energy spread and small time spread
are interchanged using induction linacs. To reduce the transverse
momentum spread, the resulting long bunch, with an average momentum of
about 250 MeV/$c$, is bunched into a 201.25 MHz bunch train and then
sent through an ionization cooling channel consisting of LH$_{2}$ energy
absorbers interspersed with rf cavities to replenish the energy lost
in the absorbers. The resulting beam is then accelerated to its final
energy using a superconducting linac to make the beam relativistic,
followed by a recirculating linear accelerator (RLA). Finally, the
muons are stored in a racetrack-shaped ring with one long straight
section aimed at a detector located at a distance of roughly 3000 km.

A list of the main ingredients of a Neutrino Factory is given below;
more details can be found in Chapter~\ref{neufact}. 
The details of the design described here are based on the specific
scenario of sending a neutrino beam from Brookhaven to a detector in
Carlsbad, New Mexico.  However, the design exemplifies a general class of
Neutrino Factories for which the two Feasibility Studies have 1)
demonstrated technical feasibility, 2) established a cost baseline, and 3)
established the expected range of physics performance.

\begin{itemize}
\item  \textbf{Proton Driver} Provides 1 MW of protons on target from an
upgraded AGS; a new booster at Fermilab would perform equivalently.

\item  \textbf{Target and Capture} A mercury-jet target immersed in a 20-T
superconducting solenoidal field to capture pions, produced in
proton-nucleus interactions. (A technically less-ambitious alternative in which
the target is made of graphite has also been considered.)

\item  \textbf{Decay and Phase Rotation} Three induction linacs, with
internal superconducting solenoidal focusing, to contain the muons
from pion decays and provide nearly non-distorting phase rotation; a
``minicooling'' absorber section is included after the first induction
linac.

\item  \textbf{Bunching and Cooling} A solenoidal focusing channel, with
high-gradient rf cavities and liquid-hydrogen absorbers, that bunches
the 250~MeV/$c$ muons into 201.25-MHz rf buckets and cools their
transverse normalized emittance from 12 mm$\cdot $rad to 2.7 mm$\cdot
$rad.

\item  \textbf{Acceleration} A superconducting linac with solenoidal
focusing to raise the muon beam energy to 2.48 GeV, followed by a
four-pass superconducting RLA to provide a 20 GeV muon beam; a second
RLA could optionally be be added to reach 50 GeV, 
if the physics requires such high energy.

\item  \textbf{Storage Ring} A compact racetrack-shaped superconducting
storage ring in which $\approx$ 35\% of the stored muons decay 
toward a detector
located $\approx$ 2900 km from the ring.

\end{itemize}

\subsection{Predicted Performance}

Complete simulations up to the start of acceleration have been
performed using the code MARS~\cite{EPP:mars} (for pion production)
followed by ICOOL~\cite{icool} 
(for transport, phase rotation, and cooling). These results
have been confirmed by GEANT4~\cite{EPP:GEANT4}. They show an average
of 0.17 final muons per initial proton on the target, \textit{i.e.},
0.0071 $\mu /p$/GeV (considering the energy of the initial beam). This can be
compared with a value of \ 0.0011 $\mu /p$/GeV produced in Study-I. 
The gain compared with Study-I (a factor of 6) comes from:

\begin{itemize}
\item  Use of mercury, instead of carbon, as a target (1.9$\times $)

\item  Use of three, instead of one, induction linacs for phase rotation (2$%
\times $)

\item  Use of a more efficient tapered cooling channel design (1.4$\times $)

\item  Use of a larger acceptance for the acceleration channel (1.2$\times $)
\end{itemize}

Based on the improvements from Study-II, the number of muons delivered
to the storage ring per Snowmass year (10$^{7}$ s) from a
1-MW proton driver would be:

\begin{eqnarray*}
\mu /\text{year} &=&10^{14}\text{ ppp}\times 2.5\text{ Hz}\times 10^{7}\text{
s/year}\times 0.17\text{ }\mu /\text{p}\times 0.81 \\ &=&3.4\times
10^{20}
\end{eqnarray*}

\noindent where the last factor (0.81) is the calculated efficiency of the
acceleration system. Note that for the case of an upgraded 4 MW proton
driver, the muon production would increase to 1.4 $\times $ $10^{21}$
$\mu $ /year. The number of muons decaying in the production straight
section
per Snowmass year would be 35\% of this number, or 1.2 $\times $ 10$
^{20}$ decays for a 1 MW proton driver (4.8 $\times $ 10$^{20}$ decays
for a 4 MW proton driver).

Though these numbers of neutrinos are potentially available for
experiments, in the current storage-ring design the angular divergence at
both ends of the production straight section is higher than desirable for
the physics program.  In any case, we anticipate that storage-ring designs
that allow 30--40\% of the muon decays to provide useful neutrinos are
feasible.

It is worthwhile noting that a $\mu^+$ storage ring with an 
average neutrino energy of 15~GeV and 
2$\times$10$^{20}$ useful muon decays will yield 
(in the absence of oscillations) $\approx$ 30,000
charged current events/kiloton-year in a detector placed 732~km away
in the $\nu_e$ channel.  In comparison, a
1.6~MW superbeam~\cite{superbeams} from the Fermilab Main Injector 
with an average neutrino energy of 15~GeV will
yield $\approx$ 13,000 $\nu_\mu$ charged current events per kiloton-year. 
However, these superbeams have a significant $\nu_e$
contamination which will be the major background in
$\nu_\mu\rightarrow\nu_e$ appearance searches.  It is much easier to
detect the oscillation $\nu_e\rightarrow\nu_\mu$ from muon storage
rings than the oscillation $\nu_\mu\rightarrow\nu_e$ from conventional
neutrino beams, since the electron final state from conventional beams
has significant background contribution from $\pi^0$'s produced in the events.

\section{Outline of Report}

In what follows, we give a scenario for a staged approach to
constructing a Neutrino Factory and eventually a Muon
Collider. Chapter~\ref{physics} discusses the physics opportunities,
starting from conventional ``superbeams'' and going to cold muon
beams, then a Neutrino Factory with its near and far detectors, and
finally a Muon Collider. In Chapter~\ref{neufact}, we describe the
components of a Neutrino Factory, based on the Study-II
design. Chapter~\ref{higgsfact} covers our present concept of an
entry-level Higgs Factory Muon Collider. Our present understanding of
the costs of a Neutrino Factory and the financial implications of
possible staging scenarios will be described in Chapter~\ref{costs}.
Before embarking on construction of a Neutrino Factory, an R\&D
program is needed to address various technical issues. A description
of the required program and its budget requirements is presented in
Chapter~\ref{r_and_d}.  Chapter~\ref{r_and_d} also describes current
thinking about a cooling demonstration experiment that would be
carried out as an international effort. Finally, in Chapter~\ref{international} 
we provide a brief overview of the international
scope of the R\&D effort for intense muon beam accelerators.

\chapter{Physics Motivation}
\label{physics}
In this chapter we cover the physics potential of the neutrino factory
accelerator complex, which includes superbeams of conventional
neutrinos that are possible using  the proton driver needed for the
factory,  and intense beams of cold
muons that become available once the muon cooling and collection
systems for the factory are in place. Once the cold muons are
accelerated and stored in the muon storage ring, we realize the full
potential of the factory in both neutrino oscillation and
non-oscillation physics.

 Cooling muons will be a learning experience. We hope that the
 knowledge gained in constructing a neutrino factory can be used to
 cool muons sufficiently to produce the first muon collider operating
 as a Higgs factory. We examine the physics capabilities of  such a
 collider, which if realized, will invariably lead to higher energy
 muon colliders with exciting physics opportunities.
\section{Neutrino Oscillation Physics}
Here we discuss~\cite{study2} the current evidence for neutrino 
oscillations, and hence
neutrino masses and lepton mixing, from solar and atmospheric data.  A review
is given of some theoretical background including models for neutrino masses
and relevant formulas for neutrino oscillation transitions.  We next mention
the near-term and mid-term experiments in this area and comment on what they
hope to measure.  We then discuss the physics potential of a muon storage ring
as a neutrino factory in the long term. 

\subsection{Evidence for Neutrino Oscillations} 

In a modern theoretical context, one generally expects nonzero neutrino masses
and associated lepton mixing.  Experimentally, there has been accumulating
evidence for such masses and mixing.  All solar neutrino experiments
(Homestake, Kamiokande, SuperKamiokande, SAGE, GALLEX and SNO) 
show a significant
deficit in the neutrino fluxes coming from the Sun~\cite{sol}. This deficit
can be explained by oscillations of the $\nu_e$'s into other weak
eigenstate(s), with $\Delta m^2_{sol}$ of the order $10^{-5}$ eV$^2$ for
solutions involving the Mikheev-Smirnov-Wolfenstein (MSW) resonant matter
oscillations~\cite{wolf}-\cite{ms} 
or of the order of $10^{-10}$ eV$^2$ for vacuum
oscillations.  Accounting for the data with vacuum oscillations (VO) requires
almost maximal mixing.  The MSW solutions include one for small mixing angle
(SMA) and one for large mixing angle (LMA).

Another piece of evidence for neutrino oscillations is the atmospheric neutrino
anomaly, observed by Kamiokande~\cite{kam}, IMB~\cite{imb}, SuperKamiokande
~\cite{sk} with the highest statistics, and by Soudan~\cite{soudan2} and MACRO
~\cite{macro}.  These data can be fit by the inference of $\nu_{\mu} \rightarrow
\nu_x$ oscillations with $\Delta m^2_{atm}\sim 3.5 \times 10^{-3}$ eV$ ^2$
~\cite{sk} and maximal mixing $\sin^2 2 \theta_{atm} = 1$.  The identification
$\nu_x = \nu_\tau$ is preferred over $\nu_x=\nu_{sterile}$, and the
identification $\nu_x=\nu_e$ is excluded by both the Superkamiokande data and
the Chooz experiment~\cite{chooz}.

In addition, the LSND experiment~\cite{lsnd} has reported 
$\bar\nu_\mu \to \bar \nu_e$ and $\nu_{\mu} \to \nu_e$ oscillations with
$\Delta m^2_{LSND} \sim 0.1 - 1$ eV$^2$ and a range of possible mixing angles.
This result is not confirmed, but also not completely ruled out, by a similar
experiment, KARMEN~\cite{karmen}.  The miniBOONE experiment at Fermilab is
designed to resolve this issue, as discussed below.

If one were to try to fit all of these experiments, then, since they involve
three quite different values of $\Delta m^2_{ij}=m(\nu_i)^2-m(\nu_j)^2$, which
could not satisfy the identity for three neutrino species, 
\begin{equation}
\Delta m^2_{32} + \Delta m^2_{21} + \Delta m^2_{13}=0
\label{mident}
\end{equation}
it would follow that one would have to introduce further neutrino(s).
Since one knows from the measurement of the $Z$ width that there are
only three leptonic weak doublets with associated light neutrinos, it
follows that such further neutrino weak eigenstate(s) would have to be
electroweak singlet(s) (``sterile'' neutrinos).  Because the LSND
experiment has not been confirmed by the KARMEN experiment, we choose
here to use only the (confirmed) solar and atmospheric neutrino data
in our analysis, and hence to work in the context of three active
neutrino weak eigenstates.

\subsection{Neutrino Oscillation Formalism} 

     In this theoretical context, consistent with solar and atmospheric data,
there are three electroweak-doublet neutrinos and the neutrino mixing 
matrix is described by,

\begin{equation}
U=\left(
\begin{array}{ccc}
c_{12} c_{13}&c_{13} s_{12}&s_{13} e^{-i\delta}\cr
-c_{23}s_{12}-s_{13}s_{23}c_{12}e^{i\delta}
&c_{12}c_{23}-s_{12}s_{13}s_{23}e^{i\delta}&c_{13}s_{23}\cr
s_{12}s_{23}-s_{13}c_{12}c_{23}e^{i\delta}
&-s_{23}c_{12}-s_{12}c_{23}s_{13}e^{i\delta}&c_{13}c_{23}
\end{array}
\right)K^\prime
\end{equation}
where $c_{ij}=\cos\theta_{ij}$, $s_{ij}=\sin\theta_{ij}$, and $K^\prime =
diag(1,e^{i\phi_1},e^{i\phi_2})$.  The phases $\phi_1$ and $\phi_2$ do not
affect neutrino oscillations.  Thus, in this framework, the neutrino mixing
relevant for neutrino oscillations depends on the four angles $\theta_{12}$,
$\theta_{13}$, $\theta_{23}$, and $\delta$, and on two independent differences
of squared masses, $\Delta m^2_{atm.}$, which is $\Delta m^2_{32} =
m(\nu_3)^2-m(\nu_2)^2$ in the favored fit, and $\Delta m^2_{sol.}$, which may
be taken to be $\Delta m^2_{21}=m(\nu_2)^2- m(\nu_1)^2$.  Note that these
quantities involve both magnitude and sign; although in a two-species neutrino
oscillation in vacuum the sign does not enter, in the 
three-species-oscillation, that includes  both matter effects and CP violation,
the signs of the $\Delta m^2$ quantities enter and can, in principle, be
measured.

For our later discussion it will be useful to record the formulas for the
various neutrino-oscillation transitions.  In the absence of any matter effect, the probability that a (relativistic) weak neutrino eigenstate
$\nu_a$ becomes $\nu_b$ after propagating a distance $L$ is
\begin{eqnarray}
P(\nu_a \to \nu_b) &=& \delta_{ab} - 4 \sum_{i>j=1}^3
Re(K_{ab,ij}) \sin^2  \Bigl ( \frac{\Delta m_{ij}^2 L}{4E} \Bigr ) +
\nonumber\\&+& 4 \sum_{i>j=1}^3 Im(K_{ab,ij})
 \sin \Bigl ( \frac{\Delta m_{ij}^2 L}{4E} \Bigr )
\cos \Bigl ( \frac{\Delta m_{ij}^2 L}{4E} \Bigr )
\label{pab}
\end{eqnarray}
where
\begin{equation}
K_{ab,ij} = U_{ai}U^*_{bi}U^*_{aj} U_{bj}
\label{k}
\end{equation}
and
\begin{equation}
\Delta m_{ij}^2 = m(\nu_i)^2-m(\nu_j)^2
\label{delta}
\end{equation}
Recall that in vacuum, CPT invariance implies
$P(\bar\nu_b \to \bar\nu_a)=P(\nu_a \to \nu_b)$ and hence, for $b=a$,
$P(\bar\nu_a \to \bar\nu_a) = P(\nu_a \to \nu_a)$.  For the
CP-transformed reaction $\bar\nu_a \to \bar\nu_b$ and the T-reversed
reaction $\nu_b \to \nu_a$, the transition probabilities are given by the
right-hand side of (\ref{pab}) with the sign of the imaginary term reversed.
(Below we shall assume CPT invariance, so that CP violation is equivalent to T
violation.) 

In most cases there is only one mass scale
relevant for long-baseline neutrino oscillations, $\Delta m^2_{atm} \sim {\rm
few} \times 10^{-3}$ eV$^2$, and one possible neutrino mass spectrum is the
hierarchical one 
\begin{equation} 
\Delta m^2_{21}
= \Delta m^2_{sol} \ll \Delta m^2_{31} \approx \Delta m^2_{32}=\Delta m^2_{atm}
\label{hierarchy}
\end{equation}
In this case, CP (T) violation effects are negligibly small, so that in
vacuum
\begin{equation}
P(\bar\nu_a \to \bar\nu_b) = P(\nu_a \to \nu_b)
\label{pcp}
\end{equation}
and
\begin{equation}
P(\nu_b \to \nu_a) = P(\nu_a \to \nu_b)
\label{pt}
\end{equation}
In the absence of T violation, the second equality (\ref{pt}) would still hold
in uniform matter, but even in the absence of CP violation, the first equality
(\ref{pcp}) would not hold.  With the hierarchy (\ref{hierarchy}), the
expressions for the specific oscillation transitions are
\begin{eqnarray}
P(\nu_\mu \to \nu_\tau) & = & 4|U_{33}|^2|U_{23}|^2
\sin^2 \Bigl ( \frac{\Delta m^2_{atm}L}{4E} \Bigr ) \cr\cr
& = & \sin^2(2\theta_{23})\cos^4(\theta_{13})
\sin^2 \Bigl (\frac{\Delta m^2_{atm}L}{4E} \Bigr )
\label{pnumunutau}
\end{eqnarray}
\begin{eqnarray}
P(\nu_e \to \nu_\mu) & = & 4|U_{13}|^2 |U_{23}|^2
\sin^2 \Bigl ( \frac{\Delta m^2_{atm}L}{4E} \Bigr ) \cr\cr
& = & \sin^2(2\theta_{13})\sin^2(\theta_{23})
\sin^2 \Bigl (\frac{\Delta m^2_{atm}L}{4E} \Bigr )
\label{pnuenumu}
\end{eqnarray}

\begin{eqnarray}
P(\nu_e \to \nu_\tau) & = & 4|U_{33}|^2 |U_{13}|^2
\sin^2 \Bigl ( \frac{\Delta m^2_{atm}L}{4E} \Bigr ) \cr\cr
& = & \sin^2(2\theta_{13})\cos^2(\theta_{23})
\sin^2 \Bigl (\frac{\Delta m^2_{atm}L}{4E} \Bigr )
\label{pnuenutau}
\end{eqnarray}

In neutrino oscillation searches using reactor antineutrinos,
i.e. tests of $\bar\nu_e \to \bar\nu_e$, the two-species mixing hypothesis used
to fit the data is
\beqs 
P(\nu_e \to \nu_e) & = & 1 - \sum_x P(\nu_e \to \nu_x) \cr\cr
                   & = & 1 - \sin^2(2\theta_{reactor})
\sin^2 \Bigl (\frac{\Delta m^2_{reactor}L}{4E} \Bigr )
\label{preactor}
\eeqs
where $\Delta m^2_{reactor}$ is the squared mass difference relevant for
$\bar\nu_e \to \bar\nu_x$.  In particular, in the upper range of values of
$\Delta m^2_{atm}$, since the transitions $\bar\nu_e \to \bar\nu_\mu$ and
$\bar\nu_e \to \bar\nu_\tau$ contribute to $\bar\nu_e$ disappearance, one has
\begin{equation}
P(\nu_e \to \nu_e) = 1 - \sin^2(2\theta_{13})\sin^2 \Bigl
(\frac{\Delta m^2_{atm}L}{4E} \Bigr )
\label{preactoratm}
\end{equation}
i.e., $\theta_{reactor}=\theta_{13}$, and, for the value $|\Delta m^2_{32}| = 3
\times 10^{-3}$ eV$^2$ from SuperK, the CHOOZ experiment on $\bar\nu_e$ disappearance
yields the upper limit~\cite{chooz}
\begin{equation}
\sin^2(2\theta_{13}) < 0.1
\label{chooz}
\end{equation}
which is also consistent with conclusions from the SuperK data analysis
~\cite{sk}.

Further, the quantity ``$\sin^2(2\theta_{atm})$'' often used to fit
the data on atmospheric neutrinos with a simplified two-species mixing
hypothesis, is, in the three-generation case,
\begin{equation}
\sin^2(2\theta_{atm}) \equiv \sin^2(2\theta_{23})\cos^4(\theta_{13})
\label{thetaatm}
\end{equation}
The SuperK experiment finds that the best fit to their data is to infer
$\nu_\mu \to \nu_\tau$ oscillations with maximal mixing, and hence
$\sin^2(2\theta_{23})=1$ and $|\theta_{13}| << 1$.  The various solutions of
the solar neutrino problem involve quite different values of $\Delta m^2_{21}$
and $\sin^2(2\theta_{21})$: (i) large mixing angle solution, LMA: $\Delta
m^2_{21} \simeq {\rm few} \times 10^{-5}$ eV$^2$ and $\sin^2(2\theta_{21})
\simeq 0.8$; (ii) small mixing angle solution, SMA: $\Delta m^2_{21} \sim
10^{-5}$ and $\sin^2(2\theta_{21}) \sim 10^{-2}$, (iii) LOW: $\Delta m^2_{21}
\sim 10^{-7}$, $\sin^2(2\theta_{21}) \sim 1$, and (iv) ``just-so'': $\Delta
m^2_{21} \sim 10^{-10}$, $\sin^2(2\theta_{21}) \sim 1$.  The SuperK experiment
favors the LMA solutions~\cite{sol}; for other global fits, see, e.g.,
Gonzalez-Garcia et al.~\cite{sol}. 

We have reviewed the three neutrino oscillation phenomenology that is
consistent with solar and atmospheric neutrino oscillations. In what
follows, we will examine the neutrino experiments planned for the
immediate future that will address some of the relevant physics. We
will then review the physics potential of the Neutrino Factory.

\subsection{Relevant Near- and Mid-Term Experiments} 

There are currently intense efforts to confirm and extend the evidence for
neutrino oscillations in all of the various sectors -- solar, atmospheric, and
accelerator.  Some of these experiments are running; in addition to
SuperKamiokande and Soudan-2, these include the Sudbury Neutrino Observatory,
SNO, and the K2K long baseline experiment between KEK and Kamioka.  Others are
in development and testing phases, such as 
miniBOONE, MINOS, the CERN - Gran Sasso
program, KamLAND, and Borexino~\cite{anl}.  Among the long baseline neutrino
oscillation experiments, the approximate distances are $L \simeq 250$ km for
K2K, 730 km for both MINOS (from Fermilab to Soudan) and the proposed CERN-Gran
Sasso experiments.  

K2K is a $\nu_\mu$ disappearence experiment with a
conventional neutrino beam having a mean energy of about 1.4 GeV, going from
KEK 250 km to the SuperK detector.  It has a near detector for beam
calibration.  It has obtained results consistent with the SuperK experiment,
and has reported that its data disagree by $2\sigma$ with the no-oscillation
hypothesis~\cite{k2k}.  

MINOS is another conventional neutrino beam experiment
that takes a beam from Fermilab 730 km to a detector in the Soudan mine in
Minnesota.  It again uses a near detector for beam flux measurements and has
opted for a low-energy configuration, with the flux peaking at about 3 GeV.
This experiment is scheduled to start taking data in early 2004 and, after some
years of running, to obtain higher statistics than the K2K experiment and to
achieve a sensitivity down to the level $|\Delta m^2_{32}| \sim
10^{-3} $eV$^2$.  

The CERN - Gran Sasso program will come on later, around
2005.  It will use  a higher-energy neutrino beam from CERN to the
Gran Sasso deep underground laboratory in Italy.  This program will emphasize
detection of the $\tau$'s produced by the $\nu_\tau$'s that result from the
inferred neutrino oscillation transition $\nu_\mu \to \nu_\tau$.  The OPERA
experiment will do this using emulsions~\cite{opera}, while the ICARUS proposal
uses a liquid argon chamber~\cite{icanoe}.  

Plans for the Japan Hadron Facility (JHF),
also called the High Intensity Proton Accelerator (HIPA), include the use of 
a 1 MW
proton driver to produce a high-intensity conventional neutrino beam with a
pathlength 300 km to the SuperK detector~\cite{jhf}.  Moreover, at Fermilab,
the miniBOONE experiment is scheduled 
to start data taking in the near future and to confirm or
refute the LSND claim after a few years of running.

There are several relevant solar neutrino experiments.  The SNO
experiment is currently running and has recently reported their first
results that confirm solar neutrino oscillations~\cite{snolatest}.
These involve measurement of the solar neutrino flux and energy
distribution using the charged current reaction on heavy water, $\nu_e
+ d \to e + p + p$.  They are expected to report on the neutral
current reaction $\nu_e + d \to \nu_e + n + p$ shortly. The neutral
current rate is unchanged in the presence of oscillations that involve
standard model neutrinos, since the neutral current channel is equally
sensitive to all the three neutrino species.  If however, sterile
neutrinos are involved, one expects to see a depletion in the neutral
current channel also.

The KamLAND experiment in Japan is scheduled to  begin
taking data in late 2001.  This is a reactor antineutrino experiment using
baselines of  100 - 250 km and will search for $\bar\nu_e$ disappearance
and is sensitive to the solar neutrino oscillation scale.
On a similar time scale, the Borexino experiment in Gran Sasso is scheduled 
to turn
on and  measure the $^7$Be neutrinos from the sun.  These experiments
should help us determine which of the various solutions to the solar neutrino
problem is preferred, and hence the corresponding values of $\Delta m^2_{21}$
and $\sin^2(2\theta_{12})$.

This, then, is the program of relevant experiments during the period
2000-2010.  By the end of this period, we may expect that much will be learned
about neutrino masses and mixing.  However, there will remain several
quantities that will not be well measured and which can be measured by a
neutrino factory. 

\subsection{Oscillation Experiments at a Neutrino Factory } 
\label{neuf}
Although a neutrino factory based on a muon storage ring will turn on
several years after this near-term period in which K2K, MINOS, and the
CERN-Gran Sasso experiments will run, we believe that it has a
valuable role to play, given the very high-intensity neutrino beams of
fixed flavor-pure content, including, uniquely, $\nu_e$ and
$\bar\nu_e$ beams in addition to $\nu_\mu$ and $\bar\nu_\mu$ beams. A
conventional positive charge selected neutrino beam  is primarily
$\nu_\mu$ with some admixture of $\nu_e$'s and other flavors from $K$
decays(O(1\%) of the total charged current rate) and  the fluxes of these
neutrinos can only be fully understood after measuring the charged
particle spectra from the target with high accuracy.  In contrast,
the potential of
the neutrino beams from a muon storage ring is that the neutrino beams
would be of extremely high purity: $\mu^-$ beams would yield 50 \%
$\nu_\mu$ and 50 \% $\bar\nu_e$, and $\mu^+$ beams, the charge
conjugate neutrino beams.  Furthermore, these could be produced with
high intensities and low divergence that make it possible to go 
to longer baselines.

In what follows, we shall take the design values from Study-II
of $10^{20}$ $\mu$ decays per ``Snowmass year'' ($10^7$ sec) as being typical.
The types of neutrino oscillations that can be searched for with the Neutrino
Factory based on the muon storage ring are listed in 
table~\ref{tab:nu-osc-ratings} for the 
case of $\mu^-$ which decays to $ \nu_\mu e^- \bar\nu_e$: 
\begin{table}
\caption[Neutrino Oscillation Modes]{Neutrino-oscillation modes that can be studied with conventional
neutrino beams or with beams from a Neutrino Factory, with ratings as to
degree of difficulty in each case; * = well or easily measured, $\surd$ =
measured poorly or with difficulty, --- = not
measured.\label{tab:nu-osc-ratings}}
\begin{center}
\begin{tabular}{|llll|}
\hline
& & Conventional & Neutrino \\
\raisebox{1.5ex}[0pt]{Measurement } & \raisebox{1.5ex}[0pt]{Type} & beam &
Factory \\
\hline
$\nu_\mu\to\nu_\mu,\,\nu_\mu\to\mu^-$ & survival & $\surd$ & *\\
$\nu_\mu\to\nu_e,\,\nu_e\to e^-$ & appearance & $\surd$ & $\surd$\\
$\nu_\mu\to\nu_\tau,\,\nu_\tau\to\tau^-,\,\tau^-\to(e^-,\mu^-)...$  &
appearance & $\surd$ & $\surd$ \\ \hline
$\bar \nu_e\to\bar \nu_e,\,\bar\nu_e\to e^+$ & survival  & --- & $*$\\
$\bar\nu_e\to\bar\nu_\mu,\,\bar\nu_\mu\to\mu^+$  & appearance & --- &  $*$
\\
$\bar\nu_e\to\bar\nu_\tau,\,\bar\nu_\tau\to\tau^+,\,\tau^+\to(e^+,\mu^+)...$
& appearance & ---& $\surd$ \\
\hline
\end{tabular}
\end{center}
\end{table}

It is clear from the 
processes listed that since the beam contains both neutrinos and
antineutrinos, the only way to determine  the flavor of the 
parent neutrino  is to
determine the identity of the final state charged lepton and measure its
charge.  

A capability unique to the Neutrino Factory will be the 
measurement of the oscillation $\bar\nu_e \to \bar\nu_\mu$,
giving a wrong-sign $\mu^+$.  Of greater difficulty would be the measurement of
the transition $\bar\nu_e \to \bar\nu_\tau$, giving a $\tau^+$ which will decay
part of the time to $\mu^+$.  These physics goals mean that a detector must
have excellent capability to identify muons and measure their charges.
Especially in a steel-scintillator detector, the oscillation $\nu_\mu \to
\nu_e$ would be difficult to observe, since it would be 
difficult to distinguish
an electron shower from a hadron shower.  From the above formulas for
oscillations, one can see that, given a knowledge of $|\Delta m^2_{32}|$ and
$\sin^2(2\theta_{23})$ that one will have by the time a neutrino factory is
built, the measurement of the $\bar\nu_e \to \bar\nu_\mu$ transition yields the
value of $\theta_{13}$.

To get a rough idea of how the sensitivity of 
an oscillation experiment would scale with energy and baseline length, 
recall that the event rate in the absence of oscillations is 
simply the neutrino flux times the cross section.  
First of all, neutrino cross sections in the region above
about 10 GeV (and slightly higher, for $\tau$ production) grow linearly with
the neutrino energy.  Secondly, the beam divergence is
a function of the initial muon storage ring energy; 
this divergence yields a flux, as a
function of $\theta_d$, the angle of deviation from the forward direction, that
goes like $1/\theta_d^2 \sim E^2$.  Combining this with the linear $E$
dependence of the neutrino cross section 
and the overall $1/L^2$ dependence of the flux far from the
production region, one finds that the event rate goes like 
\begin{equation} 
\frac{dN}{dt} \sim \frac{E^3}{L^2}
\label{eventrate}
\end{equation}
We base our discussion on the event rates given in the 
Fermilab Neutrino Factory study~\cite{INTRO:ref9}. For
a stored muon energy of 20 GeV,  and a distance of 
$L=2900$ to the WIPP Carlsbad site in New Mexico, these event rates amount to 
several thousand events per kton of detector per year, i.e. they are
satisfactory for the physics program.   This is also true for the other
pathlengths under consideration, namely $L=2500$ km from BNL to Homestake and
$L=1700$ km to Soudan.  A usual racetrack design would only allow a single
pathlength $L$, but a bowtie design could allow two different pathlengths
(e.g.,~\cite{zp}). 

\begin{figure}[tbh!]
\centerline{\includegraphics[width=4.0in]{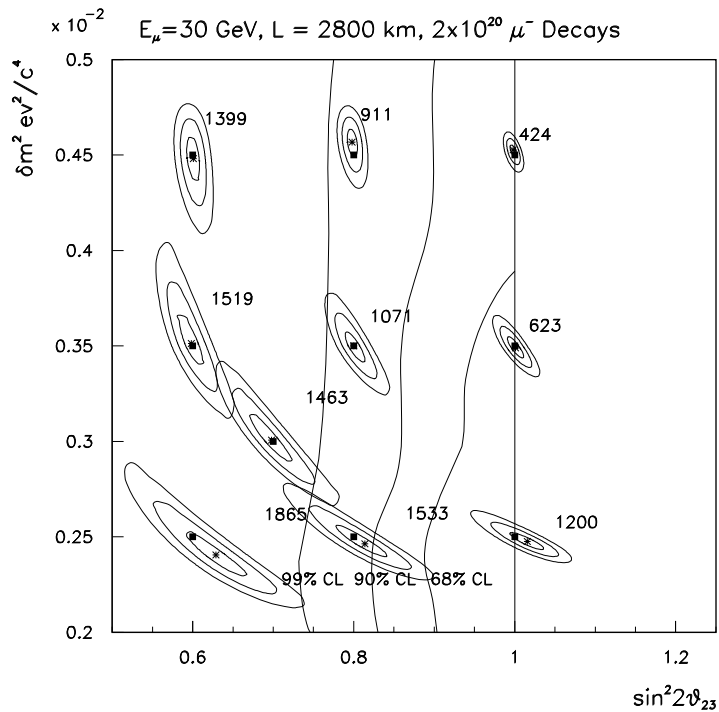}}
\bigskip
\caption[Error ellipses in  $\delta m^2$ sin$^2 2\theta$ space for a neutrino factory]
{ \label{fig:30gev_disap_fit}
Fit to muon neutrino survival distribution for $E_\mu=30$ GeV and $L=2800$~km for 10
pairs of sin$^2 2\theta$, $\delta m^2$ values. For each fit, the
1$\sigma$,\ 2$\sigma$
and 3$\sigma$ contours are shown. The generated points are indicated by the
dark
rectangles and the fitted values by stars. The SuperK 68\%, 90\%, and 99\% 
confidence
levels are superimposed. Each point is labelled by the predicted number of 
signal events for that point.}
\end{figure}

One could estimate that at a time when the neutrino factory turns on, $|\Delta
m^2_{32}|$ and $\sin^2(2\theta_{23})$ would be known at perhaps the 20 \% level
(we emphasize that future projections such as this are obviously uncertain).
The neutrino factory will significantly improve precision in these parameters,
as can be seen from figure~\ref{fig:30gev_disap_fit} which shows the error ellipses possible for a 30~GeV muon storage ring.
In addition the neutrino factory can contribute to the measurement of: (i) 
$\theta_{13}$, as discussed above; (ii) measurement of the sign of $\Delta
m^2_{32}$ using matter effects; and (iii) possibly a measurement of CP
violation in the leptonic sector, if $\sin^2(2\theta_{13})$,
$\sin^2(2\theta_{21})$, and $\Delta m^2_{21}$ are sufficiently large.  To
measure the sign of $\Delta m^2_{32}$, one uses the fact that matter effects
reverse sign when one switches from neutrinos to antineutrinos, and carries out
this switch in the charges of the stored $\mu^\pm$.  We elaborate on this next.

\subsection{Matter Effects} 

With the advent of the muon storage ring, the distances at which one 
can place detectors
are large enough so that for the first time matter effects can be exploited in
accelerator-based oscillation experiments.  Simply put, matter effects are the
matter-induced oscillations which neutrinos undergo along their flight path
through the Earth from the source to the detector.  Given the typical density
of the earth, matter effects are important for the neutrino energy range $E
\sim O(10)$ GeV and $\Delta m^2_{32} \sim 10^{-3}$ eV$^2$ values relevant for
the long baseline experiments. Matter effects in neutrino propagation were 
first pointed out by Wolfenstein~\cite{wolf} and Mikheyev and Smirnov~\cite{ms}. 
(See the papers~\cite{dgh}--\cite{cpv} for details  of the MSW effect
and its relevance to neutrino factories.) In brief,
the transition
probabilities
in the leading oscillation approximation for propagation through
matter of constant density are
\begin{eqnarray}
P(\nu_e\to \nu_\mu) &=& s_{23}^2 \sin^2 2\theta_{13}^m
\sin^2\Delta_{32}^m \,,
\nonumber\\
P(\nu_e\to \nu_\tau) &=& c_{23}^2 \sin^2 2\theta_{13}^m
\sin^2\Delta_{32}^m \,,
\label{eq:probs}\\
P(\nu_\mu\to \nu_\tau) &=& \sin^2 2\theta_{23} \left[
(\sin\theta_{13}^m)^2 \sin^2\Delta_{21}^m +(\cos\theta_{13}^m)^2
\sin^2\Delta_{31}^m -(\sin\theta_{13}^m\cos\theta_{13}^m)^2
\sin^2\Delta_{32}^m \right]
\,. \nonumber
\end{eqnarray}
 The oscillation arguments are given by
\begin{equation}
\Delta_{32}^m = \Delta_0 S \,,\qquad
\Delta_{31}^m = \Delta_0 {1\over2} \left[ 1+{A\over\delta m^2_{32}}+S \right]
\,, \qquad
\Delta_{21}^m = \Delta_0 {1\over2} \left[ 1+{A\over\delta m^2_{32}}-S \right]
\,, \label{eq:arg}
\end{equation}
where $S$ is given by
\begin{equation}
S \equiv \sqrt{ \left( {A\over\delta m^2_{32}}-\cos2\theta_{13}
\right)^2 + \sin^2 2\theta_{13}} \,,
\label{eq:S}
\end{equation}
and
\begin{equation}
\Delta_0 = {\delta m^2_{32} L\over 4E} = 1.267 {\delta m^2_{32}
{\rm\,(eV^2)} \; L {\rm\ (km)} \over E_\nu {\rm\ (GeV)}}
\,,
\label{eq:arg0}
\end{equation}
\begin{equation}
\sin^2 2\theta_{13}^m = {\sin^2 2\theta_{13}\over
\left({A\over\delta m^2_{32}} - \cos 2\theta_{13} \right)^2
+ \sin^2 2\theta_{13}} \,. \label{eq:sin}
\end{equation}
The amplitude $A$ for $\nu_e e$ forward scattering in matter is given
by
\begin{equation}
A = 2\sqrt2 G_F N_e E_\nu = 1.52 \times 10^{-4}{\rm\,eV^2} Y_e
\rho({\rm\,g/cm^3}) E({\rm\,GeV}) \,.
\label{eq:A}
\end{equation}
Here $Y_e$ is the electron fraction and $\rho(x)$ is the matter
density. For neutrino trajectories that pass through the earth's
crust, the average density is typically of order 3~gm/cm$^3$ and $Y_e
\simeq 0.5$.  The oscillation probability $P(\nu_e\to \nu_\mu)$ is
directly proportional to $\sin^2 2\theta_{13}^m$, which is
approximately proportional to $\sin^2 2\theta_{13}$. There is a
resonant enhancement for
\begin{equation}
  \cos2\theta_{13} = \frac{A}{\delta m^2_{32}} \,.
\end{equation}
For electron neutrinos, $A$ is positive and the resonance enhancement
occurs for positive values of $\delta m^2_{32}$ for $\cos2\theta_{13}>0$.
 The reverse is true
for electron anti-neutrinos and the enhancement occurs for negative
values of $\delta m^2_{32}$.  Thus for a neutrino factory operating
with positive stored muons (producing a $\nu_e$ beam) one expects an
enhanced production of opposite sign ($\mu^-$) charged-current events
as a result of the oscillation $\nu_e\to \nu_\mu$ if $\delta m^2_{32}$
is positive and vice versa for stored negative
beams.

\begin{figure}[tbh!]
\centerline{\includegraphics[width=4.0in]{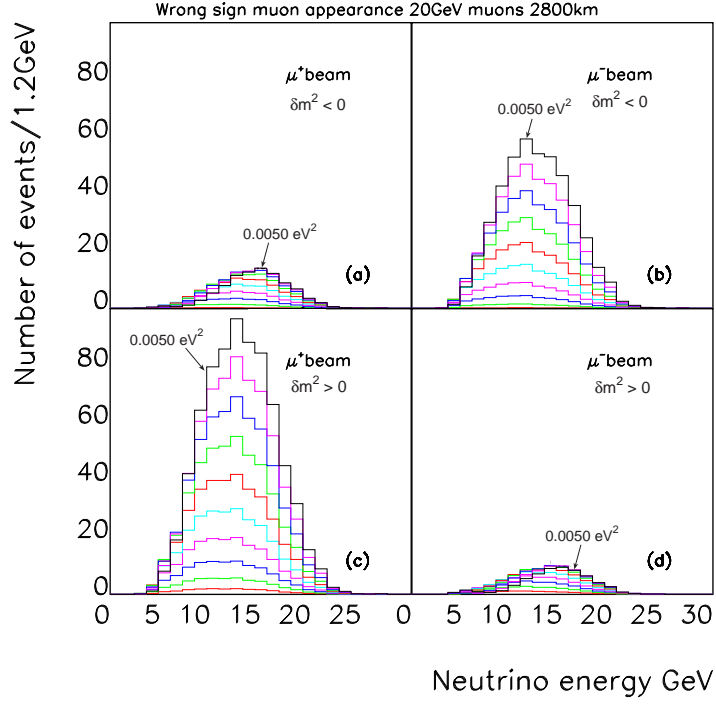}}
\caption[Wrong sign muon appearance rates and sign of  $\delta m^2_{32}$]
{The wrong sign muon appearance rates for a 20 GeV muon storage ring at
a baseline of 2800~km with 10$^{20}$ decays and a 50 kiloton detector
for (a)~$\mu^+$ stored and negative $\delta m^2_{32}$\,, (b)~$\mu^-$ stored
and negative $\delta m^2_{32}$\,, (c)~$\mu^+$ stored and positive $\delta
m^2_{32}$\,,
(d)~$\mu^-$ stored and positive $\delta m^2_{32}$. The values of $|\delta
m^2_{32}|$ range from 0.0005 to 0.0050 eV$^2$ in steps of 0.0005~eV$^2$.  
Matter enhancements are evident in (b) and (c).
\label{fig:hists}}
\end{figure}

Figure~\ref{fig:hists}~\cite{barger-raja} shows the wrong-sign 
muon appearance spectra 
 as function of $\delta m^2_{32}$ for both $\mu^+$ and
$\mu^-$ beams for both signs of $\delta m^2_{32}$ at a baseline of
2800~km. The resonance enhancement in wrong sign muon production is
clearly seen in Fig.~\ref{fig:hists} (b) and (c).

\begin{figure}[tbh!]
\centerline{\includegraphics[width=4.0in]{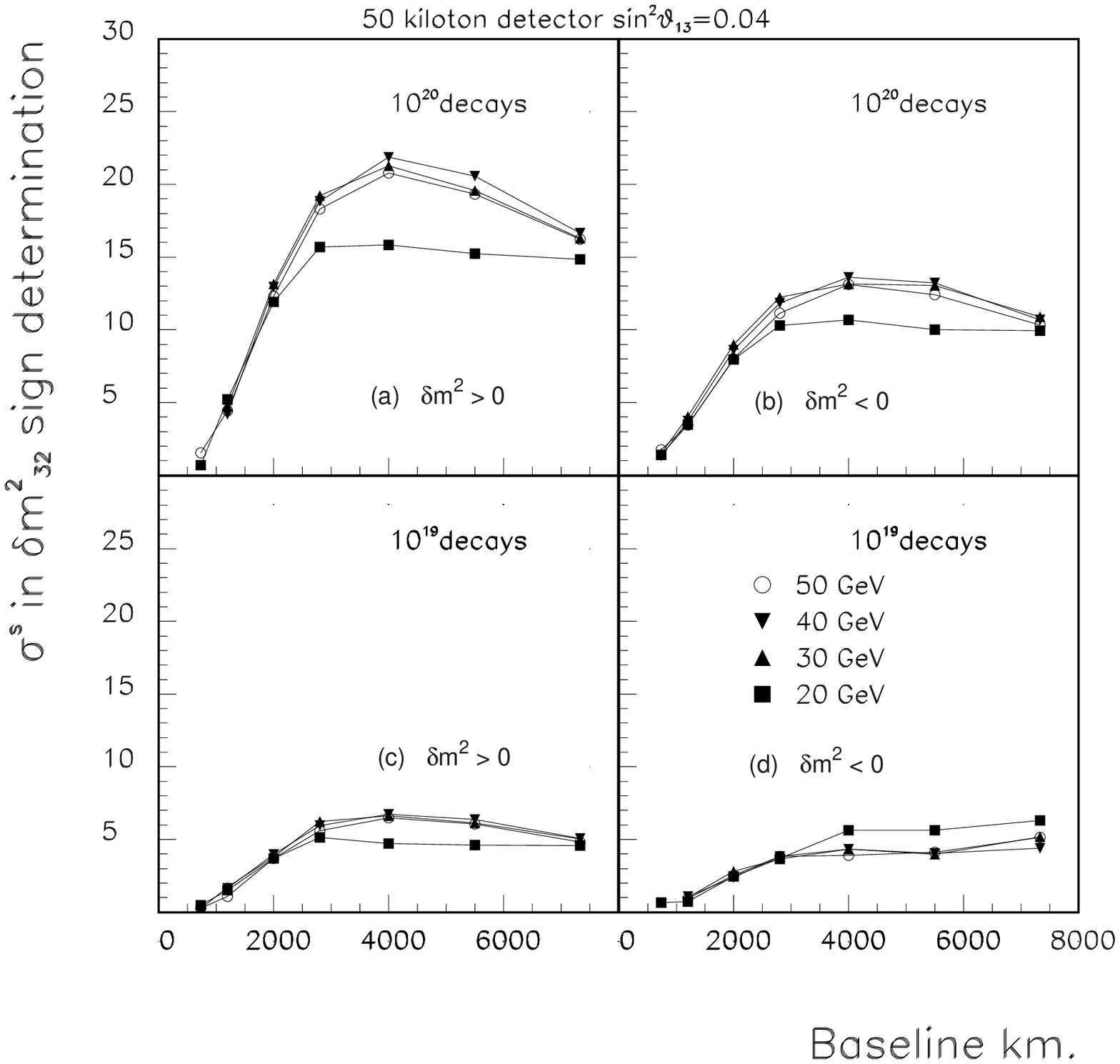}}
\caption[$\delta m_{32}^2$ sign determination at a Neutrino Factory]
{The statistical significance (number of standard deviations) 
with which the
sign of $\delta m_{32}^2$ can be determined versus baseline length for
various muon storage ring energies. The results are shown for 
a 50~kiloton detector, and (a)~10$^{20}$
$\mu^+$ and $\mu^-$ decays and positive values of $\delta m_{32}^2$;
(b)~10$^{20}$ $\mu^+$ and $\mu^-$ decays and
negative values of $\delta m_{32}^2$; (c)~10$^{19}$ $\mu^+$ and 
$\mu^-$ decays and positive values of $\delta
m_{32}^2$; (d)~10$^{19}$ $\mu^+$ and $\mu^-$
decays and negative values of $\delta m_{32}^2$.
\label{fig:sigmas}}
\end{figure}

By comparing these (using first a stored $\mu^+$ beam and then a stored $\mu^-$
beam) one can thus determine the sign of $\Delta m^2_{32}$ as well as the value
of $\sin^2(2\theta_{13})$.  
Figure~\ref{fig:sigmas}~\cite{barger-raja} shows the difference in negative
log-likelihood between a
correct and wrong-sign mass hypothesis expressed as a number of
equivalent Gaussian standard deviations versus baseline length for
muon storage ring energies of 20, 30, 40 and 50~GeV. The values of the
oscillation parameters are for the LMA scenario with 
$\sin^22\theta_{13}=0.04$. 
Figure~\ref{fig:sigmas}(a) is for 10$^{20}$ decays
for each sign of stored energy and a 50 kiloton detector and positive
$\delta m^2_{32}$ , (b) is for negative $\delta m^2_{32}$ for various
values of stored muon energy. Figures~\ref{fig:sigmas} ;(c) and (d)
show the corresponding curves for 10$^{19}$ decays and a 50 kiloton
detector. An entry-level machine would permit one to perform a
5$\sigma$ differentiation of the sign of $\delta m^2_{32}$ at a
baseline length of $\sim$2800~km.

For the Study II design, in accordance with the previous
Fermilab study~\cite{INTRO:ref9}, 
one estimates  that it is possible to determine the sign of $\delta m^2_{32}$
even if 
$\sin^2(2\theta_{13})$ is as small as $\sim 10^{-3}$.

\subsection{CP Violation}

CP violation is measured by the (rephasing-invariant) Jarlskog product
\beqs
J & =& Im(U_{ai}U_{bi}^* U_{aj}^* U_{bj}) \cr\cr
& & = \frac{1}{8}
\sin(2\theta_{12})\sin(2\theta_{13})\cos(\theta_{13})\sin(2\theta_{23})\sin
\delta 
\eeqs
Leptonic CP violation also requires that each of the leptons in each charge
sector be nondegenerate with any other leptons in this sector; this is,
course, true of the charged lepton sector and, for the neutrinos, this requires
$\Delta m^2_{ij} \ne 0$ for each such pair $ij$.  In the quark sector, $J$ is 
known to be small: $J_{CKM} \sim O(10^{-5})$.  
A promising asymmetry to measure is $P(\nu_e \to \nu_\mu)-P(\bar\nu_e - 
\bar\nu_\mu)$.  As an illustration, in the absence of matter effects, 
\beqs
P(\nu_e \to \nu_\mu) - P(\bar\nu_e \to \bar\nu_\mu) & = & -4J(\sin 2\phi_{32}+
\sin 2\phi_{21} + \sin 2\phi_{13}) \cr\cr
& = & -16J \sin \phi_{32} \sin \phi_{13} \sin \phi_{21} 
\label{pnuenumudif}
\eeqs
where
\begin{equation}
\phi_{ij} = \frac{\Delta m^2_{ij}L}{4E}
\label{phiijdef}
\end{equation}
In order for the CP violation in eq. (\ref{pnuenumudif}) to be large enough to
measure, it is necessary that $\theta_{12}$, $\theta_{13}$, and $\Delta
m^2_{sol} = \Delta m^2_{21}$ not be too small. From atmospheric neutrino data,
we have $\theta_{23}\simeq \pi/4$ and $\theta_{13} << 1$.  If LMA describes
solar neutrino data, then $\sin^2(2\theta_{12}) \simeq 0.8$, so $J \simeq
0.1\sin(2\theta_{13})\sin \delta$.  For example, if
$\sin^2(2\theta_{13})=0.04$, then $J$ could be $>> J_{CKM}$.  Furthermore, for
 parts of  the LMA phase space where  
$\Delta m^2_{sol} \sim 4 \times 10^{-5}$ eV$^2$ 
the CP violating effects might be observable. In the absence of matter, one
would measure the asymmetry
\begin{equation}
\frac{P(\nu_e \to \nu_\mu) - P(\bar\nu_e \to \bar\nu_\mu)}{
P(\nu_e \to \nu_\mu) + P(\bar\nu_e \to \bar\nu_\mu)} =
-\frac{\sin(2\theta_{12})\cot(\theta_{23})\sin\delta \sin \phi_{21}}{
\sin \theta_{13}}
\end{equation}
However, in order to optimize this ratio, because of the smallness of $\Delta
m^2_{21}$ even for the LMA, one must go to large pathlengths $L$, and here
matter effects are important.  These make leptonic CP violation challenging to
measure, because, even in the absence of any intrinsic CP violation, these
matter effects render the rates for $\nu_e \to \nu_\mu$ and $\bar\nu_e \to
\bar\nu_\mu$ unequal since the matter interaction is opposite in sign for $\nu$
and $\bar\nu$.  One must therefore subtract out the matter effects in order to
try to isolate the intrinsic CP violation.  Alternatively, one might think of
comparing $\nu_e \to \nu_\mu$ with the time-reversed reaction $\nu_\mu \to
\nu_e$.  Although this would be equivalent if CPT is valid, as we assume, and
although uniform matter effects are the same here, the detector response is
quite different and, in particular, it is quite difficult to identify $e^\pm$.
Results from SNO and KamLAND testing the LMA will help further planning.

\begin{figure}[tbh!]
\centerline{\includegraphics[width=4.0in]{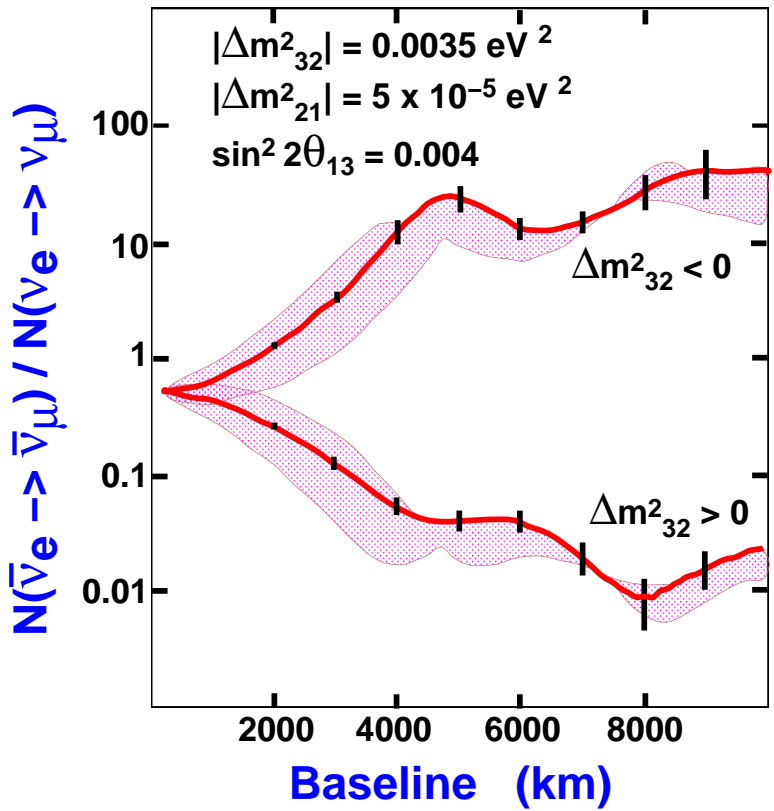}}
\bigskip
\caption[CP violation effects in a neutrino factory]
{ \label{cpfig}
Predicted ratios of wrong-sign muon event rates when positive and
negative muons are stored in a 20~GeV neutrino factory, shown as a
function of baseline.  A muon measurement threshold of 4~GeV is
assumed. The lower and upper bands correspond respectively to negatve
and positive $\delta m^2_{32}$. The widths of the bands show how the
predictions vary as the CP violating phase $\delta$ is varied from
$-\pi$/2 to $\pi$/2, with the thick lines showing the predictions for
$\delta$=0. The statistical error bars correspond to a
high-performance neutrino factory yielding a data sample of 10$^{21}$
decays with a 50~kiloton detector. Figure is based on calculations
presented in~\cite{barger-entry} }

\end{figure}

The Neutrino Factory provides an ideal set of controls to measure CP
violation effects since we can fill the storage ring with both $\mu^+$
and $\mu^-$ particles and measure the ratio of the number of events
$\anti\nu_e\rightarrow \anti\nu_\mu$/$\nu_e\rightarrow\nu_\mu$.
Figure~\ref{cpfig} shows this ratio for a Neutrino Factory with
10$^{21}$ decays and a 50~kilo-ton detector as a function of the
baseline length. The ratio depends on the sign of $\delta
m^2_{32}$. The shaded band around either curve shows the variation of
this ratio as a function of the CP violating phase $\delta$. The
number of decays needed to produce the error bars shown is directly
proportional to $sin^2\theta_{13}$, which for the present example is
set to 0.004. Depending on the magnitude of $J$, one may be driven to
build a Neutrino Factory just to understand CP violation in the lepton
sector, which could have a significant role in explaining the baryon
asymmetry of the Universe~\cite{yanag}.

\section{Physics Potential of Superbeams}

It is possible to extend the reach of the current conventional
neutrino experiments by enhancing the capabilities of the proton
sources that drive them. These enhanced neutrino beams have been
termed ``superbeams'' and form an intermediate step on the way to a
neutrino factory. Their capabilities have been explored in  recent
papers~\cite{superbeams,bargersuperbeam}. These articles consider
the capabilities of enhanced proton drivers at (i) the proposed
0.77~MW 50~GeV proton synchrotron at the Japan Hadron Facility
(JHF)~\cite{jhfloi}, (ii) a 4~MW upgraded version of the JHF, (iii) a
new $\sim 1$~MW 16~GeV proton driver~\cite{brighter} that would replace
the existing 8~GeV Booster at Fermilab, or (iv) a fourfold intensity
upgrade of the 120~GeV Fermilab Main Injector (MI) beam (to 1.6~MW)
that would become possible once the upgraded (16~GeV) Booster was
operational.  Note that the 4~MW 50~GeV JHF and the 16~GeV upgraded
Fermilab Booster are both suitable proton drivers for a neutrino
factory. The conclusions of both reports are that superbeams will
extend the reaches in the oscillation parameters of the current
neutrino experiments but ``the sensitivity at a neutrino factory to
$CP$ violation and the neutrino mass hierarchy extends to values of
the amplitude parameter $\sin^2 2\theta_{13}$ that are one to two
orders of magnitude lower than at a superbeam''~\cite{bargersuperbeam}.

\begin{figure}[tbh!]
\centerline{\includegraphics[width=4.0in]{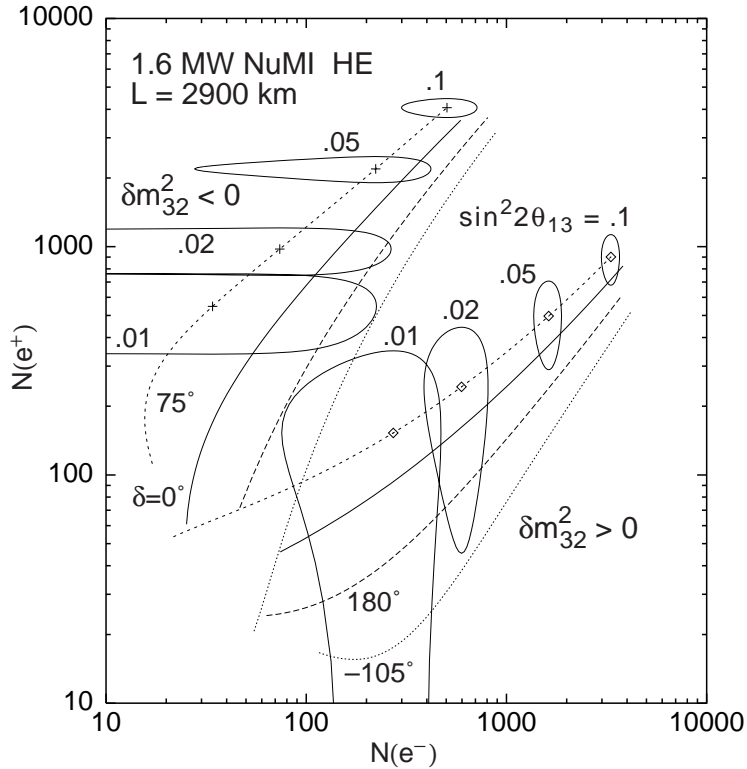}}
\caption[Error ellipses for superbeams for electron appearance]
{Three--sigma error ellipses in the 
$\left(N(e^+), N(e^-)\right)$--plane, shown for 
$\nu_\mu \to \nu_e$ and $\bar\nu_\mu \to \bar\nu_e$ oscillations 
in a NuMI--like 
high energy neutrino beam driven by a 1.6~MW proton driver. 
The calculation assumes a liquid argon detector with the parameters 
listed in \cite{superbeams}, a baseline of 2900~km, 
and 3~years of running with neutrinos, 6~years running 
with antineutrinos. 
Curves are shown for different CP--phases $\delta$ (as labelled), and 
for both signs of $\delta m^2_{32}$ with 
$|\delta m^2_{32}| = 0.0035$~eV$^2$, and 
the sub--leading scale $\delta m^2_{21} = 10^{-4}$~eV$^2$. 
Note that $\sin^22\theta_{13}$ varies along the curves from
0.0001 to 0.01, as indicated~\cite{bargersuperbeam}.
}
\label{fig:signdm2}
\end{figure}

To illustrate these points, we choose one of the most favorable superbeam 
scenarios 
studied: a 1.6~MW NuMI--like high energy beam with $L = 2900$~km, detector 
parameters corresponding to the liquid argon scenario in 
~\cite{bargersuperbeam}, and oscillation parameters 
$|\delta m^2_{32}| = 3.5 \times 10^{-3}$~eV$^2$ and 
$\delta m^2_{21} = 1 \times 10^{-4}$~eV$^2$. 
The calculated three--sigma error ellipses in the 
$\left(N(e^+), N(e^-)\right)$--plane are shown in Fig.~\ref{fig:signdm2}
for both signs of $\delta m^2_{32}$, with the curves corresponding to 
various CP--phases $\delta$ (as labelled). The magnitude of 
the $\nu_\mu \to \nu_e$ oscillation amplitude parameter 
$\sin^2 2\theta_{13}$ varies along each curve, as indicated. The 
two groups of curves, which correspond to the two signs of $\delta m^2_{32}$, 
are separated by more than $3\sigma$ provided 
$\sin^2 2\theta_{13} \gsim 0.01$. Hence the mass heirarchy can be determined 
provided the $\nu_\mu \to \nu_e$ oscillation amplitude is not less than an 
order of magnitude below the currently excluded region. Unfortunately, within 
each group of curves, the CP--conserving predictions are separated from the 
maximal CP--violating predictions by at most $3\sigma$. Hence, it will 
be difficult to conclusively establish CP violation in this scenario.

\begin{figure}[tbh!]
\centerline{\includegraphics[width=4.0in]{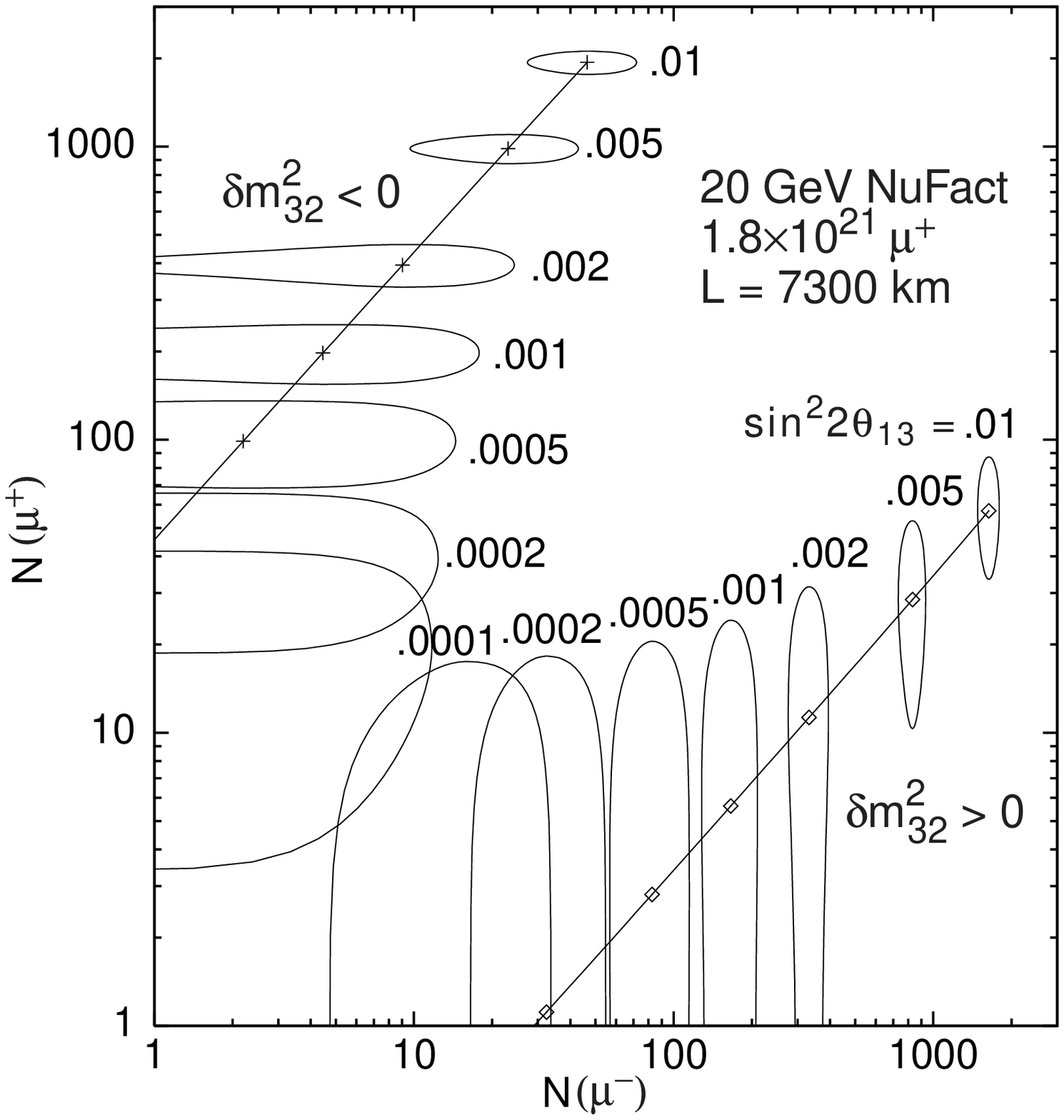}}
\caption[Error ellipses for neutrino factory for muon appearance]
{Three--sigma error ellipses in the 
$\left(N(\mu+), N(\mu-)\right)$--plane, shown for a 20~GeV neutrino 
factory delivering $3.6\times10^{21}$ useful muon decays and
$1.8\times10^{21}$ antimuon decays, with a 50~kt
detector at $L = 7300$~km, $\delta m^2_{21} = 10^{-4}$~eV$^2$, 
and $\delta = 0$. Curves are shown for both signs of
$\delta m^2_{32}$; $\sin^22\theta_{13}$ varies along the curves from
0.0001 to 0.01, as indicated~\cite{bargersuperbeam}.
}
\label{fig:nufact}
\end{figure}

Note for comparison that a very long baseline experiment at a neutrino 
factory would be able to observe $\nu_e \to \nu_\mu$ oscillations and 
determine the sign of $\delta m^2_{32}$ for values of $\sin^2 2\theta_{13}$ 
as small as O(0.0001)~! This is illustrated in Fig.~\ref{fig:nufact}.
A Neutrino Factory, thus outperforms a conventional superbeam in its ability to
determine the sign of  $\delta m^2_{32}$.
Comparing Fig.~\ref{fig:signdm2} and Fig.~\ref{fig:nufact} one sees that 
the value of   $\sin^2 2\theta_{13}$, which has yet to be measured, 
will determine the parameters of the first neutrino factory.

\begin{figure}[tbh!]
\centerline{\includegraphics[width=4.0in]{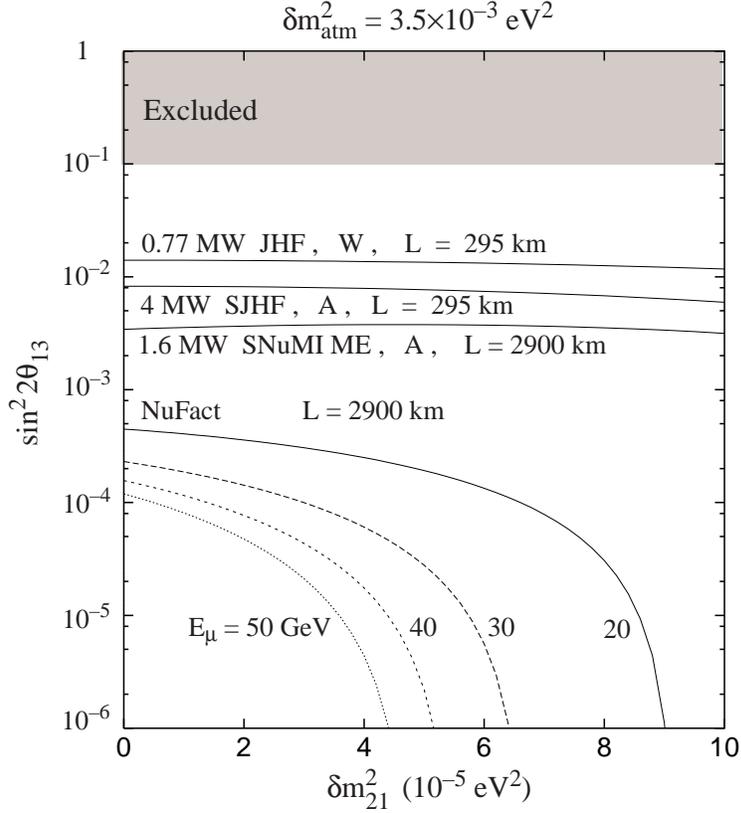}}
\caption[Comparison of superbeams and neutrino factories]
{Summary of the $3\sigma$ level sensitivities for the 
observation of $\nu_\mu \to \nu_e$ at various MW--scale superbeams 
(as indicated) with liquid argon ``A'' and water cerenkov ``W'' detector 
parameters, and the observation of $\nu_e \to \nu_\mu$ in a 50~kt detector 
at 20, 30, 40, and 50~GeV neutrino factories delivering $2 \times 10^{20}$ 
muon decays in 
the beam forming straight section. The limiting $3\sigma$ contours are 
shown in the $\delta m^2_{21}, \sin^2 2\theta_{13}$--plane. All curves 
correspond to 3~years of running. The grey shaded 
area is already excluded by current experiments.
}
\label{fig:reach} 
\end{figure}

Finally, we compare the superbeam $\nu_\mu \to \nu_e$ reach with the 
corresponding neutrino factory $\nu_e \to \nu_\mu$ reach in 
Fig.~\ref{fig:reach}, which shows the $3\sigma$ sensitivity contours in 
the $(\delta m^2_{21}, \sin^2 2\theta_{13})$--plane. The superbeam 
$\sin^2 2\theta_{13}$ reach of a few $\times 10^{-3}$ is almost independent 
of the sub--leading scale $\delta m^2_{21}$. However, since the neutrino 
factory probes oscillation amplitudes $O(10^{-4})$ the sub--leading effects 
cannot be ignored, and $\nu_e \to \nu_\mu$ events 
would be observed at a neutrino factory 
over a significant range 
of $\delta m^2_{21}$ even if $\sin^2 2\theta_{13} = 0$.

\section{Non-oscillation physics at a Neutrino Factory}
The study of the utility of intense neutrino beams from a muon storage ring in
determining the parameters governing non-oscillation physics was begun
in 1997~\cite{rajageer}. More complete studies can be found in
~\cite{INTRO:ref9} and recently a European group has brought out an
extensive study on this topic~\cite{cern-nonosc}. We quote their
conclusions here verbatim.  

``In the case of determinations of the
partonic densities of the nucleon, we proved that the \nufact\ could
significantly improve the already good knowledge we have today. In the
unpolarized case, the knowledge of the valence distributions would
improve by more than one order of magnitude, in the kinematical region
$x\gsim 0.1$, which is best accessible with 50~GeV muon beams. The
individual components of the sea ($\bar{u}$, $\bar{d}$, ${s}$ and
$\bar{s}$), as well as the gluon, would be measured with relative
accuracies in the range of 1--10\%, for $0.1\lsim x \lsim 0.6$. The
high statistics available over a large range of $Q^2$ would
furthermore allow the accurate determination of higher-twist
corrections, strongly reducing the theoretical systematics that affect
the extraction of $\as$ from sum rules and global fits.

``In the case of polarized densities, we stressed the uniqueness of the
\nufact\ as a means of disentangling quark and antiquark distributions,
and their first moments in
particular. These can be determined at the level of few per cent for
up and down, and 10\% for the strange, sufficient to distinguish
between theoretical scenarios, and thus allowing a full understanding
of the proton spin structure. A potential ability
to pin down the shapes of individual flavour components with accuracies at the
level of few per cent is  limited by the mixing with the 
polarized gluon. To identify this possible
weakness of the \nufact\ polarized-target programme, it was crucial to
perform our analysis at the NLO; we showed in fact that any study
based on the LO formalism would have resulted in far too optimistic
conclusions. This holds true both in the
case of determinations based on global fits and on direct
extractions using flavour tagging in the final state. 
Our conclusion here is that a full exploitation of the \nufact\
potential for polarized measurements of the shapes of
individual partonic densities requires an a-priori knowledge of
the polarized gluon density. It is hoped that the new information
expected to arise from the forthcoming set of polarized DIS
experiments at CERN, DESY and RHIC will suffice.

``The situation is also very bright for measurements of C--even
distributions. Here, the first moments of singlet, triplet and octet
axial charges can be measured with
accuracies which are up to one order of magnitude better than the
current uncertainties. In particular, the improvement in the
determination of the singlet axial charge would allow a definitive
confirmation or refutation of the anomaly scenario compared to the
`instanton' or `skyrmion' scenarios, at least if the theoretical
uncertainty originating from the small--$x$ extrapolation can be kept under
control. The measurement of the octet axial charge with a few percent
uncertainty will allow a determination of the strange contribution to
the proton spin better than 10\%, and allow stringent tests of models
of $SU(3)$ violation when compared to the direct determination from
hyperon decays.

``The measurement of two fundamental constants of nature, $\as(M_Z)$ and
$\sin^2\theta_W$, will be possible using a variety of techniques. At
best the
accuracy of these measurements will match or slightly improve
the accuracy available
today, although the
measurements at the \nufact\ are subject to different systematics and
therefore provide an important consistency check of current data.
In the case of $\as(M_Z)$, the dependence of the results on the
modeling of higher-twist corrections both in the structure function
fits and in the GLS sum rule is significantly reduced relative to
current measurements, as mentioned above. 
In the case of $\sin^2\theta_W$, its determination via $\nu e$
scattering at the \nufact\ has an uncertainty of approximately
$2\times 10^{-4}$, dominated by the statistics and the luminosity
measurement. This error is comparable to what already known
today from EW measurements in $Z^0$ decays. Compared to these,
however, this determination would
improve current low-energy extractions, and be 
subject to totally different systematic uncertainties. It would also be
sensitive to different classes of new-physics contributions.
The extrapolation
to $Q=M_Z$ is affected, at the same level of uncertainty, by the
theoretical assumptions used in the evaluation of the hadronic-loop
corrections to $\gamma$-$Z$ mixing. The determination via DIS, on the
other hand, is limited by the uncertainties on the heavy-flavour
parton densities. As shown earlier, these should be significantly
reduced using the \nufact\ data themselves.

``In several other areas, the data from the \nufact\ will allow 
quantitative studies to be made of phenomena that, so far have only
been explored at a mostly qualitative level. This is the case of the
exclusive production of charmed mesons and baryons (leading to very
large samples, suitable for precise extractions of branching ratios
and decay constants), of the study of spin-transfer
phenomena, and of the study of nuclear effects in DIS. While nuclear
effects could be bypassed at the \nufact\ by using  hydrogen
targets directly, the flavour separation of partonic densities will require
using also targets containing neutrons. This calls for an accurate
understanding of nuclear effects. The ability to run with both $H$ and
heavier targets will in turn provide rich data sets useful for
quantitative studies of nuclear models.
The study of $\Lambda$ polarization both in the target and in the
fragmentation regions, will help clarifying the intriguing problem of
spin transfer. We reviewed several of the existing models, and
indicated how semi-inclusive neutrino DIS will allow the
identification of the
right ones, as well as providing input for the measurement of
polarized fragmentation functions.

``Finally, we presented some cases of exploration for physics beyond the
SM using the \nufact\ data. Although the neutrino beam energies
considered in our work are well below any reasonable threshold for new
physics, the large statistics makes it possible to search for
manifestations of virtual effects. The exchange of new gauge bosons
decoupled from the first generation of quarks and leptons can be seen
via enhancements of the inclusive charm production rate, with a
sensitivity well beyond the present limits. Rare
lepton-flavour-violating decays of muons in the ring could be tagged
in the DIS final states
through the detection of wrong-sign electrons and muons, or of prompt
taus. Once again, the sensitivity at the \nufact\ goes well beyond
existing limits...''

\section{Physics that can be done with Intense Cold Muon Beams}
Experimental studies of muons at low and medium energies have had a
long and distinguished history, starting with the first search for
muon decay to electron plus gamma-ray~\cite{Hincks-Pontecorvo}, and
including along the way the 1957 discovery of the nonconservation of
parity, in which the $g$ value and magnetic moment of the muon were
first measured~\cite{Garwinetal}.  The years since then have brought
great progress: limits on the standard-model-forbidden decay $\mu\to
e\gamma$ have dropped by nine orders of magnitude, and the muon
anomalous magnetic moment $a_\mu=(g_\mu-2)/2$ has yielded one of the
more precise tests ($\approx1$ ppm) of physical theory, as well as a
possible hint of physics beyond the standard model~\cite{BNLg-2}.

The front end of a neutrino factory has the potential to provide
$\sim10^{21}$ muons per year, five orders of magnitude beyond the most
intense beam currently available.\footnote{The $\pi$E5 beam at PSI,
Villigen, providing a maximum rate of $10^9$ muons/s~\cite{Edgecock}.}
Such a facility could enable precision measurements of the muon
lifetime $\tau_\mu$ and Michel decay parameters as well as sensitive
searches for lepton-flavor nonconservation (LFV), a possible ($P$- and
$T$-violating) muon electric dipole moment (EDM)
$d_\mu$~\cite{HIMUS99}, and $P$ and $T$ violation in muonic atoms. It
could also lead to an improved direct limit on the mass of the muon
neutrino~\cite{numass}. Of these possibilities,
Marciano~\cite{Marciano97} has suggested that muon LFV (especially
coherent muon-to-electron conversion in the field of a nucleus) is the
``best bet" for discovering signatures of new physics using low-energy
muons; measurement of $d_\mu$ could prove equally exciting but is not
yet as well developed, being only at the Letter of Intent stage at
present~\cite{EDMLOI}.\footnote{Experimentalists might argue that
extending such measurements as $\tau_\mu$ and the Michel parameters is
worthwhile whenever the state of the art allows substantial
improvement. However, their comparison with theory is dominated by
theoretical uncertainties. Thus, compared to Marciano's ``best bets,"
they represent weaker arguments for building a new facility.}

The search for $\mu\to e \gamma$ is also of great interest. The MEGA
experiment recently set an upper limit $B(\mu^+\to
e^+\gamma)<1.2\times10^{-11}$~\cite{MEGA}. Ways to extend sensitivity
to the $10^{-14}$ level have been
discussed~\cite{Cooper97}. Sensitivity greater than this may be
possible but will be difficult since at high muon rate there will be
background due to accidental coincidences; a possible way around this
relies on the correlation between the electron direction and the
polarization direction using a polarized muon beam. The
$\mu$-to-$e$-conversion approach does not suffer from this drawback
and has the additional virtue of sensitivity to possible new physics
that does not couple to the photon.

In the case of precision measurements ($\tau_\mu$, $a_\mu$, etc.),
new-physics effects can appear only as small corrections arising from
the virtual exchange of new massive particles in loop diagrams. In
contrast, LFV and EDMs are forbidden in the standard model, thus their
observation at any level constitutes evidence for new physics. The
current status and prospects for advances in these areas are
summarized in Table~\ref{tab:LEmuons}. It is worth recalling that LFV
as a manifestation of neutrino mixing is suppressed as $(\delta
m^2)^2/m_W^4$ and is thus entirely negligible. However, a variety of
new-physics scenarios predict observable effects.
Table~\ref{tab:newmuphys} lists some examples of limits on new physics
that would be implied by nonobservation of $\mu$-to-$e$ conversion
($\mu^-N\to e^-N$) at the $10^{-16}$ level~\cite{Marciano97}.

\begin{table}
\caption[Current and future tests in low energy muons] {Some current
and future tests for new physics with low-energy muons
(from~\protect\cite{Marciano97}, \protect\cite{PDG}, and
\protect\cite{Aoki01}). Note that the ``Current prospects" column
refers to anticipated sensitivity of experiments currently approved or
proposed; ``Future" gives estimated sensitivity with Neutrino Factory
front end. (The $d_\mu$ measurement is still at the Letter of Intent
stage and the reach of experiments is not yet entirely
clear.)\label{tab:LEmuons}}
\begin{center}
\begin{tabular}{|lccc|}
\hline
Test & Current bound & Current prospects & Future \\
\hline
$B(\mu^+\to e^+\gamma)$ & $<1.2\times10^{-11}$ &
$\approx5\times10^{-12}$ & $\sim10^{-14}$\\ $B(\mu^-{\rm Ti}\to
e^-{\rm Ti})$ & $<4.3\times10^{-12}$ & $\approx2\times10^{-14}$ &
$<10^{-16}$\\ $B(\mu^-{\rm Pb}\to e^-{\rm Pb})$ & $<4.6\times10^{-11}$
& & \\ $B(\mu^-{\rm Ti}\to e^+{\rm Ca})$ & $<1.7\times10^{-12}$ & & \\
$B(\mu^+\to e^+e^-e^+)$ & $<1\times10^{-12}$ & & \\ $d_\mu$ &
$(3.7\pm3.4)\times10^{-19}\,e\cdot$cm & $10^{-24}\,e\cdot$cm? & ? \\
\hline
\end{tabular}
\end{center}
\end{table}

\begin{table}
\caption[New physics probed by $\mu\rightarrow e$ experiments]
{Some examples of new physics probed by the nonobservation of 
$\mu\rightarrow e$ conversion at the $10^{-16}$ level 
(from~\protect\cite{Marciano97}).\label{tab:newmuphys}}
\begin{center}
\begin{tabular}{|lc|}
\hline
New Physics & Limit \\
\hline
Heavy neutrino mixing & $|V_{\mu N}^*V_{e N}|^2<10^{-12}$\\ Induced
$Z\mu e$ coupling & $g_{Z_{\mu e}}<10^{-8}$\\ Induced $H\mu e$
coupling & $g_{H_{\mu e}}<4\times10^{-8}$\\ Compositeness &
$\Lambda_c>3,000\,$TeV\\
\hline
\end{tabular}
\end{center}
\end{table}

Precision studies of atomic electrons have provided notable tests of
QED ({\it e.g.}\ the Lamb shift in hydrogen) and could in principle be
used to search for new physics were it not for nuclear corrections.
Studies of muonium ($\mu^+e^-$) are free of such corrections since it
is a purely leptonic system. Muonic atoms also can yield new
information complementary to that obtained from electronic atoms. A
number of possibilities have been enumerated by Kawall {\it et
al.}~\cite{Kawall97} and Molzon~\cite{Molzon97}. As an example we
consider the hyperfine splitting of the muonium ground state, which
has been measured to 36 ppb~\cite{Mariam} and currently furnishes the
most sensitive test of the relativistic two-body bound state in
QED~\cite{Kawall97}. The precision could be further improved with
increased statistics.  The theoretical error is 0.3 ppm but could be
improved by higher-precision measurements in muonium and muon spin
resonance, also areas in which the Neutrino Factory front end could
contribute. Another interesting test is the search for
muonium-antimuonium conversion, possible in new-physics models that
allow violation of lepton family number by two units. The current
limit is $R_g \equiv G_C / G_F< 0.0030$~\cite{PDG}, where $G_C$ is the
new-physics coupling constant and $G_F$ is the Fermi coupling
constant. This sets a lower limit of $\approx 1 \,$TeV$/c^2$ on the
mass of a grand-unified dileptonic gauge boson and also constrains
models with heavy leptons~\cite{Abela}.

\section[Physics Potential of a Higgs Factory Muon Collider]
{Physics potential of a Low energy Muon Collider 
operating as a Higgs Factory}

Muon colliders~\cite{bargersnow,clinehanson} have a number of unique
features that make them attractive candidates for future
accelerators~\cite{INTRO:ref5}.  The most important and fundamental of
these derive from the large mass of the muon in comparison to that of
the electron.  This leads to: a) the possibility of extremely narrow
beam energy spreads, especially at beam energies below $100\gev$; b)
the possibility of accelerators with very high energy; c) the
possiblity of employing storage rings at high energy; d) the
possibility of using decays of accelerated muons to provide a high
luminosity source of neutrinos as discussed in section~\ref{neuf}; e)
increased potential for probing physics in which couplings increase
with mass (as does the SM $\hsm f\anti f$ coupling).

The relatively large mass of the muon compared to the mass of the electron
means that the coupling of Higgs bosons to $\mu^+\mu^-$ is very 
much larger than to $e^+e^-$, implying much larger $s$-channel Higgs
production rates at a muon collider as compared to an electron collider.
For Higgs bosons with a very small (MeV-scale) width, 
such as a light SM Higgs boson,
production rates in the $s$-channel 
are further enhanced by the 
muon collider's ability to achieve beam energy spreads
comparable to the tiny Higgs width. 
In addition, there is little bremsstrahlung, 
and the beam energy can be tuned to one part
in a million through continuous spin-rotation measurements~\cite{Raja:1998ip}.
Due to these important qualitative difference
between the two types of machines, only muon colliders can be 
advocated as potential $s$-channel
Higgs factories capable of determining the mass and decay width
of a Higgs boson to very high precision~\cite{Barger:1997jm,Barger:1995hr}.
High rates of Higgs production at $\epem$ colliders rely on
substantial $VV$Higgs coupling for the
$Z+$Higgs (Higgstrahlung) or $WW\to$Higgs ($WW$ fusion) reactions.
In contrast, a $\mupmum$ collider can provide a factory for producing
a Higgs boson with little or no $VV$ coupling so long as it
has SM-like (or enhanced) $\mupmum$ couplings.

Of course, there is a tradeoff between small beam energy spread,
$\delta E/E=R$, and luminosity. Current estimates for yearly
integrated luminosities (using
$\call=1\times 10^{32}$cm$^{-2}$s$^{-1}$ as implying $ L=1\fbi/{\rm yr}$) are:
$\lyear\gsim 0.1,0.22,1 \fbi$ at $\rts\sim 100\gev$
for beam energy resolutions of $R=0.003\%,0.01\%,0.1\%$, respectively;
$\lyear\sim 2,6,10 \fbi$ at $\rts\sim 200,350,400\gev$, respectively, for 
$R\sim 0.1\%$.
Despite this, studies show that for small Higgs width the $s$-channel
production rate (and statistical significance over background) is maximized
by choosing $R$ to be such that $\srts\lsim \gamhtot$. In particular,
in the SM context for $\mhsm\sim 110\gev$ this corresponds to $R\sim 0.003\%$.

If the $\mh\sim 115\gev$ LEP signal is real or if the 
interpretation of the precision
electroweak data as an indication of a light Higgs boson (with
substantial $VV$ coupling) is valid,
then both $\epem$ and $\mupmum$ colliders will be valuable.
In this scenario the Higgs boson would have been discovered at a previous 
higher energy collider (possibly a muon collider
running at high energy), and then the Higgs factory
would be built with a center-of-mass energy 
precisely tuned to the Higgs boson mass.
The most likely scenario is that the Higgs boson 
is discovered at the LHC via gluon fusion
($gg\to H$) or perhaps 
earlier at the Tevatron via associated production 
($q\bar{q}\to WH, t\overline{t}H$), and its mass is determined to an 
accuracy of about 100~MeV. If a linear collider has also observed the Higgs
via the Higgs-strahlung process ($e^+e^-\to ZH$), one might know the Higgs 
boson mass to better than 50~MeV with an integrated luminosity of 
$500$~fb$^{-1}$.
The muon collider would be optimized to run at $\sqrt{s}\approx m_H$, and this
center-of-mass energy would be varied over a narrow range
so as to scan over the Higgs resonance (see Fig.~\ref{mhsmscan} below). 

\subsection{Higgs Production}

The production of a Higgs boson (generically denoted $\h$)
in the $s$-channel with interesting rates is  
a unique feature of a muon collider~\cite{Barger:1997jm,Barger:1995hr}. 
The resonance cross section is
\begin{equation}
\sigma_h(\sqrt s) = {4\pi \Gamma(h\to\mu\bar\mu) \, \Gamma(h\to X)\over
\left( s - m_h^2\right)^2 + m_h^2 \left(\Gamma_{\rm tot}^h \right)^2}\,.
\label{rawsigform}
\end{equation}
In practice, however, there is a Gaussian spread ($\srts$) to
the center-of-mass energy and one must compute the
effective $s$-channel Higgs cross section after convolution 
assuming some given central value of $\rts$:
\begin{eqnarray}
\anti\sigma_h(\sqrt s) & =& {1\over \sqrt{2\pi}\,\srts} \; \int
\sigma_h  
(\sqrt{\what s}) \; \exp\left[ -\left( \sqrt{\what s} - \sqrt s\right)^2
\over  
2\sigma_{\sqrt s}^2 \right] d \sqrt{\what s}\\
&&\stackrel{\rts=\mh}{\simeq} {4\pi\over m_h^2} \; {\br(h\to\mu\bar\mu)
\,
\br(h\to X) \over \left[ 1 + {8\over\pi} \left(\srts\over\gamhtot 
\right)^2 \right]^{1/2}} \,.
%
\label{sigform}
\end{eqnarray}
\begin{figure}[tbh!]
\centering\leavevmode
\centerline{\includegraphics[width=4.0in]{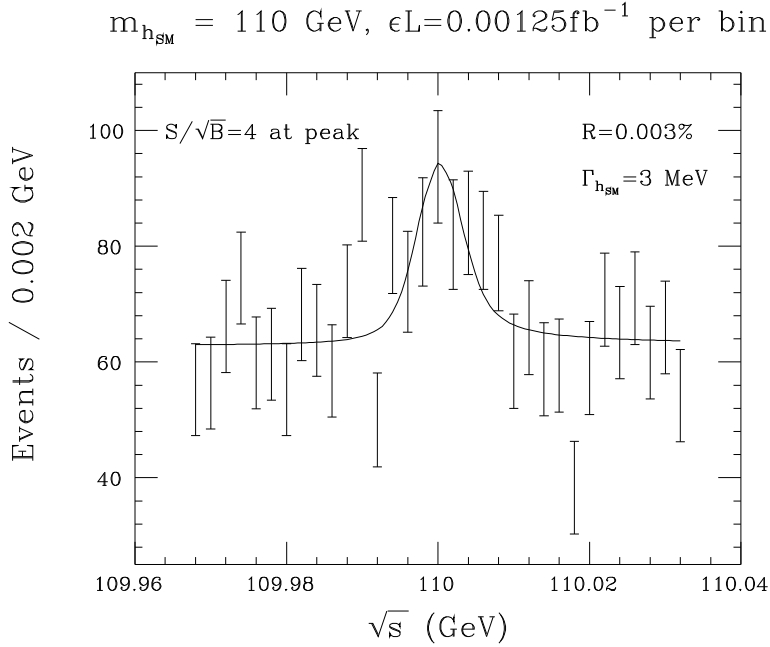}}
\caption[Scan of the Higgs resonance using a muon collider]{
Number of events and statistical errors in the $b\overline{b}$
final state as a function
of $\protect\rts$ in the vicinity of $\mhsm=110\gev$,
assuming $R=0.003\%$,
and $\epsilon L=0.00125$~fb$^{-1}$ at each data point.
\label{mhsmscan}}
\end{figure}

It is convenient to express $\srts$ in 
terms of the root-mean-square (rms) Gaussian spread
of the energy of an individual beam, $R$: 
\begin{equation}
\srts = (2{\rm~MeV}) \left( R\over 0.003\%\right) \left(\sqrt s\over  
100\rm~GeV\right) \,.
\end{equation}
From Eq.~(\ref{rawsigform}), it is apparent that a
resolution $\srts \lsim \gamhtot$ is needed to be
sensitive to the Higgs width. Further, Eq.~(\ref{sigform}) implies that
$\anti\sigma_h\propto 1/\srts$ for $\srts>\gamhtot$ {\it and}
that large event rates are only possible if $\gamhtot$ is not so large
that $\br(\h\to \mu\anti\mu)$ is extremely suppressed.
The width of a light SM-like Higgs is very small (e.g. a few MeV
for $\mhsm\sim 110\gev$), implying the need for $R$
values as small as $\sim 0.003\%$ for studying a light SM-like $\h$.
Fig.~\ref{mhsmscan} illustrates the result for the SM Higgs boson 
of an initial centering scan over $\rts$ values
in the vicinity of $\mhsm=110\gev$.
This figure dramatizes: a)  that the beam energy spread must be very small
because of the very small $\gamhsmtot$ (when $\mhsm$ is small
enough that the $WW^\star$ decay
mode is highly suppressed); b) that we require
the very accurate {\it in situ} determination 
of the beam energy to one part in a million through the spin 
precession of the muon noted earlier in order to perform the scan
and then center on $\rts=\mhsm$ with a high degree of stability.

If the $\h$ has SM-like couplings to $WW$, its width will
grow rapidly for $\mh>2m_W$ and its $s$-channel production cross
section will be severely suppressed by the resulting 
decrease of $\br(\h\to\mu\mu)$. 
More generally, any $\h$ with SM-like or larger $\h\mu\mu$ coupling
will retain a large $s$-channel production rate when 
$\mh>2m_W$ only if the $\h WW$ coupling becomes 
strongly suppressed relative to the $\hsm WW$ coupling.

The general theoretical prediction within supersymmetric models is that the 
lightest supersymmetric Higgs boson $\hl$ will
be very similar to the $\hsm$ when the other Higgs bosons are
heavy.  This `decoupling limit' is very likely to arise if the
masses of the supersymmetric particles are large (since the Higgs
masses and the superparticle masses are typically similar in
size for most boundary condition choices).
Thus, $\hl$ rates will be very similar to $\hsm$ rates.
In contrast, the heavier Higgs bosons in a typical supersymmetric model
decouple from $VV$ at large mass  and remain reasonably
narrow. As a result, their $s$-channel production rates remain large.


For a SM-like $\h$, at $\sqrt s = \mh \approx 115$~GeV
and $R=0.003\%$, the $b\bar b$ rates are
\vspace{-.05in}
\begin{eqnarray}
\rm signal &\approx& 10^4\rm\ events\times L(fb^{-1})\,,\\
\rm background &\approx& 10^4\rm\ events\times L(fb^{-1})\,.
\end{eqnarray}

\subsection{What the Muon Collider Adds to LHC and LC Data}
An assessment of the need for a Higgs factory requires that one detail the 
unique capabilities of a muon collider versus the other possible future 
accelerators as well as comparing the abilities of all the machines to 
measure the same Higgs properties. 
Muon colliders and a Higgs factory in particular
would only become operational after the LHC physics program is well-developed 
and quite possibly after a linear collider program is mature as well. So one
important question is the following: if
a SM-like Higgs boson and, possibly, important
physics beyond the Standard Model have been discovered at the LHC and perhaps 
studied at a linear collider, what new information could a Higgs factory 
provide?

The $s$-channel production process allows one to determine the mass, 
total width, and the cross sections
$\overline \sig_h(\mupmum\to\h\to X)$ 
for several final states $X$ 
to very high precision. The Higgs mass, total width and the cross sections 
can be used to constrain the parameters of the Higgs sector. 
For example, in the MSSM their precise values will
constrain the Higgs sector parameters
$\mha$ and $\tanb$ (where $\tanb$ is 
the ratio of the two vacuum expectation values (vevs) of the 
two Higgs doublets of the MSSM). The main question is whether these
constraints will be a valuable addition to LHC and LC constraints.

The expectations for the luminosity available at linear colliders has risen 
steadily. The most recent studies assume an integrated luminosity of some
$500$~fb$^{-1}$ corresponding to 1-2 years of running at a 
few$\times100$~fb$^{-1}$ 
per year. This luminosity results in the production of greater than $10^4$
Higgs bosons per year through the Bjorken Higgs-strahlung process, 
$e^+e^-\to Z\h$, provided the Higgs boson is kinematically accessible. This is 
comparable or even better than can be achieved with the current machine
parameters for a muon collider operating at the Higgs resonance; in fact, 
recent studies have described high-luminosity linear colliders as ``Higgs
factories,'' though for the purposes of this report, we will reserve this term
for muon colliders operating at the $s$-channel Higgs resonance. 

A linear collider with such high luminosity can certainly perform quite 
accurate measurements of certain Higgs parameters such as the Higgs mass, 
couplings to gauge bosons, couplings to heavy quarks, 
etc.~\cite{Battaglia:2000jb}.
Precise measurements of the couplings of the Higgs boson to the Standard 
Model particles is an important test of the mass generation mechanism.
In the Standard Model with one Higgs doublet, this coupling is proportional 
to the particle mass. In the more general case there can be mixing angles
present in the couplings. Precision measurements of the couplings can 
distinguish the Standard Model Higgs boson from one from a more general model
and can constrain the parameters of a more general Higgs sector.

\begin{table*}[h!]
\begin{center}
\caption[Comparison of a Higgs factory muon collider with LHC and LC]
{Achievable relative
uncertainties for a SM-like $\mh=110$~GeV for measuring the
Higgs boson mass and total width
for the LHC, LC (500~fb$^{-1}$), and the muon collider (0.2~fb$^{-1}$). 
}\label{unc-table}
\protect\protect
\begin{tabular}{cccc}
\hline
\ & LHC & LC & $\mu^+\mu^-$\\
$\mh$ & $9\times 10^{-4}$ & $3\times 10^{-4}$ & $1-3\times 10^{-6}$ \\
$\gamhtot$ & $>0.3$ & 0.17 & 0.2 \\
\hline
\end{tabular}
\end{center}
\end{table*}

The accuracies possible at different colliders
for measuring $\mh$ and $\gamhtot$ of
a SM-like $\h$ with $\mh\sim 110\gev$ are given in Table~\ref{unc-table}.
Once the mass is determined to about 1~MeV at the LHC and/or LC, 
the muon collider would employ a
three-point fine scan~\cite{Barger:1997jm} near the resonance peak.
Since all the couplings of the Standard Model are known, $\gamhsmtot$
is known. Therefore a precise determination of 
$\gamhtot$ is an important test of the Standard Model, and any deviation
would be evidence for a nonstandard Higgs sector. 
For a SM Higgs boson with a mass sufficiently below the $WW^\star$ 
threshold, the Higgs total width is very small (of order several MeV), and the 
only process where it can be measured {\it directly} is in the $s$-channel
at a muon collider. Indirect determinations at the LC can have
higher accuracy once $\mh$ is large enough that the $WW^\star$ mode
rates can be accurately measured, requiring $\mh>120\gev$.
This is because at the LC the total width must be determined 
indirectly by measuring a partial width and a branching fraction, and then 
computing the total width,
\begin{eqnarray}
&&\Gamma _{tot}={{\Gamma(h\to X)}\over {BR(h\to X)}}\;,
\end{eqnarray} 
for some final state $X$. For a Higgs boson so light that the $WW^\star$ decay
mode is not useful, then the total width measurement would probably require
use of the $\gamma \gamma $ decays~\cite{Gunion:1996cn}. This would require
information from a photon collider as well as the LC
and a small error is not possible.
The muon collider can measure the total width of the Higgs boson directly,
a very valuable input for precision tests of the Higgs sector.

To summarize,
if a Higgs is discovered at the LHC or possibly earlier at the Fermilab 
Tevatron, attention will turn to determining  whether this Higgs has the 
properties expected of the Standard Model Higgs. If the Higgs is discovered
at the LHC, it is quite possible that supersymmetric states will be 
discovered concurrently. The next goal for a linear collider or a muon collider
will be to better measure the Higgs boson properties to determine if 
everything is consistent within a supersymmetric framework or consistent
with the Standard Model.
A Higgs factory of even modest luminosity can provide uniquely
powerful constraints on the parameter space of the supersymmetric
model via its very precise measurement of the light Higgs mass, the
highly accurate determination of the total rate for $\mupmum\to\hl\to
b\anti b$ (which has almost zero theoretical systematic uncertainty
due to its insensitivity to the unknown $m_b$ value) and the
moderately accurate determination of the $\hl$'s total width.  In
addition, by combining muon collider data with LC data, a completely
model-independent and very precise determination of the
$h^0\mu^+\mu^-$ coupling is possible. This will provide another strong
discriminator between the SM and the MSSM.  Further, the
$h^0\mu^+\mu^-$ coupling can be compared to the muon collider and LC
determinations of the $h^0\tau^+\tau^-$ coupling for a precision test
of the expected universality of the fermion mass generation mechanism.

\section{Physics Potential of a High Energy Muon Collider} Once one
learns to cool muons, it become possible to build muon colliders with
energies of $\approx$ 3 TeV in the center of mass that fit on an
existing laboratory site~\cite{INTRO:ref5,rajawitherell}. At
intermediate energies, it becomes possible to measure the W mass
\cite{bergerw} and the top quark mass~\cite{bergertop} with high
accuracy, by scanning the thresholds of these particles and making use
of the excellent energy resolution of the beams. We consider here
further the ability of a higher energy muon collider to scan higher
lying Higgs like objects such as the H$^0$ and the A$^0$ in the MSSM
that may be degenerate with each other.

\subsection{Heavy Higgs Bosons}
As discussed in the previous section, precision measurements of the
light Higgs boson properties might make it possible to not only
distinguish a supersymmetric boson from a Standard Model one, but also
pinpoint a range of allowed masses for the heavier Higgs bosons.  This
becomes more difficult in the decoupling limit where the differences
between a supersymmetric and Standard Model Higgs are
smaller. Nevertheless with sufficiently precise measurements of the
Higgs branching fractions, it is possible that the heavy Higgs boson
masses can be infered.  A muon collider light-Higgs factory might be
essential in this process.

In the context of the MSSM, $\mha$ can probably be restricted to
within $50\gev$ or better if $\mha<500\gev$.
This includes the $250-500\gev$
range of heavy Higgs boson masses for which discovery is not possible 
via $\hh\ha$ pair production 
at a $\rts=500\gev$ LC. Further, the $\ha$ and $\hh$
cannot be detected in this mass range at either the LHC or LC 
in  $b\anti b \hh,b\anti b\ha$ production
for a wedge of moderate $\tanb$ 
values. (For large enough 
values of $\tanb$ the heavy Higgs bosons are expected to be observable
in $b\anti b \ha,b\anti b \hh$ production
at the LHC via their $\tau ^+\tau ^-$ decays and also at the LC.)

A muon collider can fill some, perhaps all of this moderate $\tanb$ wedge.
If $\tanb$ is large the $\mupmum \hh$ and $\mupmum\ha$ couplings (proportional
to $\tanb$ times a SM-like value) are enhanced
thereby leading to enhanced production rates in $\mupmum$ collisions.
The most efficient procedure is to operate the muon collider
at maximum energy and produce the $\hh$ and $\ha$ (often as overlapping
resonances) 
via the radiative return mechanism. By looking for a peak
in the $b\anti b$ final state, the $\hh$ and $\ha$ 
can be discovered and, once discovered, the machine $\rts$
can be set to $\mha$ or $\mhh$ and factory-like precision studies pursued.
Note that the $\ha$ and $\hh$ are typically broad enough that $R=0.1\%$
would be adequate to maximize their $s$-channel production rates.
In particular, $\Gamma\sim 30$~MeV
if the $t\overline{t}$ decay channel is not open, and $\Gamma\sim 3$~GeV if it
is. Since $R=0.1\%$ is sufficient, much higher luminosity
($L\sim 2-10~{\rm fb}^{-1}
/{\rm yr}$) would be possible as compared to that 
for $R=0.01\%-0.003\%$ required for studying the $\hl$.
 
In short, for these moderate $\tanb$--$\mha\gsim 250\gev$
scenarios that are particularly difficult for both
the LHC and the LC, the muon collider would be the only 
place that these extra Higgs bosons can be discovered and their properties 
measured very precisely.

\begin{figure}[tbh!]
\centering\leavevmode
\centerline{\includegraphics[width=4.0in]{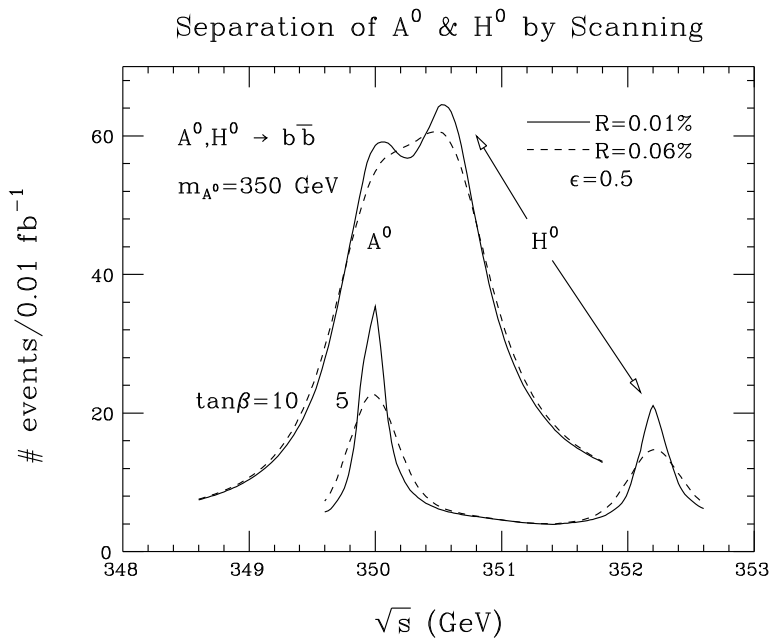}}

\caption[Separation of $A$ and $H$ signals for $\tan\beta=5$ and $10$]
{Separation of $A$ and $H$ signals for $\tan\beta=5$ and $10$. From  
Ref.~\cite{Barger:1997jm}. \label{H0-A0-sep}}
\end{figure}

In the MSSM, the heavy Higgs bosons are largely degenerate, especially in the 
decoupling limit where they are heavy. Large values of $\tan \beta$ heighten
this degeneracy.
A muon collider with sufficient energy resolution might be
the only possible means for separating out these states.
Examples showing the $H$ and $A$ resonances for $\tan \beta =5$ and $10$
are shown in Fig.~\ref{H0-A0-sep}. For the larger value of 
$\tan \beta$ the resonances are clearly overlapping. For the better energy 
resolution of $R=0.01\%$, the two distinct resonance peaks are still 
visible, but become smeared out for $R=0.06\%$.

Once muon colliders of these intermediate energies can be built,
higher energies such as 3-4~TeV in the center of mass become feasible.
Muon colliders with these energies will be complementary to hadron
colliders of the SSC class and above. The background radiation from
neutrinos from the muon decay becomes a problem at $\approx$~3~TeV in
the CoM. Ideas for ameliorating this problem have been discussed and include
optical stochastic cooling to reduce the number of muons needed for a
given luminosity, elimination of straight sections via wigglers or
undulators, or special sites for the collider such that the neutrinos
break ground in uninhabited areas.

\chapter{ Neutrino Factory}

\label{neufact}

\section{Description of Neutrino Factory}

In this Chapter we describe the various components of a Neutrino Factory.
The details here are taken from the most recent Feasibility Study (Study-II)
~\cite{EPP:studyii} that was carried out jointly by BNL and the MC. The
scheme we follow was outlined in Chapter 1. \ 

\subsection{Proton Driver}

\begin{figure}[tbh!]
\begin{center}
\includegraphics[width=5.5in]{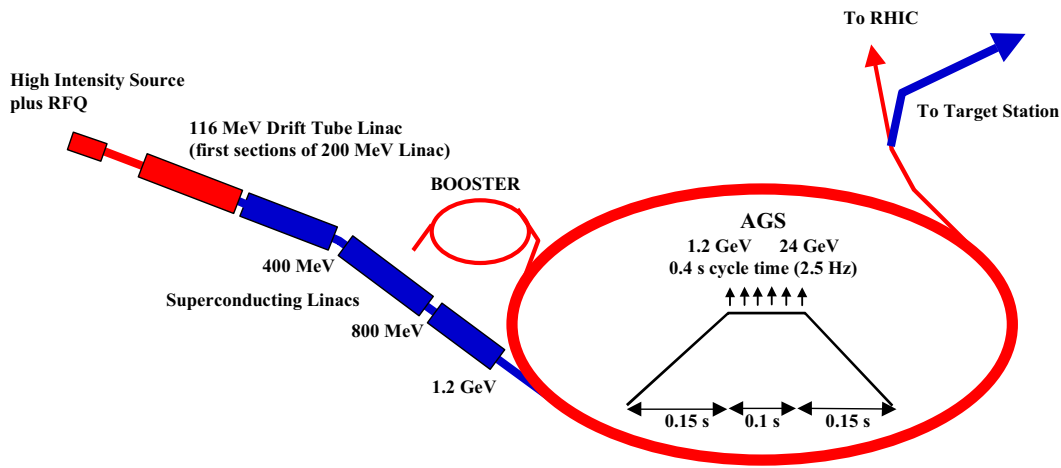}
\end{center}
\caption{AGS proton driver layout.}
\label{Proton:bnl}
\end{figure}

The proton driver considered in Study-II is an upgrade of the BNL
Alternating Gradient Synchrotron (AGS) and uses most of the existing
components and facilities; parameters are listed in Table~\ref{Proton:tb1}.
The existing booster is replaced by a 1.2-GeV superconducting proton linac.
The modified layout is shown in Fig.~\ref{Proton:bnl}. 
\begin{figure}[tbh!]
\begin{center}
\includegraphics[width=5.5in]{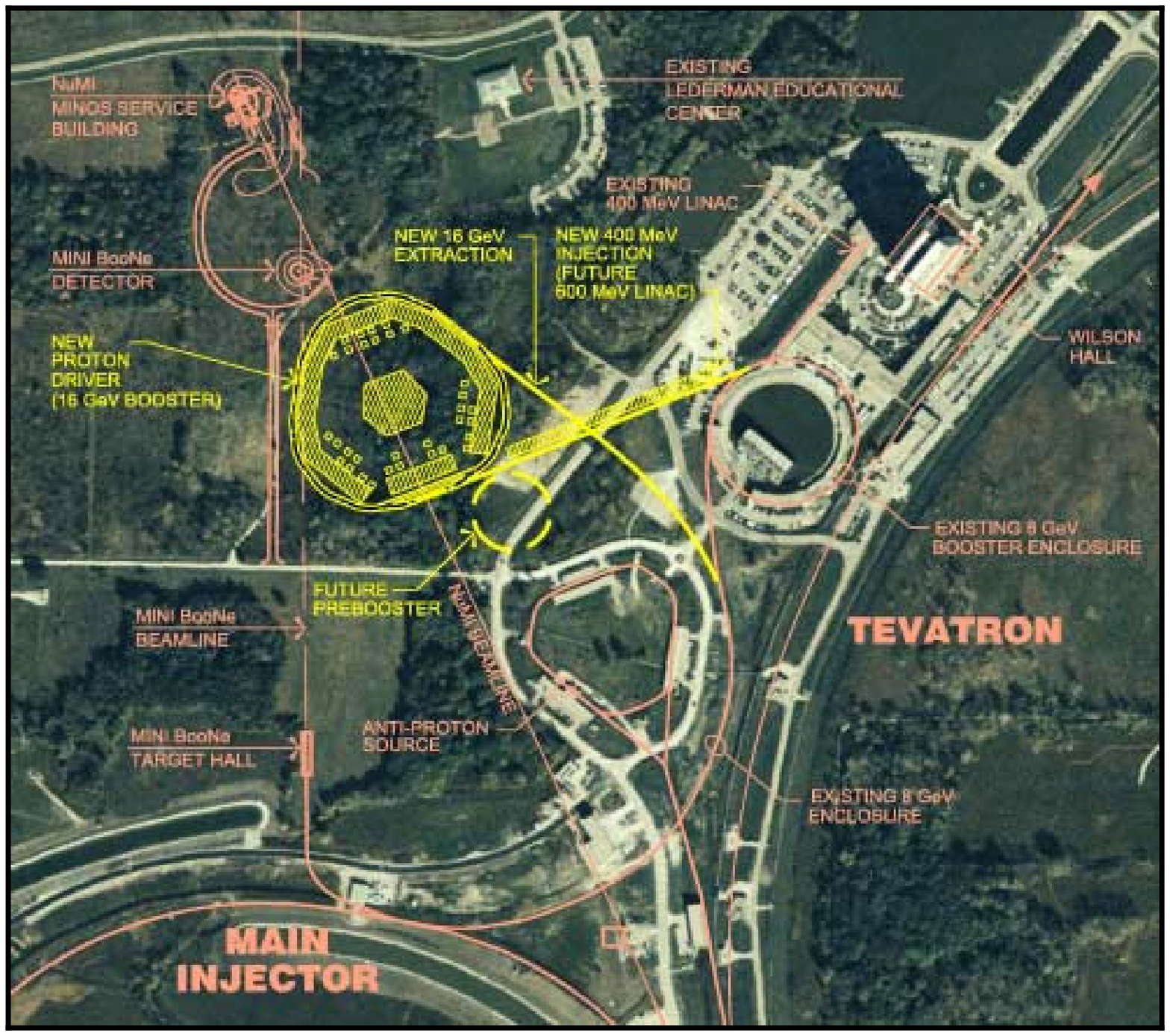}
\end{center}
\caption{FNAL proton driver layout.}
\label{Proton:fnal}
\end{figure}
The AGS repetition rate is increased from 0.5 Hz to 2.5 Hz by adding power
supplies to permit ramping the ring more quickly. No new technology is
required for this---the existing supplies are replicated and the magnets are
split into six sectors rather than the two used presently. The total proton
charge (10$^{14}$ ppp in six bunches) is only 40\% higher than the current
performance of the AGS. Nonetheless, the large increase in peak current
argues for an improved vacuum chamber; this is included in the upgrade. The
six bunches are extracted separately, spaced by 20 ms, so that the target,
induction linacs, and rf systems that follow need only deal with single
bunches at an instantaneous repetition rate of 50 Hz (average rate of 15
Hz). The average proton beam power is 1 MW. A possible future upgrade to 2 $%
\times $10$^{14}$ ppp and 5 Hz could give an average beam power of 4 MW. At
the higher intensity, a superconducting bunch compressor ring would be
needed to maintain the rms bunch length at 3 ns.

If the facility were built at Fermilab, the proton driver would be a newly
constructed 16-GeV rapid cycling booster synchrotron~\cite{FNALbooster}. The
planned facility layout is shown in Fig.~\ref{Proton:fnal}. The initial beam
power would be 1.2 MW, and a future upgrade to 4 MW would be possible. The
Fermilab design parameters are included in Table~\ref{Proton:tb1}. A less
ambitious and more cost-effective 8-GeV proton driver option has also been
considered for FNAL \cite{FNALbooster}.

\begin{table}[tbh]
\caption{Proton driver parameters for BNL and FNAL designs.}
\label{Proton:tb1}
\begin{center}
\begin{tabular}{lcc}
\hline
& BNL & FNAL \\ \cline{2-3}
Total beam power (MW) & 1 & 1.2 \\ 
Beam energy (GeV) & 24 & 16 \\ 
Average beam current ($\mu $A) & 42 & 72 \\ 
Cycle time (ms) & 400 & 67 \\ 
Number of protons per fill & $1\times 10^{14}$ & $3\times 10^{13}$ \\ 
Average circulating current (A) & 6 & 2 \\ 
No. of bunches per fill & 6 & 18 \\ 
No. of protons per bunch & $1.7\times 10^{13}$ & $1.7\times 10^{12}$ \\ 
Time between extracted bunches (ms) & 20 & 0.13 \\ 
Bunch length at extraction, rms (ns) & 3 & 1 \\ \hline
\end{tabular}
\end{center}
\end{table}

\subsection{Target and Capture}

\begin{figure}[tbh!]
\begin{center}
\includegraphics*[width=4in]{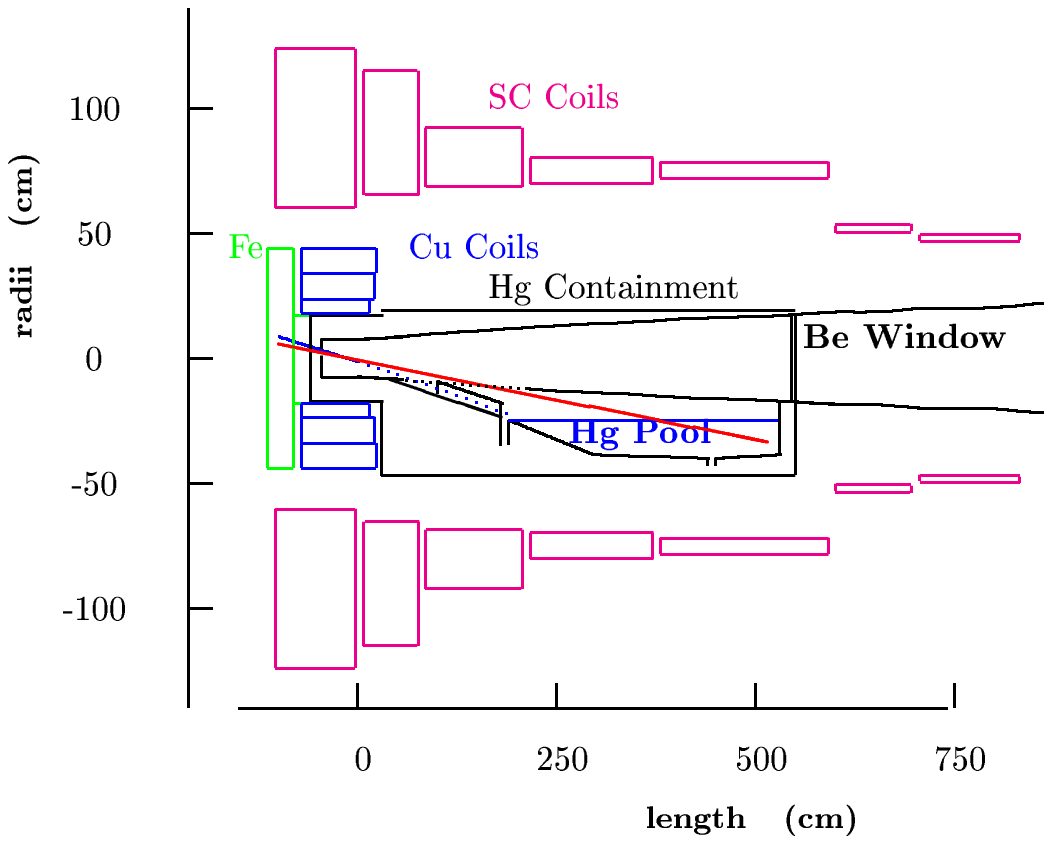}
\end{center}
\caption[Target, capture solenoids and mercury containment ]{Target, capture
solenoids and mercury containment.}
\label{tgtc}
\end{figure}

A mercury jet target is chosen to give a high yield of pions per MW of
incident proton power. The 1-cm-diameter jet is continuous, and is tilted
with respect to the beam axis. The target layout is shown in Fig.~\ref{tgtc} 
We assume that the thermal shock from the interacting proton bunch fully
disperses the mercury, so the jet must have a velocity of 30 m/s to be
replaced before the next bunch. Calculations of pion yields that reflect the
detailed magnetic geometry of the target area have been performed with the
MARS code~\cite{MARSstudyii}. To avoid mechanical fatigue problems, a
mercury pool serves as the beam dump. This pool is part of the overall
target---its mercury is circulated through the mercury jet nozzle after
passing through a heat exchanger.

Pions emerging from the target are captured and focused down the decay
channel by a solenoidal field that is 20 T at the target center, and tapers
down, over 18 m, to a periodic (0.5-m) superconducting solenoid channel ($%
B_{z}$ = 1 .25 T) that continues through the phase rotation to the start of
bunching. The 20-T solenoid, with a hollow copper conductor magnet insert
and superconducting outer coil, is similar in character to the higher field
(up to 45 T), but smaller bore, magnets existing at several laboratories ~
\cite{ITERmag}. The magnet insert is made with hollow copper conductor
having ceramic insulation to withstand radiation. MARS~\cite{MARSstudyii}
simulations of radiation levels show that, with the shielding provided, both
copper and superconducting magnets could have a lifetime greater than 15
years at 1 MW.

In Study-I, the target was a solid carbon rod. At high beam energies, this
implementation has a lower yield than the mercury jet, and is expected to be
more limited in its ability to handle the proton beam power, but should
simplify the target handling issues that must be dealt with. At lower beam
energies, say 6 GeV, the yield difference between C and Hg essentially
disappears, so a carbon target would be a competitive option with a lower
energy driver. Other alternative approaches, including a rotating Inconel
band target, and a granular Ta target are also under consideration, as
discussed in Study-II. Clearly there are several target options that could
be used for the initial facility.

\subsection{Phase Rotation}

\begin{figure}[tbh]
\begin{center}
\includegraphics*[width=4in]{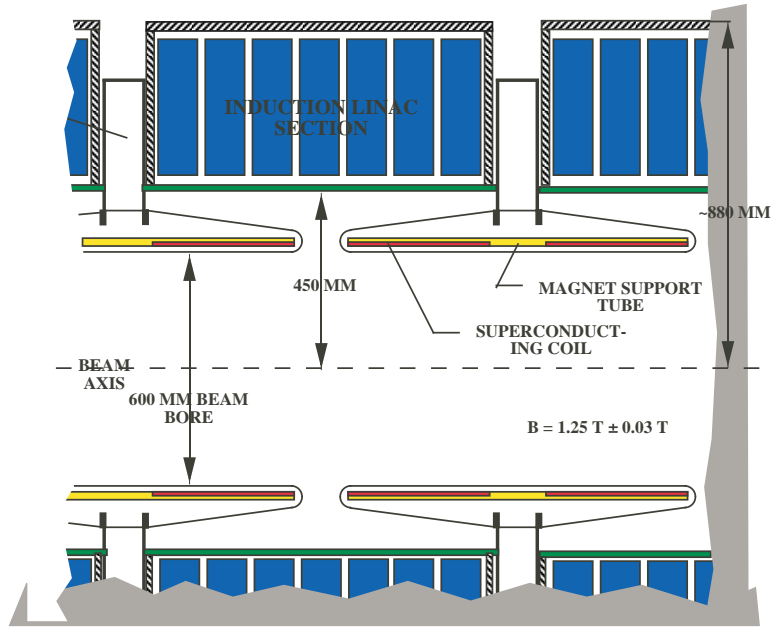}
\end{center}
\caption[Induction cell and mini-cooling solenoid]{Cross section of the
induction cell and mini-cooling solenoids.}
\label{CandPR:fg1}
\end{figure}

Pions, and the muons into which they decay, are generated in the target over
a very wide range of energies, but in a short time pulse (1--3 ns rms). This
large energy spread is ``phase rotated,'' using drifts and induction linacs, into a
pulse with a longer time duration and a lower energy spread. The muons first
drift and spread out in time, after which the induction linacs decelerate
the early ones and accelerate the later ones. Three induction linacs (with
lengths of 100, 80, and 80 m) are used in a system that reduces distortion
in the phase-rotated bunch, and permits all induction units to operate with
unipolar pulses. The 1.25-T beam transport solenoids are placed inside the
induction cores to avoid saturating the core material, as shown in Fig.~\ref
{CandPR:fg1}. 
The induction units are similar to those being built for DARHT~\cite{daarht}.

Between the first and second induction linacs, two LH$_{2}$ absorbers (each
1.7 m long and 30 cm radius), with a magnetic field reversal between them,
are introduced to reduce the transverse emittance and lower the beam energy
to a value matched to the cooling channel acceptance (``mini-cooling''). The
beam at the end of the phase rotation section has an average momentum of
about 250 MeV/$c$.

\subsection{Buncher}

The long beam pulse (400 ns) after the phase rotation is then bunched at
201.25 MHz prior to cooling and acceleration at that frequency. The bunching
is done in a lattice identical to that at the start of the cooling channel,
and is preceded by a matching section from the 1.25-T solenoids into this
lattice. The bunching has three stages, each consisting of rf (with
increasing acceleration) followed by drifts (with decreasing length). In the
first two rf sections, second-harmonic 402.5-MHz rf is used together with
the 201.25 MHz primary frequency to improve the capture efficiency. The
402.5-MHz cavities are designed to fit into the bore of the focusing
solenoids, in the location corresponding to that of the LH$_{2}$ absorber in
the downstream cooling channel.

\subsection{Cooling}

\begin{figure}[hbt!]
\begin{center}
\includegraphics[width=4in]{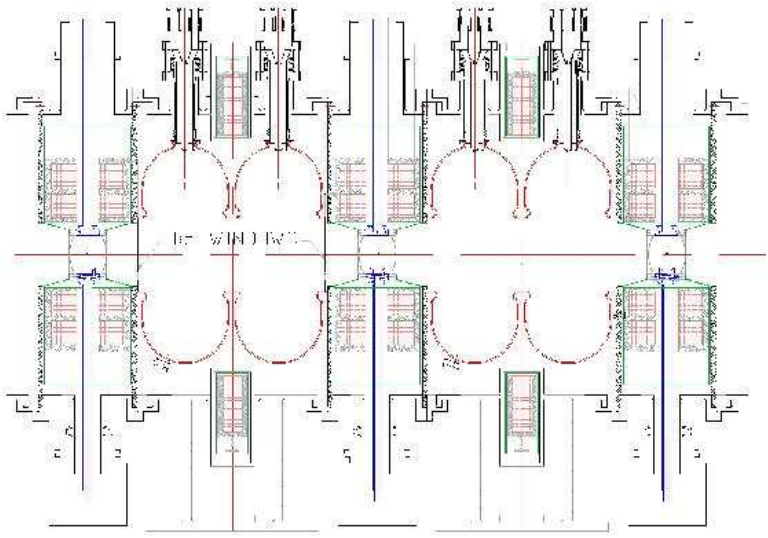}
\end{center}
\caption{Cooling channel Lattice 2, two cavities per cell.}
\label{RF:fg18.Q}
\end{figure}

Transverse emittance cooling is achieved by lowering the beam energy in LH$%
_{2}$ absorbers, interspersed with rf acceleration to keep the average
energy constant. Transverse and longitudinal momenta are lowered in the
absorbers, but only the longitudinal momentum is restored by the rf. The
emittance increase from Coulomb scattering is minimized by maintaining the
focusing strength such that the angular spread of the beam at the absorber
locations is large. This is achieved by keeping the focusing strength
inversely proportional to the emittance,\textit{\ i.e.}, increasing it as
the emittance is reduced. A modified Focus-Focus (SFOFO)~\cite{EPP:refsfofo}
lattice is employed. The solenoidal fields in each cell alternate in sign,
and the field shape is chosen to maximize the momentum acceptance ($\pm $
22\%). To maintain the tapering of the focusing, it was necessary to reduce
the cell length from 2.75 m in the initial portion of the channel to 1.65 m
in the final portion. A layout of the shorter cooling cell is shown in Fig.~%
\ref{RF:fg18.Q}.

\begin{figure}[tbh]
\begin{center}
\includegraphics*[width=100mm]{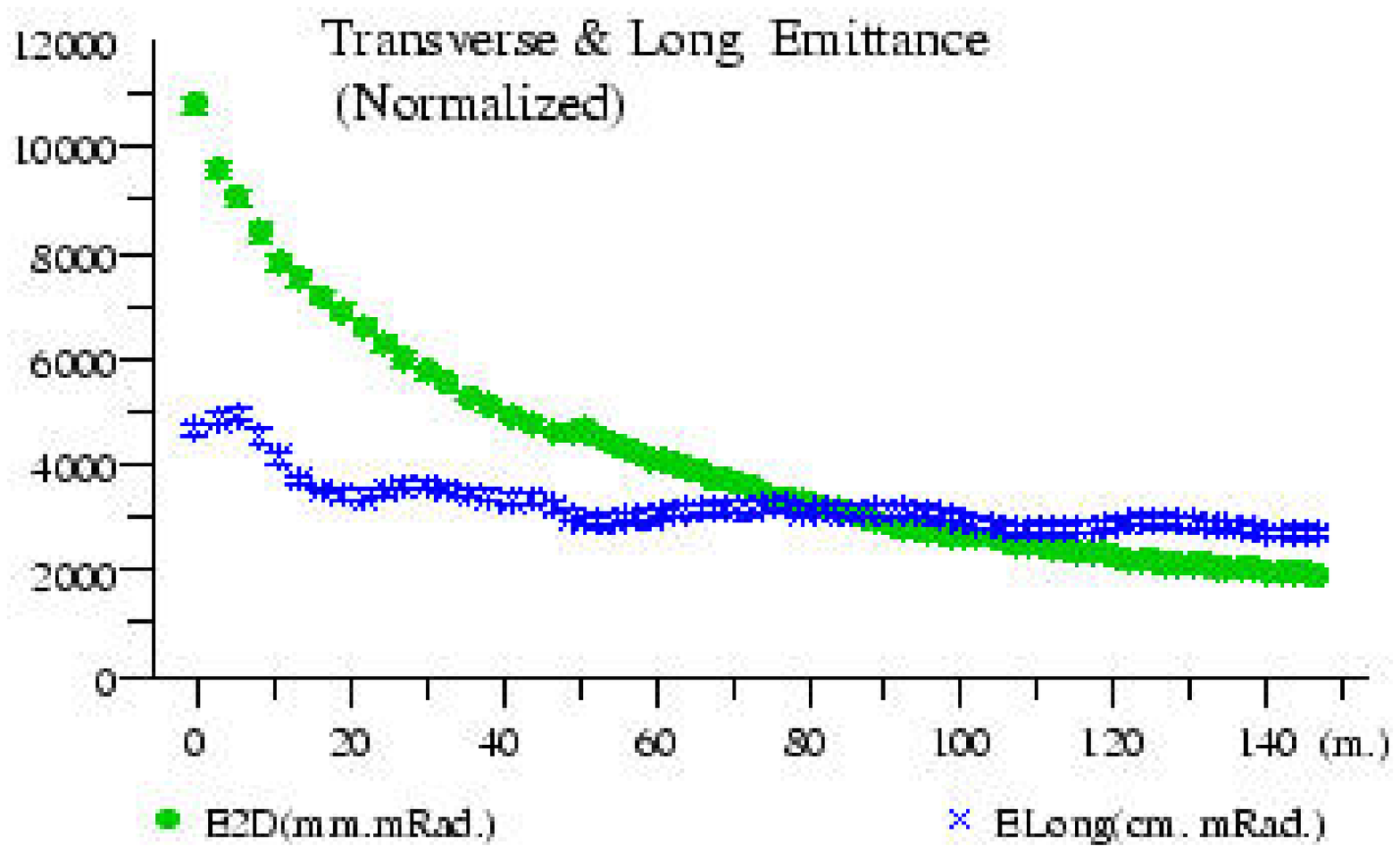}
\end{center}
\caption[The longitudinal and transverse emittances]{The longitudinal and
transverse emittances, obtained with the Geant4 simulation code, as a
function of channel length. The last lattice (2,3) was extended by $\approx $%
20~m to investigate the ultimate performance of the cooling channel.}
\label{EmittCool}
\end{figure}

\begin{figure}[tbh]
\begin{center}
\includegraphics*[width=100mm]{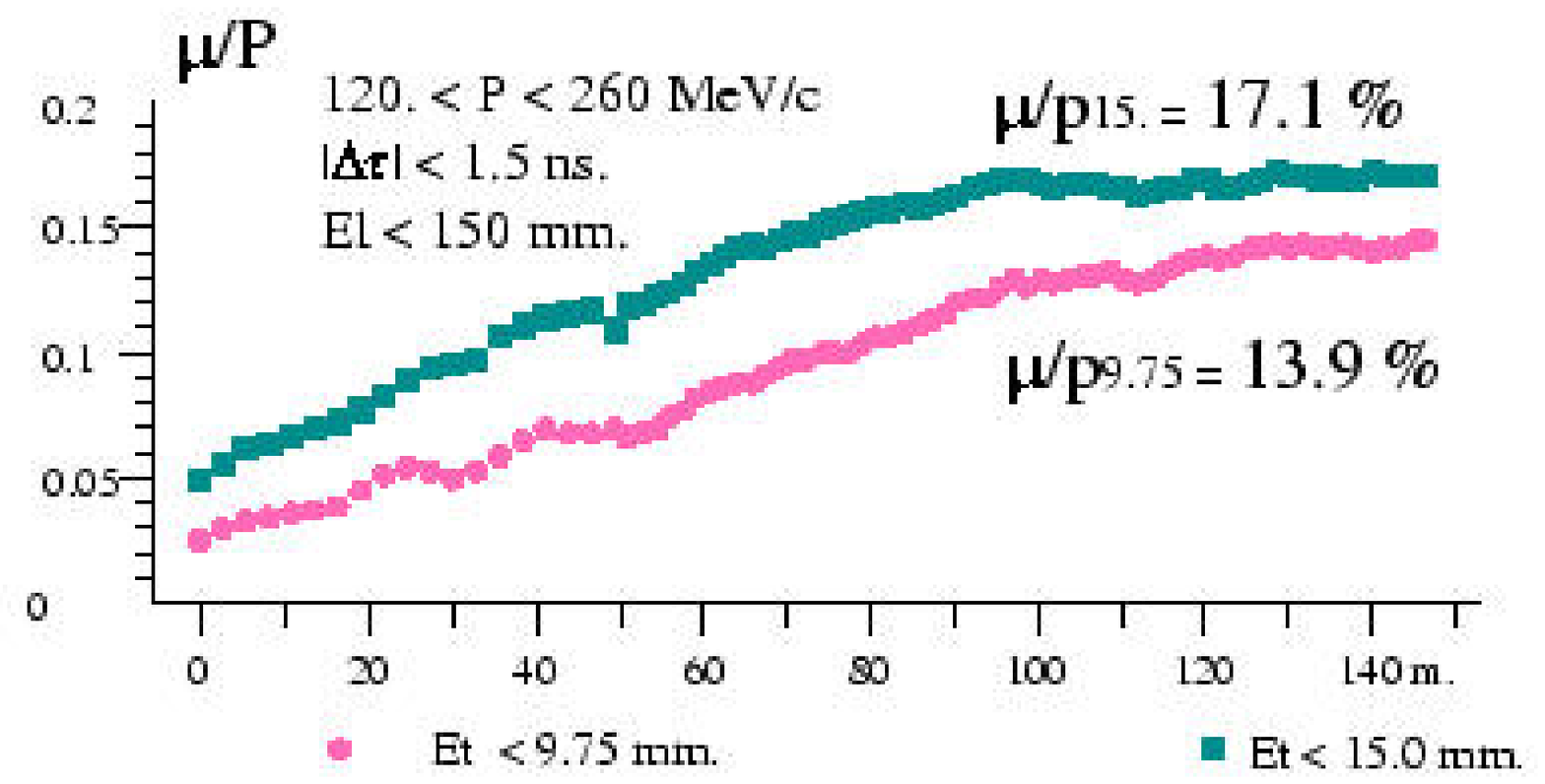}
\end{center}
\caption[$\protect\mu /p$ yield ratio for the two transverse emittance cuts]{
Geant4 simulations of the muon-to-proton yield ratio for two transverse
emittance cuts, clearly showing that the channel cools, \textit{i.e.}, the
density in the center of the phase space region increases. Since the
relevant yield $\protect\mu /$p$_{15}$ no longer increases for $z\leq 110$%
~m, the channel length was set to 108~m.}
\label{YieldCool}
\end{figure}

Figure~\ref{EmittCool} shows a simulation of cooling; the emittance falls
along the length of the channel. The increase in the number of muons that
fit within the acceptance of the downstream acceleration channel is shown in
Fig.~\ref{YieldCool}. 

\subsection{Acceleration}

\begin{figure}[tbh]
\centering\includegraphics[angle=270,width=4.0in]{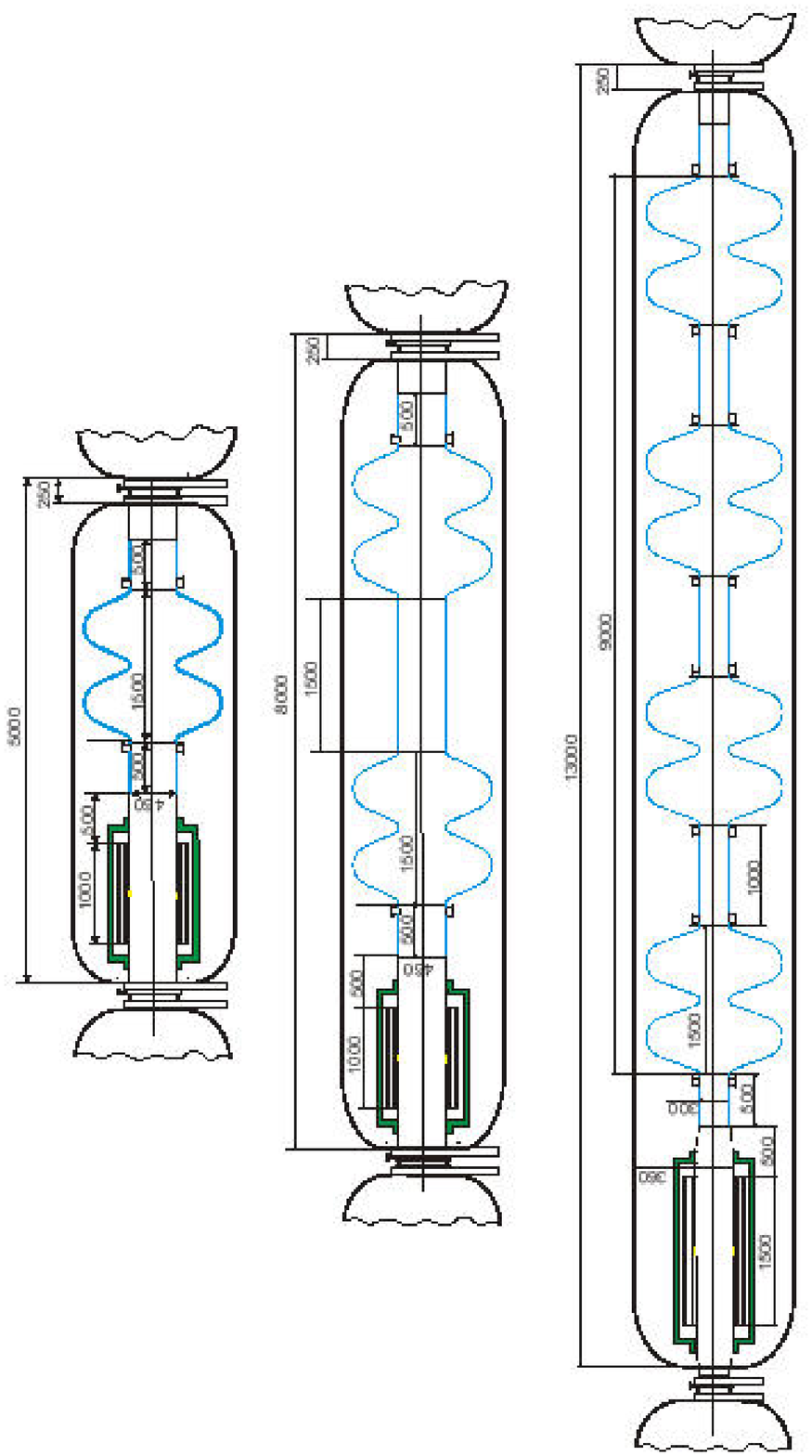}
\caption[Layouts of cryomodules.]{Layouts of short (top), intermediate
(middle) and long (bottom) cryomodules. Blue lines are the SC walls of the
cavities. Solenoid coils are indicated in red, and BPMs in yellow.}
\label{fig:acc:cryomod}
\end{figure}

Parameters of the acceleration system are listed in Table~\ref{tab:acc:parm}%
. A 20-m SFOFO matching section, using normal conducting rf systems, matches
the beam optics to the requirements of a 2.5 GeV superconducting rf linac
with solenoidal focusing. The linac is in three parts. The first part has a
single 2-cell rf cavity unit per period. The second part, as a longer period
becomes possible, has two 2-cell cavity units per period. The last section,
with still longer period, accommodates four 2-cell rf cavity units per
period. Figure~\ref{fig:acc:cryomod} shows the three cryomodule types that
make up the linac. 

\begin{figure}[!tbh]
\centering\includegraphics[angle=270,width=4.0in]{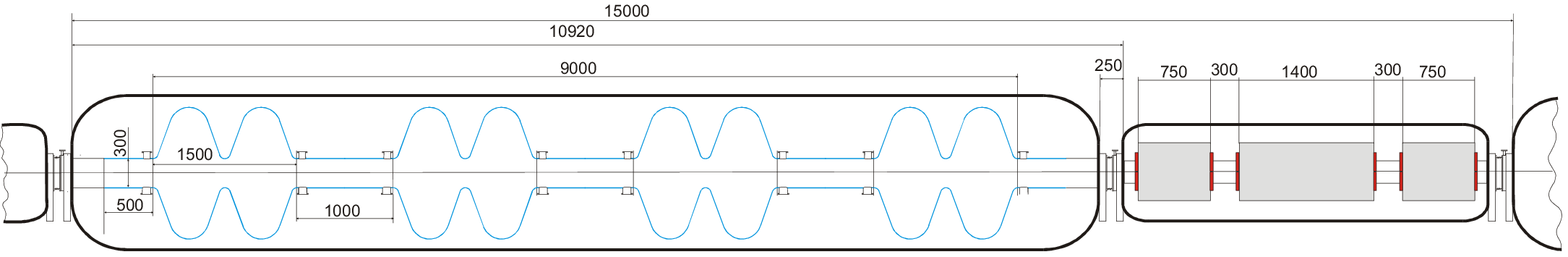}
\caption{Layout of an RLA linac period.}
\label{fig:acc:rlalinac}
\end{figure}

This linac is followed by a single, four-pass recirculating linear
accelerator (RLA) that raises the energy from 2.5 GeV to 20 GeV. The RLA
uses the same layout of four 2-cell superconducting rf cavity structures as
the long cryomodules in the linac, but utilizes quadrupole triplet focusing,
as indicated in Fig.~\ref{fig:acc:rlalinac}. The arcs have an average radius
of 62 m. The final arc has a dipole field of 2 T. 

In Study-I, where the final beam energy was chosen to be 50 GeV, a second
RLA is needed. This second RLA is similar to the first RLA but considerably
larger.

\begin{table}[tbh]
\caption{Main parameters of the muon accelerator driver.}
\label{tab:acc:parm}
\begin{center}
\begin{tabular}{lc}
\hline
Injection momentum (MeV/$c$)/Kinetic energy (MeV) & 210/129.4 \\ 
Final energy (GeV) & 20 \\ 
Initial normalized acceptance (mm-rad) & 15 \\ 
\quad rms normalized emittance (mm-rad) & 2.4 \\ 
Initial longitudinal acceptance, $\Delta pL_{b}/m_{\mu }$ (mm) & 170 \\ 
\quad momentum spread, $\Delta p/p$ & $\pm 0.21$ \\ 
\quad bunch length, $L_{b}$ (mm) & $\pm 407$ \\ 
\quad rms energy spread & 0.084 \\ 
\quad rms bunch length (mm) & 163 \\ 
Number of bunches per pulse & 67 \\ 
Number of particles per bunch\textbf{/}per pulse & $4.4\times 10^{10}$ 
\textbf{/}$3\times 10^{12}$ \\ 
Bunch frequency\textbf{/}accelerating frequency (MHz) & 201.25\textbf{/}
201.25 \\ 
Average beam power (kW) & 150 \\ \hline
\end{tabular}
\end{center}
\end{table}

\subsection{Storage Ring}

After acceleration in the RLA, the muons are injected into the upward-going
straight section of a racetrack-shaped storage ring with a circumference of
358 m. Parameters of the ring are summarized in Table~\ref{SRING:tb}.
High-field superconducting arc magnets are used to minimize the arc length
and maximize the fraction (35\%) of muons that decay in the downward-going
straight and generate neutrinos headed toward the detector located some 3000
km away.

\begin{figure}[tbh]
\begin{center}
\includegraphics[width=4.0in]{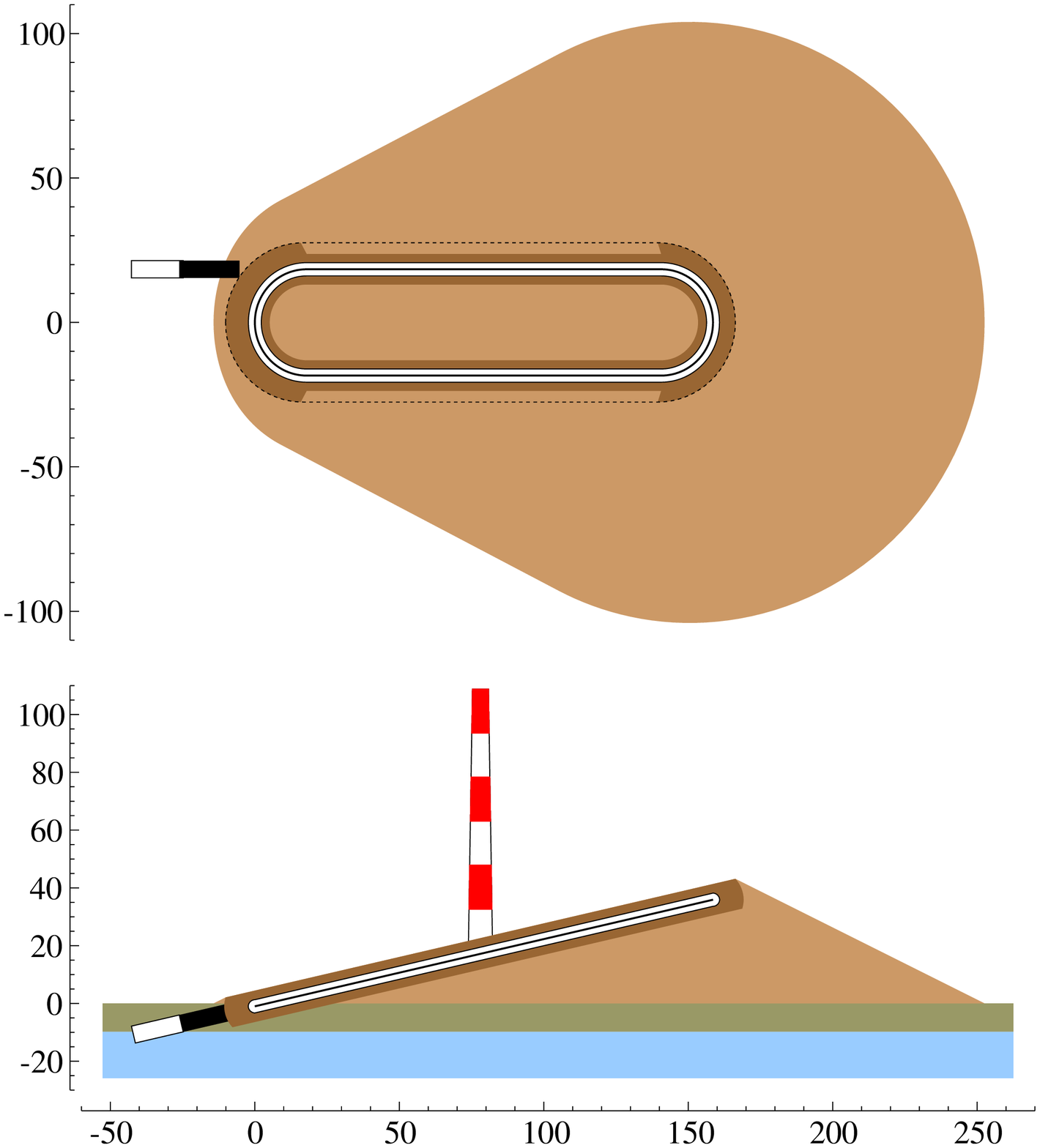}
\end{center}
\caption[Top view and cross section through ring and berm]{Top view and
cross section through 20-GeV ring and berm. The existing 110-m tower, drawn
to scale, gives a sense of the height of the ring on the BNL landscape.}
\label{EPP:fgsection}
\end{figure}

All muons are allowed to decay; the maximum heat load from their decay
electrons is 42 kW (126 W/m). This load is too high to be dissipated in the
superconducting coils. For Study-II, a magnet design has been chosen that
allows the majority of these electrons to pass out between separate upper
and lower cryostats, and be dissipated in a dump at room temperature. To
maintain the vertical cryostat separation in focusing elements, skew
quadrupoles are employed in place of standard quadrupoles. In order to
maximize the average bending field, Nb$_{3}$Sn pancake coils are employed.
One coil of the bending magnet is extended and used as one half of the
previous (or following) skew quadrupole to minimize unused space. Figure~\ref
{EPP:fgsection} shows a layout of the ring as it would be located at BNL.
(The existing 110-m-high BNL stack is shown for scale.) For site-specific
reasons, the ring is kept above the local water table and is placed on a
roughly 30-m-high berm. This requirement places a premium on a compact
storage ring.

\begin{table}[tb]
\caption{Muon storage ring parameters.}
\label{SRING:tb}
\begin{center}
\begin{tabular}{ll}
\hline
Energy (GeV) & 20 \\ 
Circumference (m) & 358.18 \\ 
Normalized transverse acceptance (mm-rad) & 30 \\ 
Energy acceptance (\%) & 2.2 \\ 
\multicolumn{2}{c}{Arc} \\ 
Length (m) & 53.09 \\ 
No. cells per arc & 10 \\ 
Cell length (m) & 5.3 \\ 
Phase advance ($\deg $) & 60 \\ 
Dipole length (m) & 1.89 \\ 
Dipole field (T) & 6.93 \\ 
Skew quadrupole length (m) & 0.76 \\ 
Skew quadrupole gradient (T/m) & 35 \\ 
$\beta _{\text{max}}$ (m) & 8.6 \\ 
\multicolumn{2}{c}{Production Straight} \\ 
Length (m) & 126 \\ 
$\beta _{\text{max}}$ (m) & 200 \\ \hline
\end{tabular}
\end{center}
\end{table}

For Study-I, a conventional superconducting ring was utilized to store the
50 GeV muon beam. The heat load from muon decay products in this scenario is
managed by having a liner inside the magnet bore to absorb the decay
products. This approach is likewise available for BNL, provided some care is
taken to keep the ring compact; acceptable solutions have been found for
this option as well.

\begin{figure}[tbh]
\begin{center}
\includegraphics[width=0.9\linewidth]{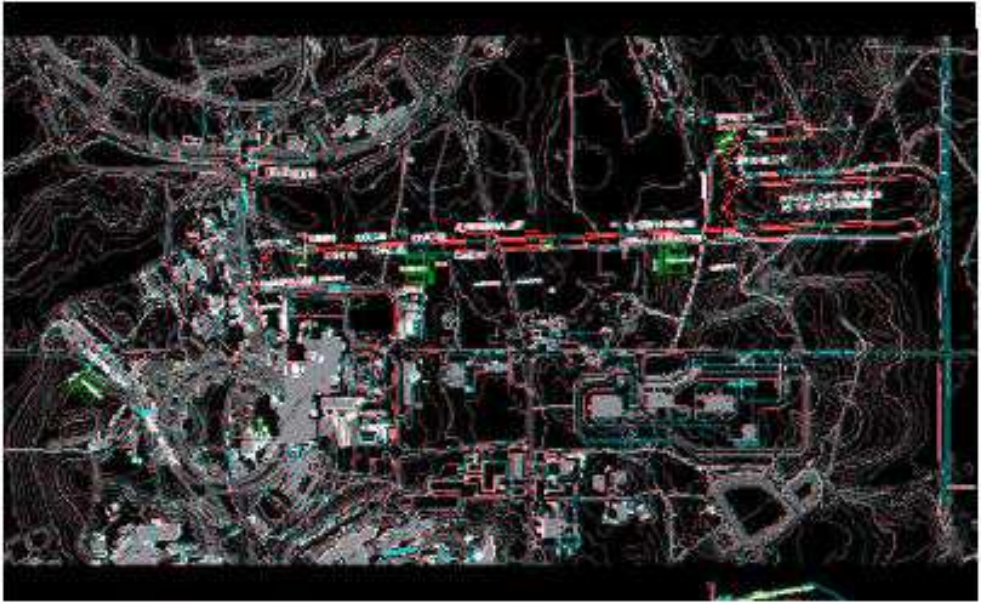}
\end{center}
\caption[Schematic of a neutrino factory at Brookhaven]{Schematic of a
20-GeV Neutrino Factory at BNL.}
\label{bnlsite}
\end{figure}

\begin{figure}[tbh]
\begin{center}
\includegraphics[width=0.9\linewidth]{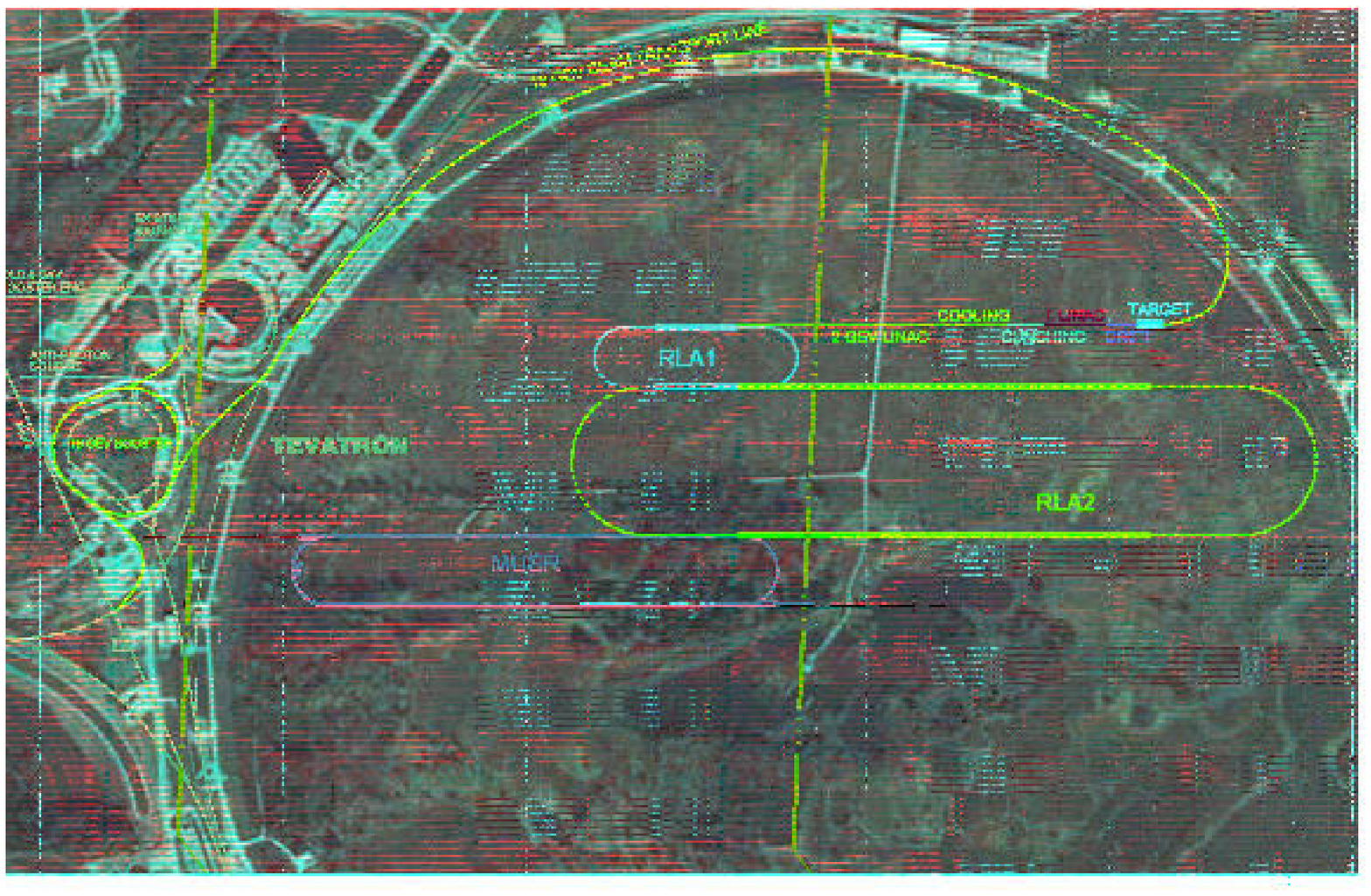}
\end{center}
\caption[Schematic of a neutrino factory at Fermilab]{Schematic of a 50-GeV
Neutrino Factory at Fermilab.}
\label{fnalsite}
\end{figure}

An overall layout of the Neutrino Factory on the BNL site is shown in Fig.~%
\ref{bnlsite}. 
Figure~\ref{fnalsite} shows the equivalent picture for a facility on the
Fermilab site. In this latter case, the layout includes the additional RLA
and longer storage ring needed to reach 50 GeV. Clearly the footprint of a
Neutrino Factory is reasonably small, and such a machine would fit easily on
the site of either BNL or Fermilab. 

\subsection{Detector}

The Neutrino Factory, plus its long-baseline detector, will have a physics
program that is a logical continuation of current and near-future neutrino
oscillation experiments in the U.S., Japan and Europe. Moreover, detector
facilities located in experimental areas near the neutrino source will have
access to integrated neutrino intensities $10^{4}$--$10^{5}$ times larger
than previously available ($10^{20}$ neutrinos per year compared with $%
10^{15}$--$10^{16}$).

The detector site taken for Study-II is the Waste Isolation Pilot Plant
(WIPP) in Carlsbad, New Mexico. The WIPP site is approximately 2900~km from
BNL. Space is potentially available for a large underground physics facility
at depths of 740--1100~m, and discussions are under way between DOE and the
UNO project~\cite{DET:uno} on the possible development of such a facility.

\subsubsection{Far Detector}

Specifications for the long-baseline Neutrino Factory detector are rather
typical for an accelerator-based neutrino experiment. However, because of
the need to maintain a high neutrino rate at these long distances, the
detectors considered here are 3--10 times more massive than those in current
neutrino experiments. Clearly, the rate of detected neutrinos depends on two
factors---the source intensity and the detector size. In the actual design
of a Neutrino Factory, these two factors must be optimized together.

\begin{figure}[tbh]
\begin{center}
\includegraphics[width=3.5in]{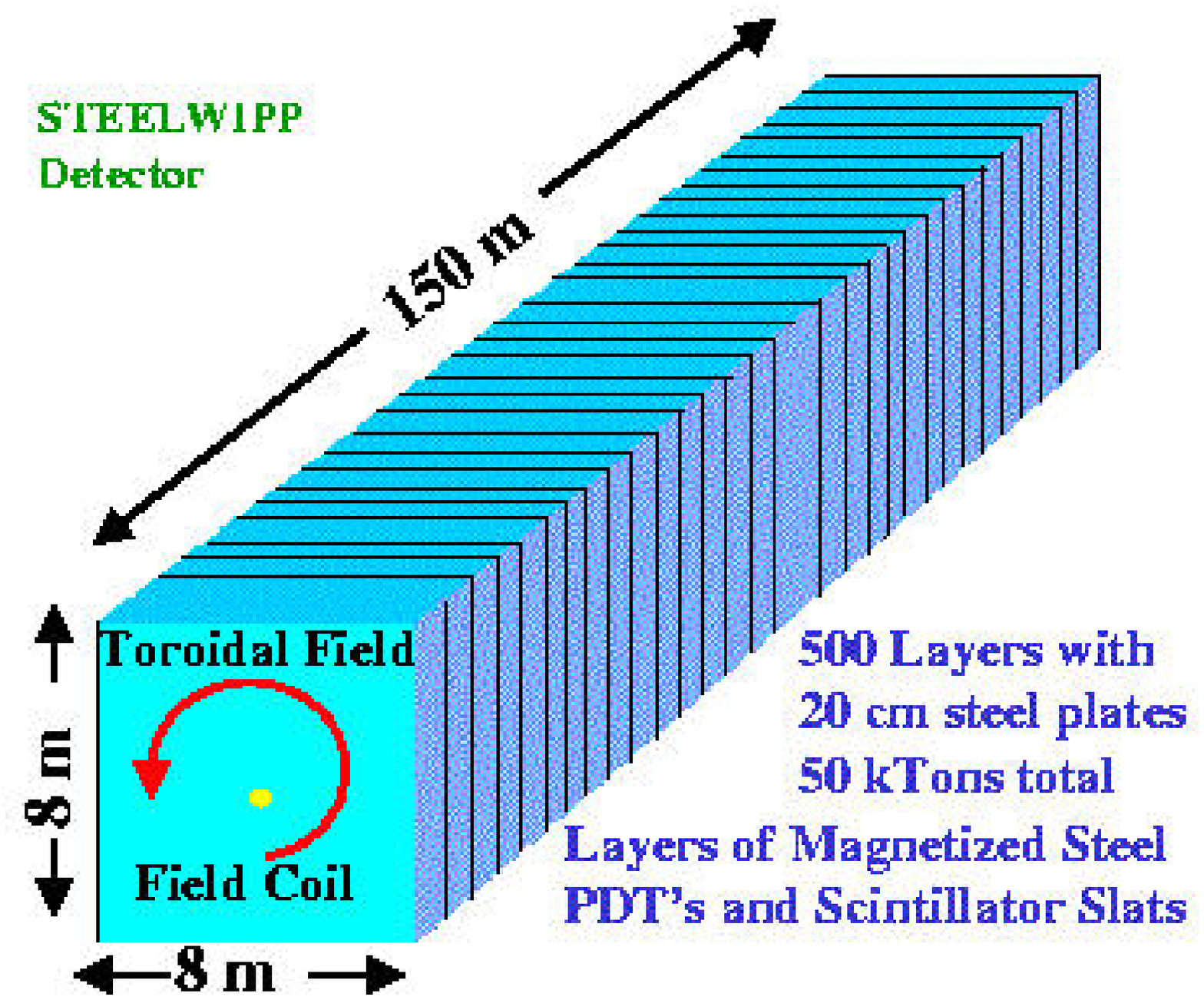}
\end{center}
\caption[A possible 50 kton Steel/Scintillator/PDT detector at WIPP]{A
possible 50-kton steel-scintillator-PDT detector at WIPP.}
\label{fg:steelwipp}
\end{figure}

Two options are considered for the WIPP site: a 50 kton
steel--scintillator--proportional-drift-tube (PDT) detector or a
water-Cherenkov detector. The PDT detector would resemble MINOS. Figure~\ref
{fg:steelwipp} shows a 50-kton detector with dimensions $8~\text{m}\times 8~%
\text{m}\times 150$~m. A detector of this size would record up to $4\times
10^{4}$ $\nu _{\mu }$ events per year. 

\begin{figure}[tbh]
\begin{center}
\includegraphics[width=3.0in]{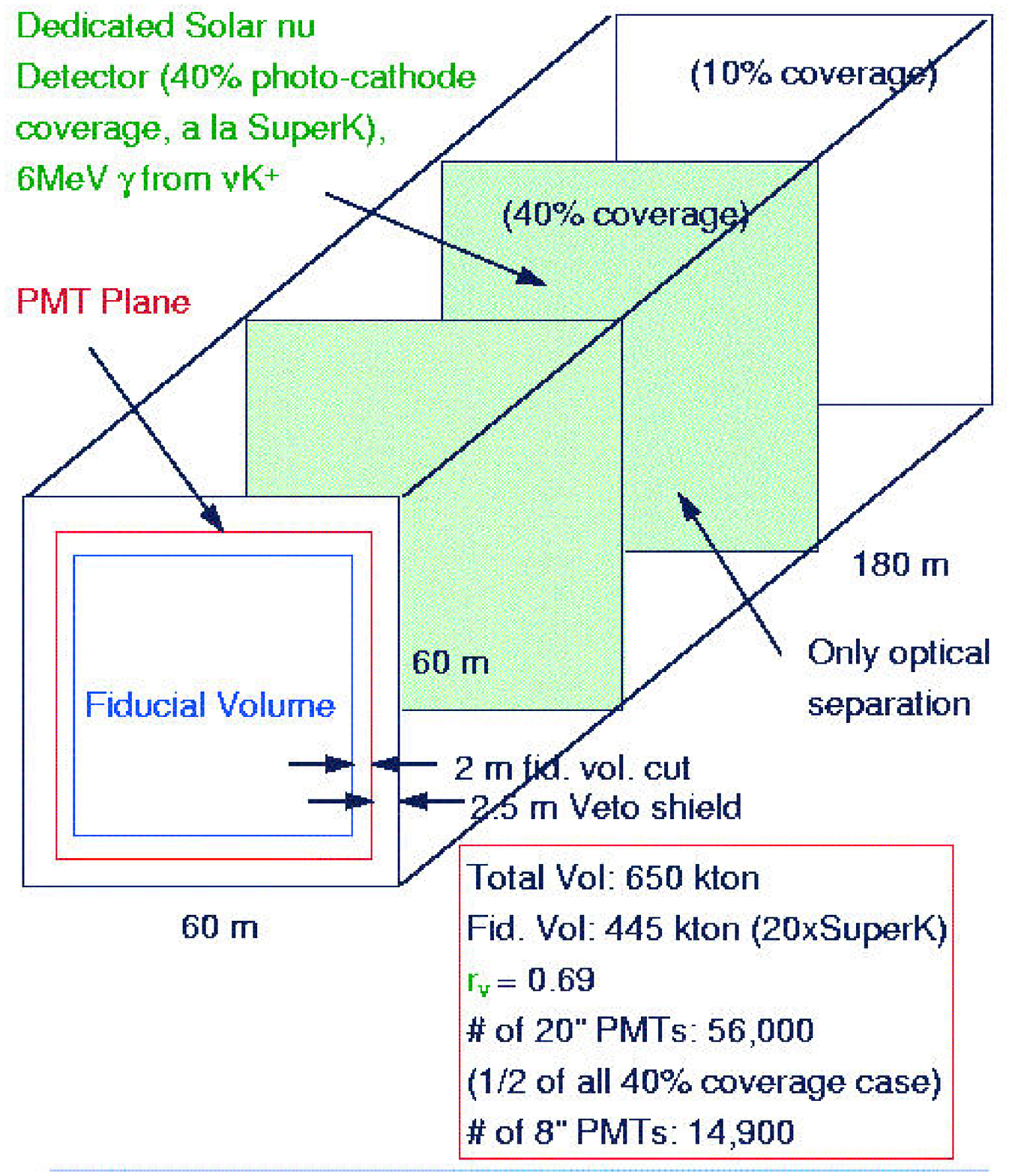}
\end{center}
\caption[Block schematic of the UNO detector]{Block schematic of the UNO
detector, including initial design parameters.}
\label{fg:unodet}
\end{figure}

A large water-Cherenkov detector would be similar to SuperKamiokande, but
with either a magnetized water volume or toroids separating smaller water
tanks. The detector could be the UNO detector~\cite{DET:uno}, currently
proposed to study both proton decay and cosmic neutrinos. UNO would be a
650-kton water-Cherenkov detector segmented into a minimum of three tanks
(see Fig.~\ref{fg:unodet}). It would have an active fiducial mass of
440~kton and would record up to $3\,\times \,10^{5}$ $\nu _{\mu }$ events
per year from the Neutrino Factory beam. 

Another proposal for a neutrino detector is a massive liquid-argon
magnetized detector~\cite{landd} that would attempt to detect proton decay,
solar and supernova neutrinos as well as serve as a Neutrino Factory
detector.

\subsubsection{Near Detector}

As noted, detector facilities located on-site at the Neutrino Factory would
have access to unprecedented intensities of pure neutrino beams. This would
enable standard neutrino physics studies such as ${\sin }^{2}\theta _{W}$,
structure functions, $\nu $ cross sections, nuclear shadowing and pQCD to be
performed with much higher precision than previously obtainable. In addition
to its primary physics program, the near detector can also provide a precise
flux calibration for the far detector, though this may not be critical given
the ability to monitor the storage ring beam intensity independently.

A compact liquid-argon TPC (similar to the ICARUS detector~\cite{ICARUS}),
cylindrically shaped with a radius of 0.5 m and a length of 1~m, would have
an active volume of $10^{3}$ kg and a neutrino event rate \textsl{O}(10~Hz).
The TPC could be combined with a downstream magnetic spectrometer for muon
and hadron momentum measurements. At these $\nu $ intensities, it is even
possible to have an experiment with a relatively thin Pb target (1~$L_{rad}$%
~), followed by a standard fixed-target spectrometer containing tracking
chambers, time-of-flight and calorimetry, with an event rate \textsl{O}%
~(1~Hz).

\chapter{Muon Colliders} 
\label{higgsfact}

\begin{figure}[bth!]
\includegraphics[width=6in,height=4.75in]{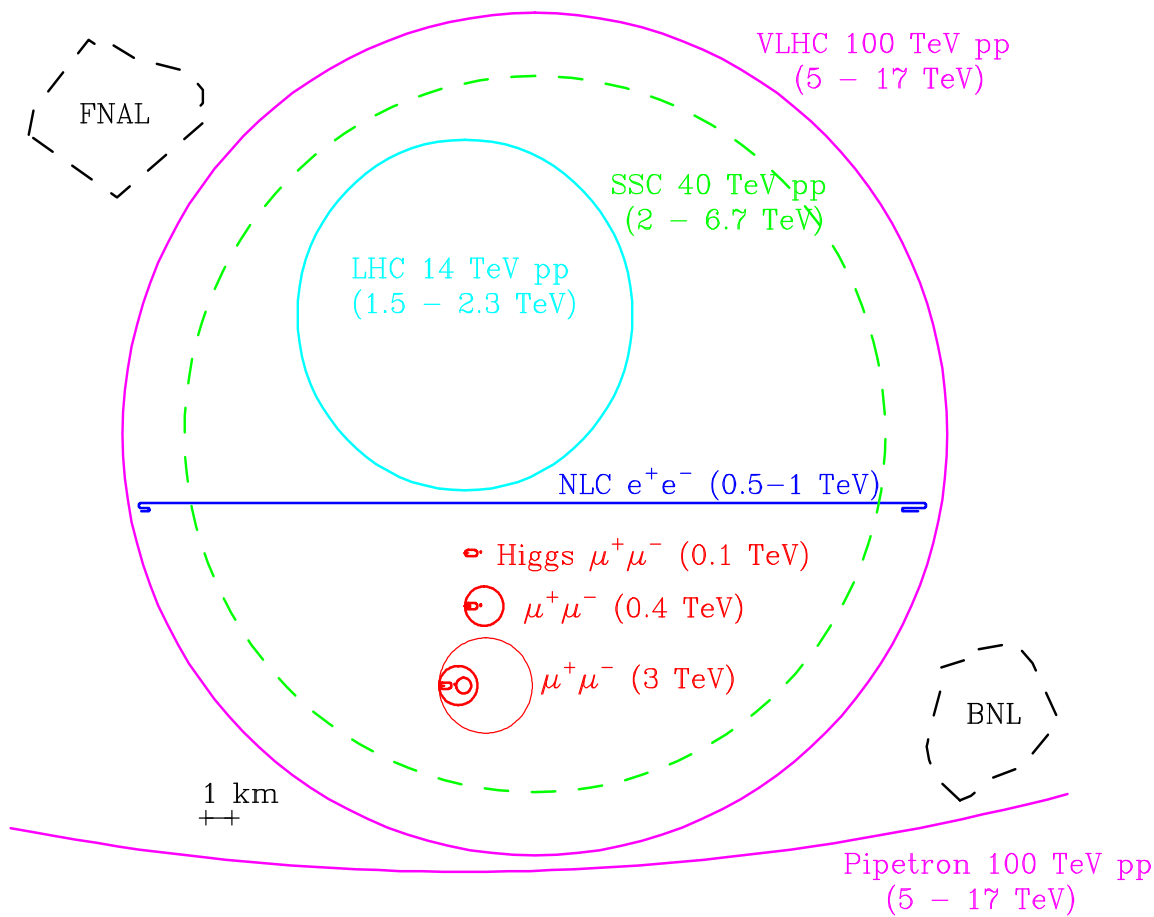}
\vspace{0.5cm}
\caption[Sizes of various proposed high energy colliders]
{Comparative sizes of various proposed high energy colliders compared
with the FNAL and BNL sites. The energies in parentheses give for
lepton colliders their CoM energies and for hadron colliders the
approximate range of CoM energies attainable for hard parton-parton
collisions.}
\label{compare}
\end{figure}

The lure of muon colliders arises from the
fact that the muon is $\approx$ 200 times heavier than the electron
and this makes it possible to accelerate the muon using circular
accelerators that are compact and fit on existing accelerator
sites. See Figure~\ref{compare} for a comparison of relative sizes of
muon colliders ranging from 500 GeV to 3 TeV center of mass energies
with respect to the LHC, SSC, and  NLC. Once we have solved the
problem of cooling a muon beam so that it can be accelerated, higher
enegies are much more easily obtained in a muon collider than in the
linear collider.
Because the muon is unstable, it becomes necessary to cool and
accelerate it before a substantial number have decayed. With typical
bending magnetic fields($\approx$ 5~Tesla) 
available with today's technology,  
the muons last $\approx$ 1000 turns before half
of them have decayed in the collider ring. This is a statement that is
independent of the energy of the collider to first order due to
relativistic time dilatation.

The muon decay also gives rise to large numbers of electrons that can
pose serious background problems for detectors in the collision
region. The 1999 Status Report~\cite{INTRO:ref5} contains an excellent
summary of the problems and possible solutions one faces on the way to
a muon collider.

\begin{figure}[bth!]
\centerline{\includegraphics[width=0.6\linewidth]{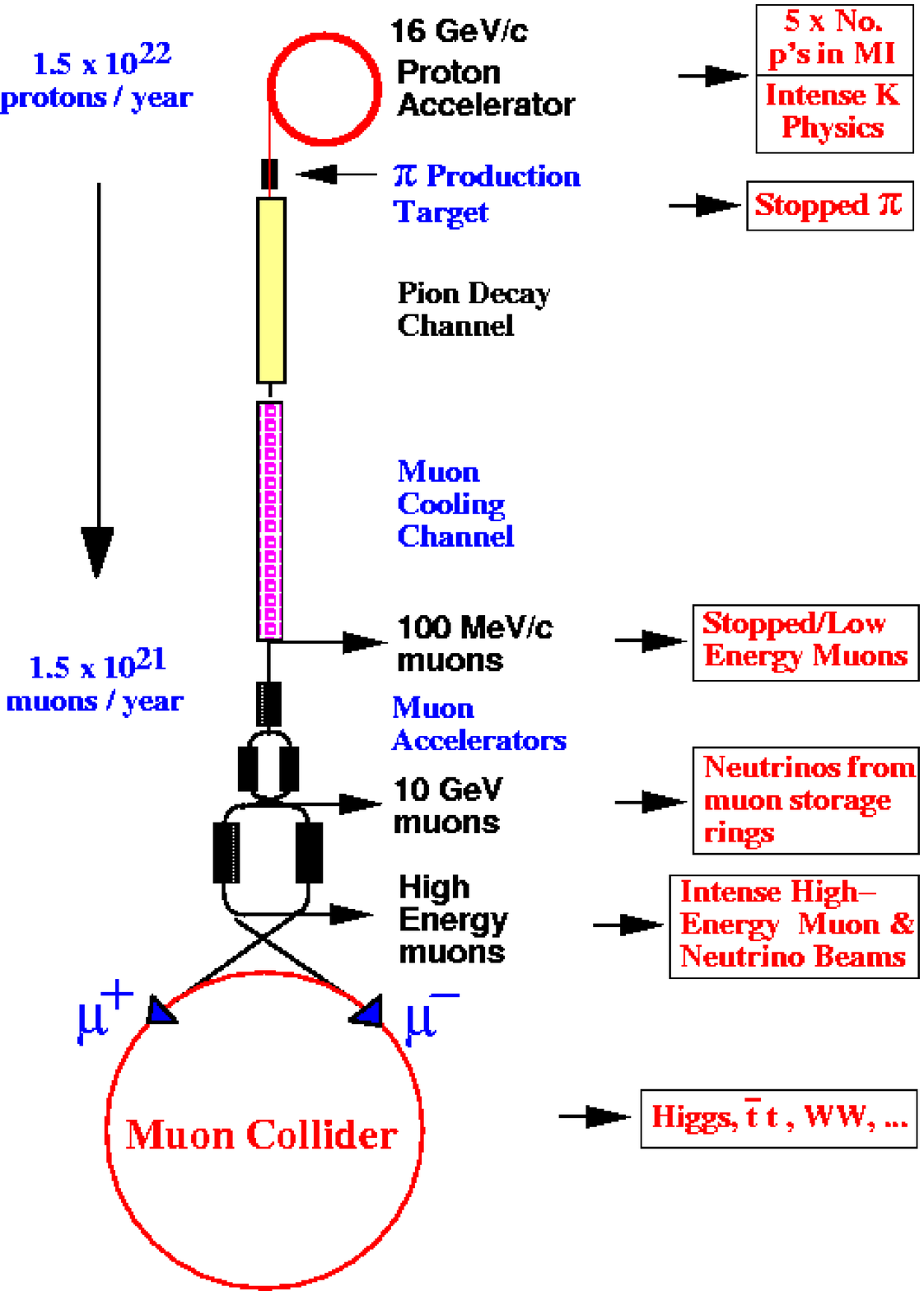}}
\vspace{0.5cm}
\caption[Schematic of a  muon collider]{Schematic of a  muon collider.}
\label{schematic}
\end{figure}

Figure~\ref{schematic} shows  a schematic of such a muon
collider, along with a depiction of the possible physics that 
can be addressed with each stage of the facility.

\section{Higgs Factory Requirements}
The emittance of the muon beam needs to be
reduced by a factor 10$^6$ from production~\cite{INTRO:ref5} 
to the point of collision
for there to be significant luminosity for experiments. This can be
achieved by ionization cooling similar to the scheme described in
chapter~\ref{neufact}. The transverse emittance is reduced during
ionization cooling, since only the longitudinal energy loss is replaced
by $rf$ acceleration. However, due to straggling, the longitudinal
emittance grows. In order to cool longitudinally, one exchanges
longitudinal and transverse emittances and proceeds to cool the
transverse emittance.

The Status report~\cite{INTRO:ref5} outlines the details of the
acceleration and collider ring for the Higgs factory. Table \ref{sum}
gives a summary of the parameters of various muon colliders including
three different modes of running the Higgs Collider that have varying
beam momentum spreads. Additional information  about the 
Muon Collider can be found at~\cite{gail,higgsreport}.

\begin{table*}[thb!]
\centering  
\caption[Baseline parameters for high- and low-energy muon colliders. ]
{Baseline parameters for high- and low-energy muon colliders.
Higgs/year assumes a cross section $\sigma=5\times 10^4$~fb; a Higgs
width $\Gamma=2.7$~MeV; 1~year = $10^7$~s.}
\label{sum}
\begin{tabular}{|l|c|c|c|c|c|}
\hline
\rr CoM energy~ TeV   &\rr 3 &\rr 0.4 &
\multicolumn{3}{c}{\rr 0.1 }  \\
$p$ energy~GeV & 16 & 16 & \multicolumn{3}{c}{16}\\ 
$p$'s/bunch & $2.5\times 10^{13}$ & $2.5\times 10^{13}$ &
\multicolumn{3}{c}{$5\times 10^{13}$  }  \\  
Bunches/fill &  4 & 4 & \multicolumn{3}{c}{2 } \\ 
Rep.~rate~Hz & 15
& 15 & \multicolumn{3}{c}{15 } \\ 
$p$ power~MW & 4 & 4 &\multicolumn{3}{c}{4} \\ 
$\mu$/bunch  & $2\times 10^{12}$ & $2\times10^{12}$ &\multicolumn{3}{c}{$4\times 10^{12}$ }  \\
\rr $\mu$ power~MW     & \rr 28 &\rr 4 & \multicolumn{3}{c}{\rr 1 }  \\
\rr Wall power~MW    &  \rr  204 &\rr 120  & \multicolumn{3}{c}{\rr
81 } \\ 
Collider circum.~m & 6000 & 1000 & \multicolumn{3}{c}{350 }\\ 
Ave bending field~T & 5.2 & 4.7 &\multicolumn{3}{c}{3 } \\
\hline
\rr Rms ${\Delta p/p}$~\%          &\rr 0.16 &\rr 0.14 &\rr
0.12 &\rr 0.01&\rr 0.003 \\
\hline
6-D $\epsilon_{6,N}$~$(\pi \textrm{m})^3$&$1.7\times
10^{-10}$&$1.7\times 10^{-10}$&$1.7\times 10^{-10}$&$1.7\times
10^{-10}$&$1.7\times 10^{-10}$\\ 
Rms $\epsilon_n$~$\pi$ mm-mrad & 50
& 50 & 85 & 195 & 290\\ 
$\beta^*$~cm & 0.3 & 2.6 & 4.1 & 9.4 &
14.1\\ 
$\sigma_z$~cm & 0.3 & 2.6 & 4.1 & 9.4 & 14.1 \\ 
$\sigma_r$spot~$\mu$m & 3.2 & 26 & 86 & 196 & 294\\ 
$\sigma_{\theta}$ IP~mrad
& 1.1 & 1.0 & 2.1 & 2.1 & 2.1\\ 
Tune shift  &0.044 &0.044 & 0.051
&0.022 & 0.015\\ 
$n_{\rm turns}$ (effective)  & 785 & 700 & 450 & 450
& 450 \\
\hline
\rr Luminosity~cm$^{-2}$s$^{-1}$&\rr $7\times 10^{34}$ & $10^{33}$ &\rr
$1.2\times 10^{32}$ &\rr $2.2\times 10^{31}$&\rr $10^{31}$ \\ 
 & & &
& & \\ 
Higgs/year  & & & $1.9\times 10^3$ & $4\times 10^3$ &
$3.9\times 10^3$ \\
\hline
\end{tabular}
\end{table*}
\begin{figure*}[tbh!]
\includegraphics[height=2.9in,width=5.7in]{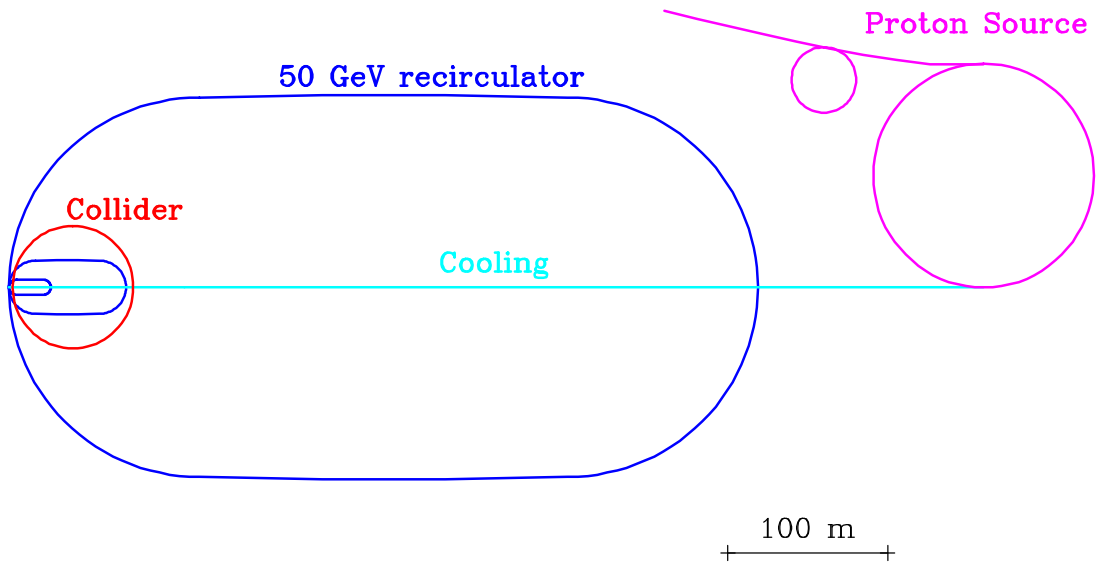}
\caption[Plan of a 0.1-TeV-CoM muon collider]
{Plan of a 0.1-TeV-CoM muon collider.}
\label{plan1}
\end{figure*}

\section{Longitudinal Cooling}

\begin{figure}[tbh!]
\centerline{\includegraphics[width=4.in,height=4in]{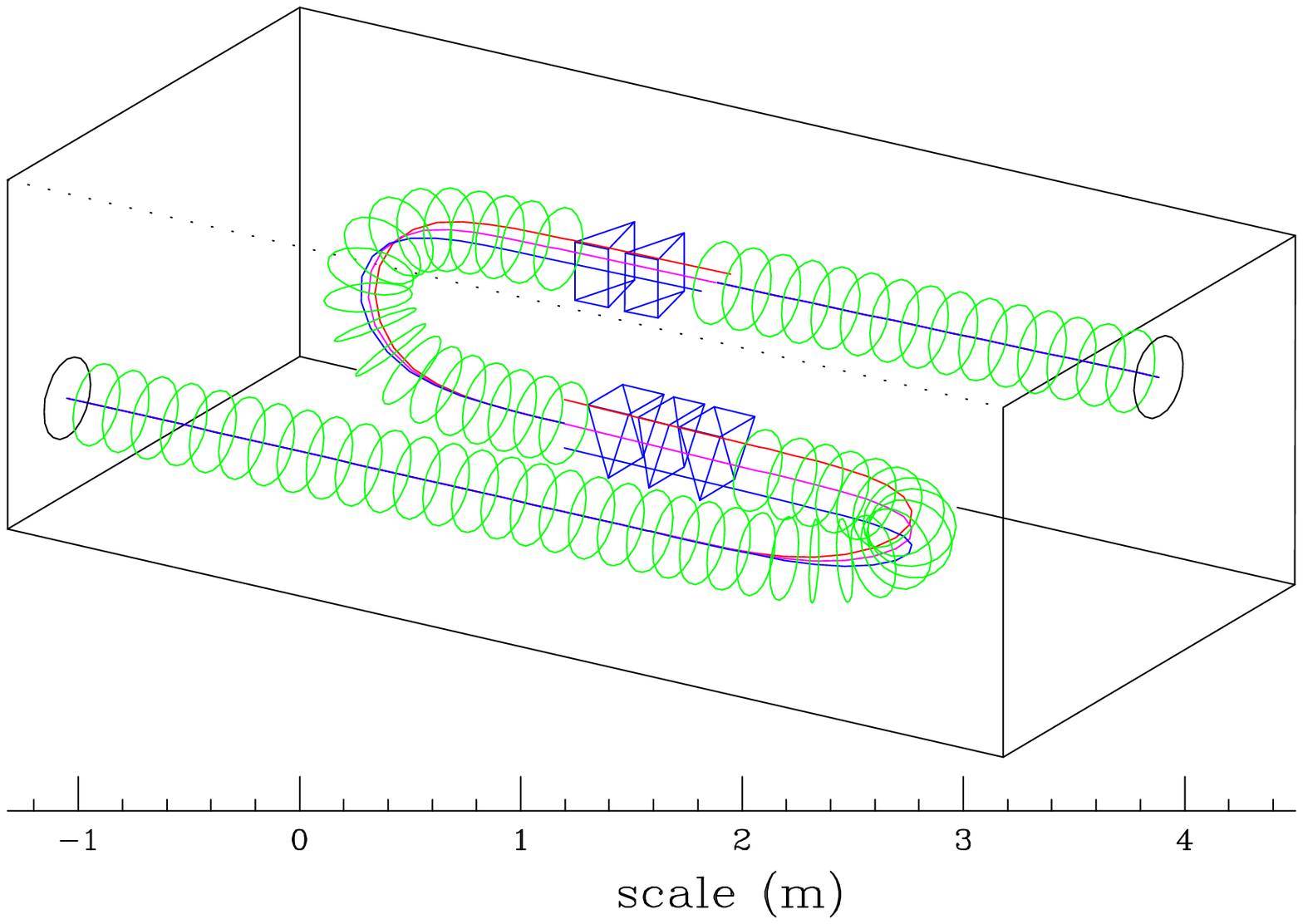}}
\caption[Bent Solenoid emittance exchange example]
{Representation of a bent solenoid longitudinal emittance exchange section.}
\label{cell_exch}
\end{figure}

\begin{figure}[bth!]
\centerline{\includegraphics[width=0.5\linewidth]{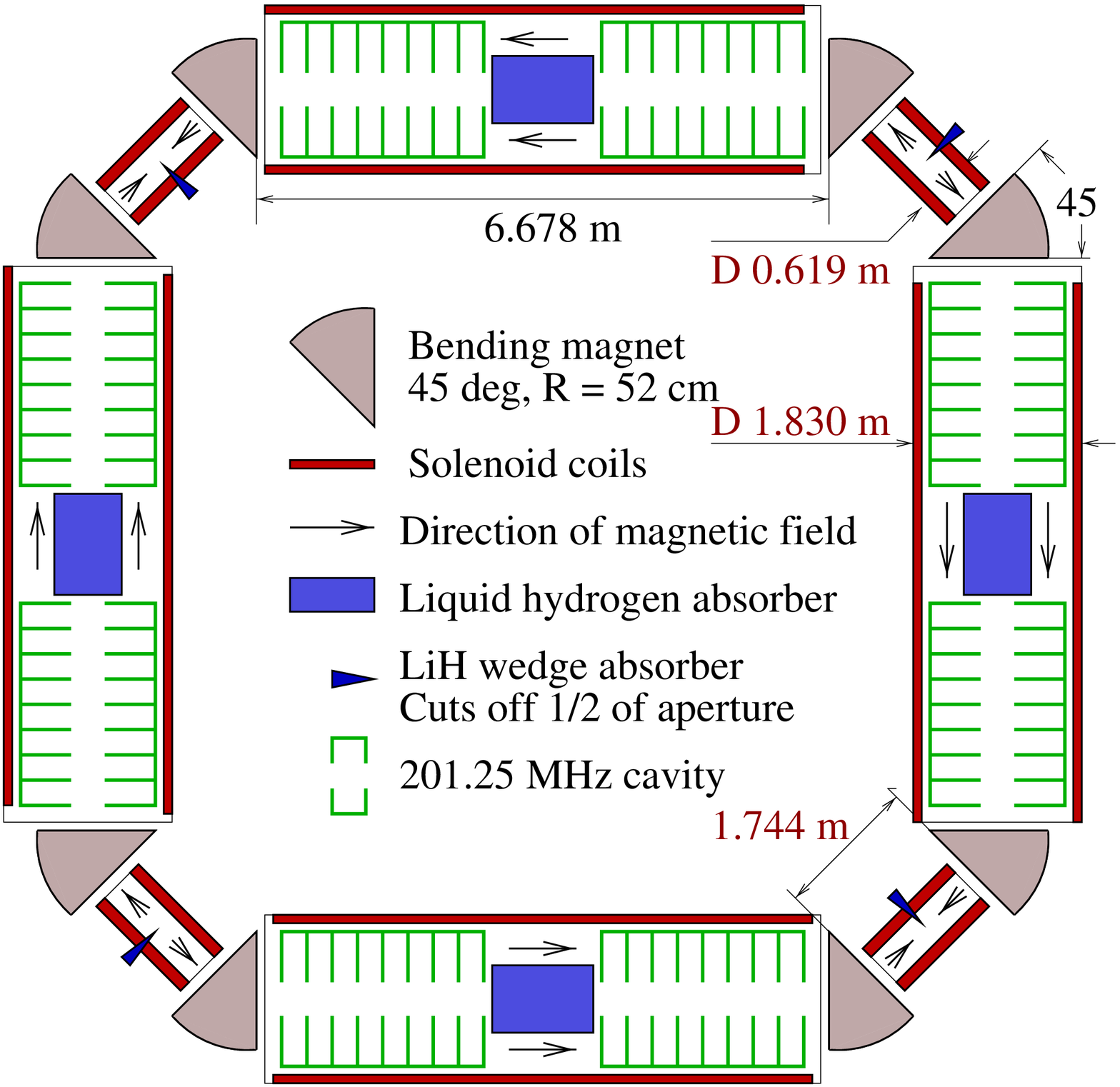}}
\caption[Example of a ring cooler]{Example of a Ring Cooler.}
\label{ring-cooler}
\end{figure}

\begin{figure}[bth!]
\centerline{\includegraphics[width=0.5\linewidth]{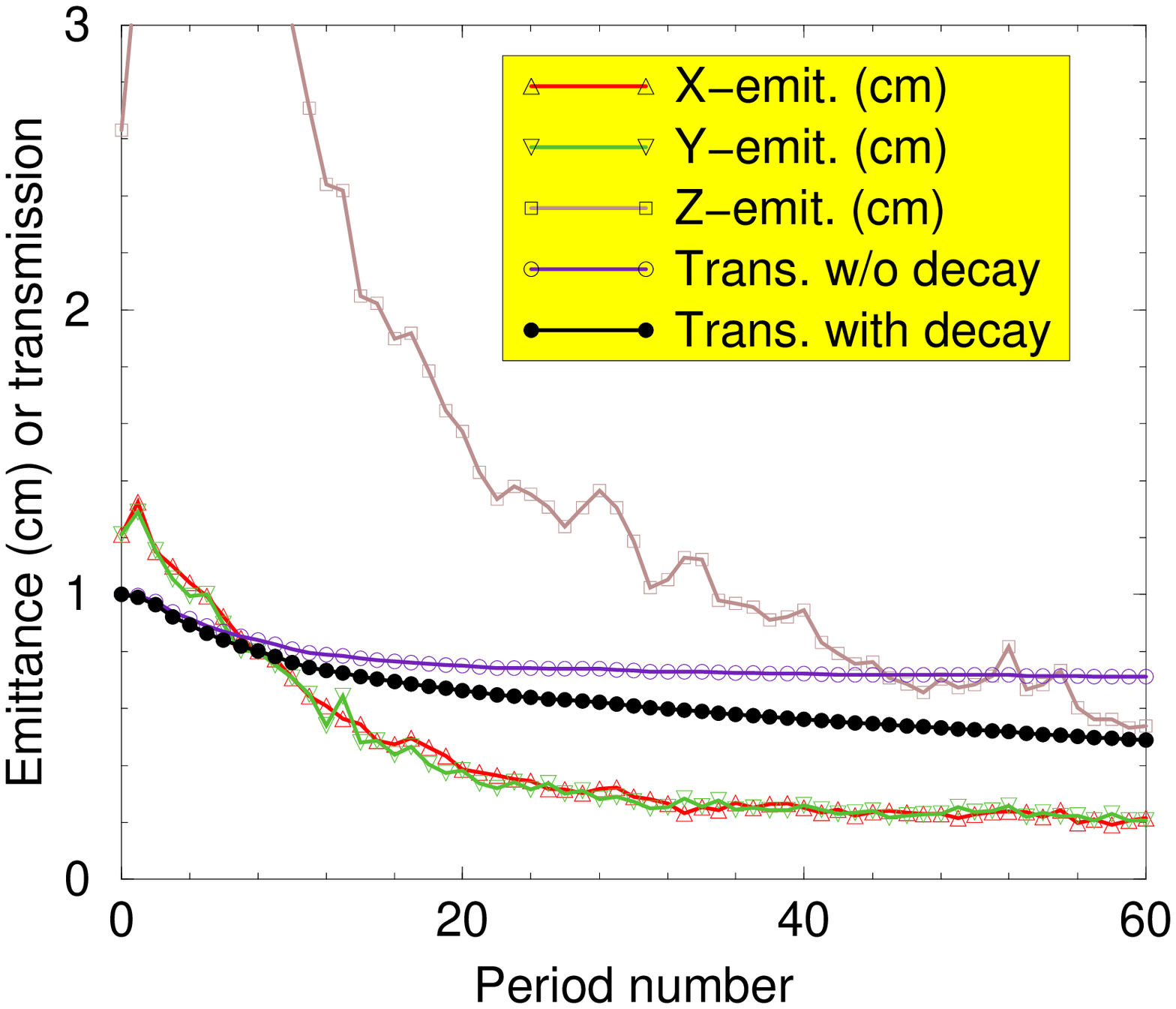}}
\caption[Cooling in a ring cooler]
{First results from a ring cooler showing 6 dimensional cooling.
 Both the transverse and longitudinal emittance plots show cooling. 
The transmission with and without muon decay is shown as a function of turn.
The horizontal axis shows period number, 4 periods making one turn.}
\label{6dcooling}
\end{figure}

Currently there is no satisfactory solution for
emittance exchange and this remains a major stumbling block towards
realizing a muon collider. Figure~\ref{cell_exch} shows one of
the schemes that are under consideration to solve the emittance
exchange problem.
 Ring Coolers have been introduced by Balbekov~\cite{BalbekovRing}.
 Figure~\ref{ring-cooler} shows a ring design currently under study.  The
 advantage of ring coolers is that one can circulate the muons in a
 cooling ring of circumference $\approx$ 30 m and circulate the muons
 many turns, thereby reusing the cooling channel elements.  It can be
 shown that such devices can cool in 6-D space. Figure~\ref{6dcooling}
 shows a simulation of the ring cooler that demonstrates cooling in 6
 dimensions. It should be emphasized that this simulation uses
 idealized magnetic fields. A study is currently under way to see if
 these results hold up with realistic magnetic fields.

 Ring coolers hold the promise to solve the emittance exchange
 problem. However, it has yet to be demonstrated that it is possible
 to inject into and extract from them high emittance beams. If these
 problems can be solved, it may be possible to cascade a number of
 ring coolers each providing cooling by a factor $\approx$ 30 to
 achieve the needed factor of 10$^6$.
Another ring cooling design by Al Garren~\cite{hansoncline} operates 
at higher energies and also holds the promise of 6-D cooling. 
If one can solve the longitudinal cooling problem,
 both neutrino factories and muon colliders will benefit. 
\section{Higher Energy Muon colliders}
Once the cooling problems have been solved to realize the first muon
collider, acceleration to higher energies becomes possible.  Colliders
with 4 TeV center of mass energy have been studied~\cite{INTRO:ref5}
and Table \ref{dntable} lists the parameters for such a collider. The
radiation from the neutrinos from the muon decay begins to become a
problem at CoM energies of 3 TeV. One may attempt to solve this by a number of
means, including optical stochastic cooling of muons in the collider, whereby
one can get the same luminosity with less intensity.

\begin{table*}[tbh!]
 \caption[Parameters of Acceleration for 4~TeV Collider]
{Parameters of Acceleration for 4~TeV Collider.}
\label{dntable}
\begin{tabular}{|l|c|c|c|c|c|}
\hline
                       & Linac & RLA1 & RLA2 & RCS1 & RCS2 \\
\hline
E (GeV) & 0.1$\rightarrow$ 1.5 & 1.5 $\rightarrow$ 10 & 10
$\rightarrow$ 70 & 70 $\rightarrow$ 250 & 250 $\rightarrow$ 2000 \\
f$_{rf}$ (MHz) & 30 $\rightarrow$ 100 & 200 & 400 & 800 & 1300 \\
N$_{turns}$ & 1 & 9 & 11 & 33 & 45 \\ V$_{rf}$(GV/turn) & 1.5 & 1.0 &
6 & 6.5 & 42 \\ C$_{turn}$(km) & 0.3 & 0.16 & 1.1 & 2.0 & 11.5 \\ Beam
time (ms) & 0.0013 & 0.005 & 0.04 & 0.22 & 1.73 \\
$\sigma_{z,beam}$(cm) & 50 $\rightarrow$ 8 & 4 $\rightarrow$ 1.7 & 1.7
$\rightarrow$ 0.5 & 0.5 $\rightarrow$ 0.25 & 0.25 $\rightarrow$ 0.12
\\ $\sigma_{E,beam}$(GeV) & 0.005 $\rightarrow$ 0.033 & 0.067
$\rightarrow$ 0.16 & 0.16 $\rightarrow$ 0.58 & 0.58 $\rightarrow$ 1.14
& 1.14 $\rightarrow$ 2.3 \\ Loss (\%) & 5 & 7 & 6 & 7 & 10 \\
\hline
\end{tabular}
\end{table*}
\section{Muon Collider Detectors} 
\begin{figure}[bth!]
\centerline{\includegraphics[width=0.5\linewidth]{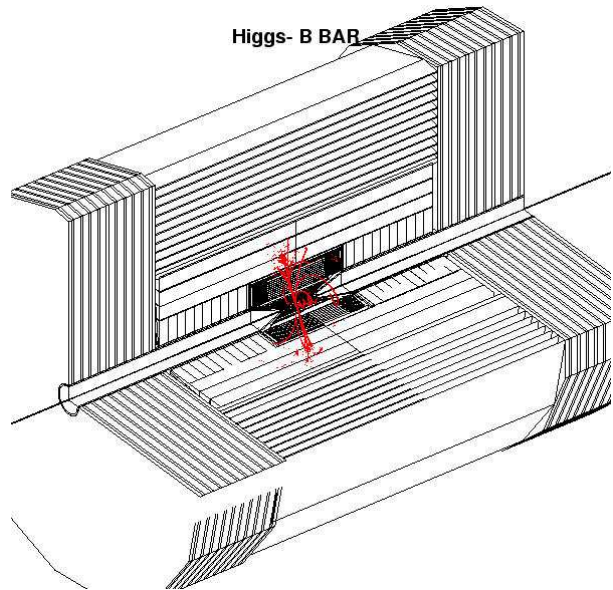}}
\caption[Strawman Geant detector for a muon collider]
{Cut view of a strawman detector in GEANT for the Higgs factory with
a Higgs$\rightarrow b\bar b$ event superimposed. No backgrounds
shown. The tungsten cones on either side of the interaction region
mask out a 20$\deg$ area.}
\label{geant}
\end{figure}

Figure~\ref{geant} shows a strawman
muon collider detector for a Higgs factory simulated in Geant. The
background from muon decay sources has been extensively 
studied~\cite{INTRO:ref5}. At the Higgs factory, 
the main sources of background are from photons generated by the
showering of muon decay electrons. At the higher energy colliders,
Bethe-Heitler muons produced in electron showers become a problem.
Work was done to optimize the shielding by using specially shaped
tungsten cones~\cite{INTRO:ref5}. 
The background rates obtained were shown to be
similar to those predicted for the LHC experiments. It still needs to
be established whether pattern recognition is possible in the presence
of these backgrounds.

\chapter{Costs and Staging Options}

\label{costs}

\section{Costs}

\subsection{Methodology}

We have specified each system of the Study-II Neutrino Factory in sufficient
detail to obtain a ``top-down'' cost estimate for it. Clearly this estimate
is not the complete and detailed cost estimate that would come from
preparing a full Conceptual Design Report (CDR). However, there is
considerable experience in designing and building accelerators with similar
components, so we have a substantial knowledge base from which costs can be
derived. The costs summarized here were obtained mainly in that way.

Where available, we have used costs from existing components---scaled as
needed to reflect essential changes---to represent the expected costs to
fabricate what we need. This applies to the Proton Driver, the
superconducting and normal conducting magnets and their power supplies, the
rf cavities, and conventional facilities and utilities. In some cases, we
were able to take advantage of experience in designing similar components in
a different context. For example, the target facility we require closely
resembles that needed for the Spallation Neutron Source (SNS) project at
ORNL, for which detailed CDR-level designs already exist and construction is
under way. The superconducting target solenoid is not a standard device, but
there is a magnet of similar size and field strength, designed for the ITER
project, that serves as a convenient scaling model. In the case of rf power
sources, we made use of the multi-beam klystron (MBK) example developed at
DESY for TESLA, along with expertise in developing other high-power tubes at
U.S. Laboratories. For devices such as the MBK, which are a significant
extrapolation from existing hardware, allowance was made for a substantial
development program, whose cost was amortized over the initial complement of
devices needed for the Neutrino Factory.

\subsection{Facility Costs}

The Neutrino Factory design we describe here favors feasibility over cost
reduction. Thus, we do not claim to present a fully cost-optimized design,
nor one that has been reviewed from the standpoint of ``value engineering.''
\ In that sense, there is hope that a detailed design study will \textit{
reduce} the costs compared with what we estimate here. We have put in an
allowance of 10\% for each of the systems to account for things we have not
considered in detail at this stage. \ Only direct costs are included here,
that is, the estimates do not contain allowances for EDIA, laboratory
overhead burdens, or contingency. The breakdown by system is summarized in
Table~\ref{costtotal}; these costs are given in FY01 dollars. However, to
facilitate comparison with the Feasibility Study-I estimate, in the last
column of Table~\ref{costtotal} we converted our costs to FY00 dollars,
using the DOE-approved inflation factor of 2.5\%.

It is interesting to compare our estimate with that of Study-I; in this
study, we have improved the performance by a factor of six over that reached
in Study-I, at a total cost (estimated in the same way for both designs) of
about 3/4 of that in the original study. This is an encouraging trend and,
as noted, we have some hope that it will continue.

\begin{table}[tbp]
\caption[Construction costs for Study-II Neutrino Factory]{Summary of
construction cost totals for Study-II Neutrino Factory. All costs are in
FY01 dollars unless otherwise noted.}
\begin{center}
\begin{tabular}{lcccc}
\hline
{\small System} & {\small Sum} & {\small Others$^{a}$} & {\small Total} & 
{\small Reconciliation$^{b}$} \\ 
& {\small (\$M)} & {\small (\$M)} & {\small (\$M)} & {\small (FY00 \$M)} \\ 
\hline
{\small Proton Driver} & \multicolumn{1}{r}{168.0} & \multicolumn{1}{r}{16.8}
& \multicolumn{1}{r}{184.8} & \multicolumn{1}{r}{180.0} \\ 
{\small Target Systems} & \multicolumn{1}{r}{92.0} & \multicolumn{1}{r}{9.2}
& \multicolumn{1}{r}{101.2} & \multicolumn{1}{r}{98.0} \\ 
{\small Decay Channel} & \multicolumn{1}{r}{4.6} & \multicolumn{1}{r}{0.5} & 
\multicolumn{1}{r}{5.1} & \multicolumn{1}{r}{5.0} \\ 
{\small Induction Linacs} & \multicolumn{1}{r}{319.0} & \multicolumn{1}{r}{
31.9} & \multicolumn{1}{r}{350.9} & \multicolumn{1}{r}{343.0} \\ 
{\small Bunching} & \multicolumn{1}{r}{69.0} & \multicolumn{1}{r}{6.9} & 
\multicolumn{1}{r}{75.9} & \multicolumn{1}{r}{74.0} \\ 
{\small Cooling Channel} & \multicolumn{1}{r}{317.0} & \multicolumn{1}{r}{
31.7} & \multicolumn{1}{r}{348.7} & \multicolumn{1}{r}{340.0} \\ 
{\small Pre-accel. linac} & \multicolumn{1}{r}{189.0} & \multicolumn{1}{r}{
18.9} & \multicolumn{1}{r}{207.9} & \multicolumn{1}{r}{203.0} \\ 
{\small RLA} & \multicolumn{1}{r}{355.0} & \multicolumn{1}{r}{35.5} & 
\multicolumn{1}{r}{390.5} & \multicolumn{1}{r}{381.0} \\ 
{\small Storage Ring} & \multicolumn{1}{r}{107.0} & \multicolumn{1}{r}{10.7}
& \multicolumn{1}{r}{117.7} & \multicolumn{1}{r}{115.0} \\ 
{\small Site Utilities} & \multicolumn{1}{r}{127.0} & \multicolumn{1}{r}{12.7
} & \multicolumn{1}{r}{139.7} & \multicolumn{1}{r}{136.0} \\ \hline\hline
{\small Totals} & \multicolumn{1}{r}{1,747} & \multicolumn{1}{r}{175} & 
\multicolumn{1}{r}{1,922} & \multicolumn{1}{r}{1,875} \\ \hline
\multicolumn{5}{l}{$^{a}${\small Others is 10\% of each system to account
for missing items,}} \\ 
\multicolumn{5}{l}{\small as was used in Study-I.} \\ 
\multicolumn{5}{l}{$^{b}${\small Reconciliation represents the Study-II
costs given in FY00}} \\ 
\multicolumn{5}{l}{\small dollars to permit direct comparison with Study-I
costs.} \\ 
\multicolumn{5}{l}{\small The inflation factor used (1/1.025) is per DOE
official rates.}
\end{tabular}
\end{center}
\label{costtotal}
\end{table}

\section{Staging Options\label{StagingOps}}

During the HEPAP sub-panel presentations on Neutrino Factory R\&D that took
place at BNL on April 19, 2001, the MC was asked to discuss the time scale
for arriving at a Muon Collider, and to outline possible staging options
that would lead to a high-performance Neutrino Factory and, at a later date,
a Muon Collider. The discussion below supplements our response to these
questions.

The collaboration has now completed two detailed ``Feasibility Studies'' of
Neutrino Factories. These studies included end-to-end simulations of the
non-conventional parts of the facilities, and sufficient engineering studies
to form the basis for defensible R\&D plans. The schedule presented
indicated that, with adequate R\&D support, construction of a Neutrino
Factory could begin in 2007. Depending upon available resources, this could
be either a complete high-performance Neutrino Factory capable of searching
for CP violation in the lepton sector, or a more modest first step that
would address many outstanding neutrino oscillation questions, and would be
upgradeable in a staged manner to a high-performance facility.

Expanding on the staging possibilities that could be considered as we
proceed toward a Neutrino Factory, we note that there are a number of
options:

\begin{enumerate}
\item  A high-intensity Proton Driver that could be used to provide an
upgraded conventional neutrino ``superbeam'' and also to support a new round
of kaon experiments if desired. We believe construction of the Proton
Driver, based on either the BNL AGS or the FNAL proton source, could begin
in 2--3 years and be completed within 4--5 years from today.

\item  A very high intensity, low energy, muon source making about 10$^{21}$
muons per year available for stopped muon experiments. This might enable $
\mu $ $\rightarrow $ e conversion, for example, to be probed with a
sensitivity exceeding that of the MECO experiment by several orders of
magnitude, and could also support a broad physics community beyond
high-energy physics. We believe construction of the muon facility could
begin 7--8 years from now, or even sooner (in parallel with construction of
the proton source) if it were deemed to have sufficiently high priority.
\end{enumerate}

In addition to the MC work on a Neutrino Factory, we are committed to
developing the technology required for a Muon Collider. Indeed, it was our
interest in Muon Colliders that led us initially to study the production of
intense muon beams. By focusing on the easier problem of the Neutrino
Factory, we have made more rapid progress and, at the same time, established
many of the technologies needed for the Collider. However, this approach has
resulted in a considerable difference in our level of understanding of the
two types of facility.

Unlike the situation for the Neutrino Factory, we do not yet have a complete
scenario, with parameters and chosen technologies, for the Collider. As a
result, we have no end-to-end simulations, little engineering, and no cost
estimates. Many of the required components are natural extensions of those
used in a Neutrino Factory, but some are specific to the Collider (e.g.,
emittance exchange and the Collider ring itself), and many involve large
extrapolations from the Neutrino Factory parameter regime (e.g., the total
amount of cooling, the final beam emittance, the charge per bunch). There
are ideas of how a Collider scenario might look, and there has been
substantial progress on possible technologies for some of these components
(e.g., a ring cooler that would give emittance exchange, and lithium lens
cooling to the required very small final emittance). However, these studies
are far from complete. Thus, we are not yet in a position to conduct the
``feasibility'' study that would be needed to make a complete R\&D plan for
this new and exciting machine.

Before we can reach the feasibility study stage, we must establish robust
technical solutions to emittance exchange, issues related to the high bunch
charges, techniques for cooling to the required final emittances, and the
design of a very low $\beta $* collider ring. We are confident that
solutions exist along the lines we have been investigating, but at present
we do not have enough R\&D support to adequately pursue both the Collider
studies and the Neutrino Factory design studies. Unless and until we obtain
such support, it is hard to predict how long it will take to solve the
emittance exchange and other collider-specific problems---it could be very
soon, or it could be a much longer time.

In view of the above, it should be clear that any estimated time scale
quoted for a Muon Collider has considerable uncertainty. The MC has
previously presented one time scale, based on particular assumptions that
the R\&D for the Collider would mainly follow that for the Neutrino Factory.
Clearly, with more optimistic assumptions a significantly faster time scale
might be possible. We have gone through the exercise of considering such
schedules, and they have the potential to reduce the time to start physics
at a Muon Collider by many years. \ On the other hand, it is also
possible---given the conceptual and technical uncertainties and the current
limits on funding---that the real time scale will be longer.

We in the MC are eager to advance to the stage of building a Muon Collider
on the earliest possible time scale. However, for that to happen there is an
urgent need to greatly increase support for our R\&D so that we can address
the vital issues.

As discussed above, it seems quite possible---perhaps even likely---that the
Neutrino Factory would be built in stages, both for programmatic and for
cost reasons. In what follows we outline a possible staging concept that
provides good physics opportunities with reasonable cost increments. The
staging scenario we consider here is not unique, nor is it necessarily
optimized. Discussions at Snowmass will serve to sharpen the thoughts of the
physics community on what is an optimal staging scenario. Depending on the
results of our technical studies and the results of continued searches for
the Higgs boson, it is hoped that the Neutrino Factory is really the
penultimate stage, to be followed later by a Muon Collider (Higgs Factory).
We assume this scenario in the staging discussion that follows.

Because the physics program would be different at different stages, it is
impractical at this time to consider detector costs. \ Therefore, none of
the costs listed in Section \ref{StagingOps} include any detector costs. To
better represent the incremental costs of a staged approach, utility costs,
which are called out separately in Table \ref{costtotal}, have been
apportioned among the various stages.

\subsection{Stage 1}

In the first stage, we envision a Proton Driver and a Target Facility to
create superbeams. The Driver could begin with a 1 MW beam level (Stage 1)
or could be designed from the outset to reach 4 MW (Stage 1a). (Since the
cost differential between 1 and 4 MW is not large, we do not consider any
intermediate options here.) Because the Proton Driver design is site
specific, cost estimates are slightly different for the BNL and FNAL
options. The Target Facility cost we take from Study-II (see Table \ref
{costtotal}). It is assumed, as was the case for both Study-I and Study-II,
that the Target Facility is built from the outset to accommodate a 4 MW
beam. \ Based on the Study-II results, a 1 MW beam would provide about $
1.2\times 10^{14}$ $\mu $/s ($1.2\times 10^{21}$ $\mu $/year) and a 4 MW
beam about $5\times 10^{14}$ $\mu $/s ($5\times 10^{21}$ $\mu $/year) into a
solenoid channel.

Costs for Stage 1 are summarized in Table \ref{costs1}. If a horn were used
in place of the Study-II solenoid capture system, the intensities may
decrease but the potential cost savings might be in the range of \$30--40M.
This option has not been explored in our Studies to date, and we have not
yet assessed the feasibility of using a horn for the 4 MW scenario. As noted
earlier, the costs in Table \ref{costs1} are only facility costs and do not
include detectors.

In addition to the neutrino program, this stage will also benefit $\pi $, $K$
, and $\overline{p}$ programs, as discussed in~\cite{proton-physics,low-pbar}

\begin{table}[tbp]
\caption[Stage 1 cost estimates]{Stage 1 cost estimates.}
\begin{center}
\begin{tabular}{lllc}
\hline
& BNL option & FNAL 16-GeV option & FNAL 8-GeV option \\ 
& \multicolumn{1}{c}{(\$M)} & \multicolumn{1}{c}{(\$M)} & (\$M) \\ 
\cline{2-4}
1 MW & \multicolumn{1}{c}{310} & \multicolumn{1}{c}{330} & 250 \\ 
4 MW & \multicolumn{1}{c}{400} & \multicolumn{1}{c}{410} & 330 \\ \hline
\end{tabular}
\end{center}
\label{costs1}
\end{table}

\subsection{Stage 2}

In Stage 2, we envision a muon beam that has been phase rotated (to give a
reasonably low momentum spread) and transversely cooled. \ In the notation
of Study-II, this stage takes us to the end of the cooling channel. Thus, we
have access to a muon beam with a central momentum of about 200 MeV/$c$, a
transverse (normalized) emittance of 2.7 mm-rad and an rms energy spread of
about 4.5\%. The intensity of the beam would be about $4\times 10^{13}$ $\mu 
$/s ($4\times 10^{20}$ $\mu $/year) at 1 MW, or $1.7\times 10^{14}$ $\mu $/s
($1.7\times 10^{21}$ $\mu $/year) at 4 MW. The \textit{incremental} cost of
this option is \$840M, based on assuming the cooling channel length adopted
in Study-II. If more intensity were needed, and if less cooling could be
tolerated, the length of the cooling channel could be reduced. \ As an
example, stopping at the end of Lattice 1 instead of the end of Lattice 2
would decrease the incremental cost by about \$180M, with the penalty of
roughly doubling the transverse emittance.

\subsection{Stage 3}

In Stage 3, we envision using the Pre-acceleration Linac to raise the beam
energy to roughly 2.5 GeV. \ The incremental cost of this option is about
\$220M. At this juncture, it may be appropriate to consider a small storage
ring, comparable to the $g-2$ ring at BNL, to be used for the next round of
muon $g-2$ experiments. This ring would, in some sense, be a throw-away
device. \ No cost estimate has been made for this ring, but it would be
expected to cost a few tens of millions. For the present purpose, we take
the cost of such a ring, which we refer to as Stage 3a, as \$30M.

\subsection{Stage 4}

At Stage 4, we envision having a complete Neutrino Factory operating with a
20 GeV storage ring. The incremental cost of this stage, which includes the
RLA and the storage ring, is \$550M. If it were necessary to provide a 50
GeV muon beam as Stage 4a, an additional RLA and a larger storage ring would
be needed. \ The incremental cost to go from Stage 4 to Stage 4a would be an
additional \$700--800M.

Table \ref{Stagecosts} summarizes the incremental costs for each stage of a
Neutrino Factory, exclusive of detector costs.

\begin{table}[tbp]
\caption[Incremental costs for Neutrino factory stages]{Estimated
incremental costs for various possible project stages leading to a Neutrino
Factory, based on the cost estimate from Table \ref{costtotal}. Utility
costs were prorated among the various stages.}
\begin{center}
\begin{tabular}{lc}
\hline
Stage & Incremental Cost \\ 
& (\$M) \\ \cline{2-2}
1 (1 MW Proton Driver) & 250--330 \\ 
1a (4 MW Proton Driver) & 80 \\ 
2 (Cooled muons, 200 MeV/$c$) & 660--840 \\ 
3 (2.5 GeV muons) & 220 \\ 
3a ($g-2$ storage ring) & 30 \\ 
4 (20-GeV Neutrino Factory) & 550 \\ 
4a (50-GeV Neutrino Factory) & 700--800 \\ \hline
\end{tabular}
\end{center}
\label{Stagecosts}
\end{table}

\subsection{Stage 5}

In Stage 5, we can envision an entry level Muon Collider to operate as a
Higgs Factory. No cost estimate has yet been prepared for this stage, so we
mention here only the obvious ``cost drivers.'' First, the initial muon beam
must be prepared as a single bunch of each charge. This may involve an
additional ring for the proton driver to coalesce proton bunches into a
single pulse. The cooling will have to be significantly augmented. First, a
much lower transverse emittance is needed, and second, it will be necessary
to provide emittance exchange to maintain a reasonable transmission of the
muons. The additional cooling will permit going to smaller solenoids and
higher frequency rf systems (402.5 or perhaps 805 MHz), which should lower
the incremental cost somewhat. Next, we will need considerably more
acceleration, though with smaller energy acceptance and aperture
requirements than at present. Lastly, we will need a very low $\beta ^{\ast
} $ lattice for the storage ring, along with mitigation of the potentially
copious background levels near the interaction point. In this case the
detector is, in effect, part of the Collider and cannot be ignored in terms
of its cost impact.

Of the items mentioned, it is likely that the additional cooling and the
additional acceleration are the most significant cost drivers. Future work
will define the system requirements better and permit a cost estimate of the
same type provided for Studies-I and -II.

\chapter{R\&D Program}

\label{r_and_d}

\section{Introduction}

Successful construction of a muon storage ring to provide a copious source
of neutrinos requires many novel approaches to be developed and
demonstrated. To construct a high-luminosity Muon Collider is an even
greater extrapolation of the present state of accelerator design. The
breadth of R\&D issues to be dealt with is beyond the resources available at
any single national laboratory or university.

For this reason, in 1995, interested members of the high-energy physics and
accelerator physics communities formed the Neutrino Factory and Muon
Collider Collaboration (MC) to coordinate the required R\&D efforts
nationally. The MC comprises three sponsoring national laboratories (BNL,
FNAL, LBNL) along with groups from other U.S. national laboratories and
universities and individual members from non-U.S. institutions. Its task is
to define and carry out R\&D required to assess the technical feasibility of
constructing initially a muon storage ring that will provide intense
neutrino beams aimed at detectors located many thousands of kilometers from
the accelerator site, and ultimately a $\mu ^{+}\mu ^{-}$ collider that will
carry out fundamental experiments at the energy frontier in high-energy
physics. The MC also serves to coordinate muon-related R\&D activities of
the NSF-sponsored University Consortium (UC) and the state-sponsored
Illinois Consortium for Accelerator Research (ICAR), and is the focal point
for defining the needs of muon-related R\&D to the managements of the
sponsoring national laboratories and to the funding agencies (both DOE and
NSF). Though the MC was formed initially to carry out R\&D that might lead
eventually to the construction of a Muon Collider, more recently its focus
has shifted mainly, but not exclusively, to a Neutrino Factory.

The MC maintains close contact with parallel efforts under way in Europe
(centered at CERN) and in Japan (centered at KEK). Through its international
members, the MC also fosters coordination of the international muon-beam
R\&D effort. Two major initiatives, a Targetry Experiment (E951) in
operation at BNL and a Muon Cooling R\&D program (MUCOOL), have been
launched by the MC. In addition, the Collaboration, working in conjunction
with the UC and ICAR in some areas, coordinates substantial efforts in
accelerator physics and component R\&D to define and assess parameters for
feasible designs of muon-beam facilities.

\section{R\&D\ Goals}

The approach taken by the MC to define the overall R\&D program was to
decide what we wished to accomplish in a five-year time span in each area
and then determine what is needed to reach that goal. For this exercise, we
assume a technology-limited schedule, that is, we assume that the required
financial resources and personnel are available. With this approach, we
expect that a five-year technology-limited plan will result in:

\begin{itemize}
\item  all optics designs being completed and self-consistent

\item  validation experiments being completed or well along

\item  all required hardware being defined

\item  prototypes of the most challenging and costly components being
completed or well along, i.e., we know how to build the ``hard parts''

\item  being ready to begin the design of, and provide cost estimates for,
most of the remaining components
\end{itemize}

At the end of the five-year period, the above goals would put the MC in
position to request permission to begin a formal Conceptual Design Report
(CDR) for a Neutrino Factory. It is expected that this CDR stage would take
1--2 years to complete. The CDR would document a complete and fully
engineered design for the facility, including a detailed bottom-up cost
estimate for all components. This document would form the basis for a full
technical, cost, and schedule review of the construction proposal,
subsequent to which construction could commence after obtaining government
approval.

As an ``intermediate milestone'' we envision preparing a Zeroth-order Design
Report (ZDR) after three years. The ZDR will examine the complete systems of
a Neutrino Factory, making sure that nothing is forgotten, and will show how
the parts merge into a coherent whole. While it will not present a fully
engineered design with a detailed cost estimate, enough detail will be
presented to ensure that the critical items are technically feasible and
that the proposed facility could be successfully constructed and operated at
its design specifications.

\section{R\&D Program Issues}

A Neutrino Factory comprises the following major systems: Proton Driver,
Target and (Pion) Capture Section, (Pion-to-Muon) Decay and Phase Rotation
Section, Bunching and Matching Section, Cooling Section, Acceleration
Section, and Storage Ring. These same categories exist for a Muon Collider,
with the exception that the Storage Ring is replaced by a Collider Ring
having a low-beta interaction point and a local detector. Parameters and
requirements for these systems are generally more severe in the case of the
Muon Collider, so a Neutrino Factory can properly be viewed as a
scientifically productive first step toward the eventual goal of a collider.
As noted earlier, the R\&D program we envision is designed to answer the key
questions needed to embark upon a ZDR after three years. After completion of
the full five-year program, it is expected that a formal Conceptual Design
Report could begin. Longer-term activities, related primarily to the Muon
Collider, are also supported and encouraged.

Each of the major systems has significant issues that must be addressed by
R\&D activities, including a mix of theoretical, simulation, modeling, and
experimental studies, as appropriate. A brief summary of the key physics and
technology issues for each major system is given below.

\bigskip \textbf{Proton Driver}

\begin{itemize}
\item  Production of intense, short proton bunches, e.g., with space-charge
compensation and/or high-gradient, low frequency rf systems
\end{itemize}

\textbf{Target and Capture Section}

\begin{itemize}
\item  Optimization of target material (low-\textit{Z} or high-\textit{Z})
and form (solid, moving band, liquid-metal jet)

\item  Design and performance of a high-field solenoid ($\approx $20 T) in a
very high radiation environment
\end{itemize}

\textbf{Decay and Phase Rotation Section}

\begin{itemize}
\item  Development of high-gradient induction linac modules having an
internal superconducting solenoid channel
\end{itemize}

\textbf{Bunching and Matching Section}

\begin{itemize}
\item  Design of efficient bunching system
\end{itemize}

\textbf{Cooling Section}

\begin{itemize}
\item  Development and testing of high-gradient normal conducting rf (NCRF)
cavities at a frequency near 200 MHz

\item  Development and testing of efficient high-power rf sources at a
frequency near 200 MHz

\item  Development and testing of LH$_{2}$ absorbers for muon cooling

\item  Development and testing of candidate diagnostics to measure emittance
and optimize cooling channel performance

\item  Design of beamline and test setup (e.g., diagnostics) needed for
demonstration of transverse emittance cooling

\item  Development of six-dimensional analytical theory to guide the design
of the cooling section
\end{itemize}

\textbf{Acceleration Section}

\begin{itemize}
\item  Optimization of acceleration techniques to increase the energy of a
muon beam (with a large momentum spread) from a few GeV to a few tens of GeV
(e.g., recirculating linacs, rapid cycling synchrotrons, FFAG rings) for a
Neutrino Factory, or even higher for a Muon Collider

\item  Development of high-gradient superconducting rf (SCRF) cavities at
frequencies near 200 MHz, along with efficient power sources (about 10 MW
peak) to drive them

\item  Design and testing of components (rf cavities, magnets, diagnostics)
that will operate in the muon-decay radiation environment
\end{itemize}

\textbf{Storage Ring}

\begin{itemize}
\item  Design of large-aperture, well-shielded superconducting magnets that
will operate in the muon-decay radiation environment
\end{itemize}

\textbf{Collider}

\begin{itemize}
\item  Cooling of 6D emittance (\textit{x}, \textit{p}$_{x}$, \textit{y}, 
\textit{p}$_{y}$, \textit{t}, \textit{E}) by up to a factor of $10^{5}-10^{6}
$

\item  Design of a collider ring with very low beta (a few mm) at the
interaction point having sufficient dynamic aperture to maintain luminosity
for about 500 turns

\item  Study of muon beam dynamics at large longitudinal space-charge
parameter and at high beam-beam tune shifts
\end{itemize}

\textbf{Detector}

\begin{itemize}
\item  Simulation studies to define acceptable approaches for both near and
far detectors at a Neutrino Factory and for a collider detector operating in
a high-background environment

\item  Develop ability to measure the sign of electrons in the Neutrino
Factory detectors
\end{itemize}

\section{FY 2001 R\&D Plans}

\subsection{Targetry}

This year, a primary effort of the Targetry experiment E951 was to carry out
initial beam tests of both a solid carbon target and a mercury target. Both
of these goals were accomplished at a beam intensity of about 4 $\times $\ 10%
$^{12}$ ppp, with encouraging results. Measurements of the velocity of
droplets emanating from the jet as it is hit with the proton beam pulse from
the AGS compare favorably with simulation estimates. High-speed photographs
indicate that the beam disruption at the present intensity does not appear
to propagate back upstream toward the jet nozzle, which will ease mechanical
design issues for this component.

\subsection{MUCOOL}

The primary efforts this year were to complete the Lab G rf test area and
begin high-power tests of the 805 MHz cavities that were completed earlier
this year. A test solenoid for the facility, capable of operating either in
solenoid mode (its two independent coils powered in the same polarity) or
gradient mode (with the two coils opposed), was tested up to its design
field of 5 T. An open-cell cavity has been installed and conditioning at
high-power is under way to explore gradient limitations we will face in a
cooling channel. A second cavity, having Be foils to close the beam iris, is
now being tuned to final frequency. This cavity will permit an assessment of
the behavior of the foils under rf heating and give indications about
multipactor effects.

Development of a prototype LH$_{2}$ absorber is in progress. A large
diameter, thin (125 $\mu $m) aluminum window has been successfully
fabricated by machining from a solid disk. A new area is being developed at
FNAL for testing the absorbers. This area, located at the end of the proton
linac, will be designed to permit beam tests of components and detectors
with 400 MeV protons. It will also have access to 201-MHz high-power rf
amplifiers for cavity testing.

Initial plans for a cooling demonstration will be firmed up this year. This
topic will be covered separately in Section \ref{Demo}.

\subsection{Feasibility Study-II}

This year the MC participated heavily in a second Feasibility Study for a
Neutrino Factory, sponsored by BNL. The results of the study were quite
encouraging (see Chapter 4 of this report), indicating that a neutrino
intensity of $1\times 10^{20}$ per Snowmass year can be sent to a detector
located 3000 km from the muon storage ring. It was clearly demonstrated by
means of these two studies that a Neutrino Factory could be sited at either
FNAL or BNL. Component R\&D needed for such a facility was identified, and
is included in the program outlined here.

\subsection{Simulation and Theory}

In addition to Study-II, this year the effort has focused on longitudinal
dynamics. We are developing theoretical tools for managing the longitudinal
aspects of cooling, with the goal of developing approaches to 6D cooling,
i.e., ``emittance exchange.'' This is a crucial aspect for the eventual
development of a Muon Collider, and would benefit a Neutrino Factory as
well. In particular, work has begun on studies of the ring cooler~\cite
{BalbekovRing} which has the potential to cool in 6D space, if the beam can
be injected and extracted from it. Avenues are being explored to see if it
can also function as a cooling demonstration where pions are produced in an
internal target.

\subsection{Component Development}

The main effort in this area is aimed at development of a high-gradient
201-MHz SCRF cavity. This year a test area of suitable dimensions is being
fabricated at Cornell. In addition, a prototype cavity is being fabricated
for the Cornell group by our CERN colleagues. Mechanical engineering studies
of microphonics and Lorentz detuning issues are being carried out. These
will lead to plans to stiffen the cavity sufficiently to avoid serious
vibration problems in these large structures.

\subsection{Collider R\&D}

Studies of possible hardware configurations to perform emittance exchange,
such as the compact ring proposed by Balbekov~\cite{BalbekovRing}, are now
getting under way. An emittance exchange workshop was held at BNL in the
fall of 2000, and a second workshop is being planned for October, 2001. In
addition to the efforts on emittance exchange, a workshop on an entry-level
Muon Collier to serve as a Higgs Factory was hosted this year by UCLA and
Indiana University. The focus of this meeting was to begin exploring the
path to get from a Neutrino Factory to a Higgs Factory. Even beyond the
cooling issues, the bunch structure required for the two facilities is very
different (the Collider demands only a single bunch), so the migration path
is not straightforward.

\section{FY2002 R\&D plans}

\subsection{Targetry}

For the targetry experiment, design of a pulsed 20-T solenoid and its 5-MW
power supply will begin. One or more selected targets will be tested with
beam this year. Simulations in support of this activity will continue.
Improvements in the AGS extraction system will be investigated, with the
goal of approaching the design single-bunch intensity of 1.7 $\times $ 10$%
^{13}$ ppp on target. An upgrade of the AGS extraction kicker to permit fast
extraction of the entire beam will be studied.

Radiation tests on selected coil materials will begin this year, to verify
behavior prior to actual magnet fabrication. As part of this work, we will
measure neutron yields from the target, to compare with predictions of the
MARS code. Systems studies of the target facility will continue to identify
and test key issues related to handling and remote maintenance, with special
attention paid to mercury-handling issues.

The next level of engineering concepts for a band target will be developed.
If its engineering aspects can be mastered, this approach might be a good
technical backup to the mercury jet .

\subsection{MUCOOL}

Testing work remaining for the 805 MHz components will continue this year in
Lab G. Completion of the linac test area at FNAL, initially to accommodate
the absorber tests and ultimately to house the 201-MHz cavity test, will
occur this year. Thermal tests of a prototype absorber will commence there.
Design and fabrication of cooling channel components required for the
initial phase of testing will begin, including a high-power 201 MHz NCRF
cavity and diagnostics that could be used for the experiment. Provisions
will be made to test both Be windows and grids for the cavity.

\subsection{Simulations and Theory}

Simulations this year will focus on iterating the front-end channel design
to be compatible with realizable component specifications. Studies of the
acceleration system and the storage ring will include errors and
fringe-field effects. From these studies will come component specifications
for the acceleration system and storage ring components. Simulation efforts
in support of a cooling demonstration program, and work on emittance
exchange, will both continue.

\subsection{Component Development}

A prototype induction linac cell, designed to operate at $\approx $1.5 MV/m
and including an internal superconducting solenoid with suitable dimensions
and field strength, will be designed. A prototype 201-MHz SCRF cavity will
be tested at low power. An initial input power coupler design will be tested
and validated. Detuning issues associated with the pulsed rf system will
also be evaluated. Finally, because of the large stored energy in a 201 MHz
SCRF cavity, a reliable quench protection system must be designed and
tested. Design of a prototype high-power rf source will begin, in
collaboration with industry.

\section{FY2003 R\&D plans}

\subsection{Targetry}

For the targetry experiment, AGS extraction kicker modifications to permit
high intensity beam tests ($\approx $1 $\times $ 10$^{14}$ protons per
pulse) will begin. The 20-T target solenoid for E951 will be fabricated this
year. Measurements of particle yield in this target geometry, but without a
solenoid will be made. These will use a much lower beam intensity (1 $\times 
$ 10$^{6}$ protons per pulse), for 6 weeks of parasitic beam time. Tests of
radiation-hard materials will be completed and selected candidate materials
will be used to begin manufacture of an actual (warm) solenoid coil for
testing in a high-radiation environment. The target simulation effort this
year will focus on understanding experimental results from the targetry
experiment.

\subsection{MUCOOL}

High-power tests of the 201-MHz NCRF cavity will begin in the linac test
area. Beam tests will be carried out with a prototype LH$_{2}$ absorber.
Fabrication of all remaining components, such as the solenoids, will
commence.

\subsection{Simulations and Theory}

Additional effort will be given to beam dynamics studies in the RLAs and
storage ring, including realistic errors. Work on finalizing the optics
design for the arcs will be done. Assessment of field-error effects on the
beam transport will be made to define acceptance criteria for the magnets.
This will require use of sophisticated tracking codes like COSY that permit
rigorous treatment of field errors and fringe-field effects. Because the
beam circulates in each RLA for only a few turns, the sensitivity to magnet
errors should not be extreme, though the large energy spread will tend to
enhance it. In many ways, the storage ring is one of the most
straightforward portions of a Neutrino Factory complex. However, beam
dynamics is an issue here as the muon beam must circulate for many hundreds
of turns. Use of a tracking code such as COSY is required to assess fringe
field and large aperture effects. As with the RLAs, the relatively large
emittance and large energy spread enhance the sensitivity to magnetic field
and magnet placement errors. Suitable magnet designs are needed, with the
main technical issue being the relatively high radiation environment.
Another lattice issue that must be studied is polarization measurement. In
the initial implementation of a Neutrino Factory it is expected that
polarization will not be considered, but its residual value may be important
in analyzing the experiment.

\subsection{Component Development}

Magnet designs suitable for the arcs of the recirculating linacs (RLAs) and
the muon storage ring will be examined this year. Both conventional and
superconducting designs will be compared where either is possible. With SC
magnets, radiation heating becomes an issue and must be assessed and dealt
with. Designs for the splitter and recombiner magnets will be developed
and---depending on how nonstandard they are---prototypes will be built.
Tests of a 201 MHz SCRF cavity will continue this year, including
demonstration of the ability to shield nearby magnetic fields in a realistic
lattice configuration. Fabrication of a prototype induction linac module and
its pulser system will begin this year.

\section{FY2004 R\&D plans}

\subsection{Targetry}

For the targetry experiment, the work this year will focus on completing
systems studies of the final target station, including issues of radiation
handling and safety. Target facility studies will progress to prototypes or
full-size models of key components, as needed, to verify the design
concepts. A (normal conducting) test solenoid coil based on the materials
tests in prior years will be completed and tested in a radiation environment.

\subsection{MUCOOL}

A full cooling cell will be assembled and bench tested. After testing, it is
anticipated that these components will be installed in a beam line and
tested with protons. The goal here is not to demonstrate cooling, but to
demonstrate operation of the components in a radiation environment. Upon
completion of the integration tests, these components will be available for
testing in a muon beam as part of the cooling demonstration experiment.

\subsection{Theory and Simulation}

Work on the Zeroth-order Design Report will commence this year. This
activity will require about two years of significant effort, leading to a
document that presents a description of all aspects of a Neutrino Factory,
in sufficient detail to demonstrate technical feasibility of the full
facility.

\subsection{Component Development}

High-power tests of the 201-MHz SCRF cavity will be carried out at FNAL. The
induction linac prototype module will be completed and tests will begin.

\section{FY 2005 R\&D plans}

This year should see the completion of the ZDR followed by a community
review of its contents. Thereafter, we will be ready to seek permission to
begin a formal CDR. Work on the Cooling Demonstration Experiment (see
Section \ref{Demo}) will be a primary activity this year.

\section{Required Budget}

Table \ref{tab:budget} summarizes the projected budget requirements for the
activities described above. The numbers represent our estimate of the
required resources \textit{from all sources}, including direct DOE funding
for the MC, DOE base program support from the sponsoring Laboratories, NSF
support to the UC, state support for the ICAR, and possible contributions by
foreign collaborators. The scale of R\&D is consistent with the experience
of other large accelerator projects, both past and present, and in that
sense is ``realistic.'' The funding profile in Table \ref{tab:budget}
corresponds to the technology-limited schedule we are considering. If this
level of funds is not available, the R\&D work would proceed more slowly.
Indeed, the amount of support available in FY2001, and that projected for
FY2002, already fall short of what is required, with the result that the
schedule shown here has begun to slip. It is worth noting that the increase
in FY2004 and FY2005 is associated mainly with the cooling demonstration
experiment. If we were to assume equal sharing among the U.S., Europe, and
Japan in this endeavor, as seems reasonable, then about \$8M of the
tabulated costs in each of these years would be borne by non-U.S. funds. In
that case, a program with roughly flat funding (from all U.S. sources) in
the \$15M range would permit us to complete the R\&D program in a timely way.

\bigskip

\begin{table}[tbh]
\caption[R\&D budget to reach a CDR]{MC\ R\&D budget to reach a CDR in a
technology-limited schedule.}
\label{tab:budget}
\begin{center}
\begin{tabular}{lllllll}
\hline
\textbf{R\&D area} & \textbf{FY01} & \textbf{FY02} & \textbf{FY03} & \textbf{%
FY04} & \textbf{FY05} & \textbf{Sum} \\ 
& \textbf{(\$M)} & \textbf{(\$M)} & \textbf{(\$M)} & \textbf{(\$M)} & 
\textbf{(\$M)} & \textbf{(\$M)} \\ \cline{2-7}
MUCOOL & 4.9 & 3.8 & 4.3 & 11.3 & 11.2 & 35.4 \\ 
Targetry & 4.7 & 3.8 & 4.1 & 3.5 & 2.1 & 18.2 \\ 
Beam Simul. & 2.3 & 2.0 & 2.0 & 2.0 & 2.0 & 10.3 \\ 
Accel./SR & 1.0 & 0.7 & 0.7 & 0.7 & 0.7 & 3.6 \\ 
Components & 1.9 & 4.5 & 7.5 & 4.3 & 4.0 & 22.2 \\ 
ZDR Prep. &  &  &  & 4.0 & 6.0 & 10.0 \\ \cline{2-7}
TOTAL & 15 & 15 & 19 & 26 & 26 & 100 \\ \hline
\end{tabular}
\end{center}
\end{table}

\section{Cooling Demonstration Experiment}

One of the more important R\&D tasks that is needed to validate the design
of a Neutrino Factory is to measure the cooling effects of the hardware we
propose. At the recent NUFACT'01 Workshop in Japan, a volunteer
organization, to be chaired initially by Alain Blondel from Geneva
University, was created to organize a cooling demonstration experiment that
would begin in 2004. See Chapter on International activities on further
details on this committee. Present membership in this group (with
representatives from the U.S., Europe, and Japan), the ``Muon Cooling
Demonstration Experiment Steering Committee'' (MCDESC), is listed in the
chapter on International Activities. The Steering Committee is responsible
for choosing a technical team to develop the proposal details, suggest a
beamline, and propose components to be tested, including absorbers, rf
cavities and power supplies, magnets, and diagnostics. This technical team
will likewise be assembled from experts from the three geographical regions.

The aim of the proposed experiment is twofold:

\begin{enumerate}
\item  To show that we can design, engineer, and build a section of cooling
channel capable of giving the desired performance for a Neutrino Factory

\item  To place it in a muon beam and measure its performance, i.e., to
validate that it performs as predicted by simulations
\end{enumerate}

\noindent It is clear that the experience gained from this experiment will
be invaluable for the design of an actual cooling channel.

The present concept is to carry out a single-particle measurement of cooling
effects, starting from a single cell of cooling hardware. We believe that
the measurement of the emittance change can be done with a precision\label%
{Demo} of about 0.5\% with standard single-particle detection techniques.
The main technical uncertainty in this concept is the ability of the
detection equipment to function properly in the large x-ray flux resulting
from very high gradient rf cavity operation. This aspect can be explored
almost immediately in the Lab G test area at FNAL, where high-gradient
cavity testing is just getting under way. (The x-ray intensity from the
cavity is expected to scale with electric field as $E^{10}$. Taking a
positive slant, this means that a small reduction in operating voltage will
lead to a large reduction in x-ray intensity.)

The plan of the Steering Committee is to create a short document ($\approx $%
10 pages) by December 15, 2001 that will define the key technology choices
and the venue for the experiment. Thereafter, a full technical proposal,
including a cost estimate suitable for presentation to the various funding
agencies, would be prepared by June, 2002. Though there obviously is no cost
estimate yet, it is expected that getting a single cooling cell into a muon
beam will require about \$10M. Thereafter, it is expected that more cells
would be added, either identical to that in the first test or perhaps of a
different design. With this in mind, it is reasonable to suppose that the
initial funding commitment needed---shared among the three regions---will be
about \$20M. \ This is consistent with the estimate presented in Table \ref
{tab:budget}. However, the profile in Table \ref{tab:budget} would have to
move forward by one year. The overall U.S. requirement of \$15M per year
would remain unchanged. (Later, it may be of interest to continue this
experimental effort, perhaps including longitudinal cooling techniques,
though that is certainly beyond the initial scope of the proposed
experiment.)

\chapter{International Activities}
\label{international}

In Europe there exists an accelerator study group mandated by CERN and a 
physics
study group  mandated by the European committee for Future Accelerators
(ECFA)~\cite{int:1}. 
These activities are co-ordinated by European Muon Steering Committee (MUG) 
whose membership~\cite{int:2} is: 
Alain Blondel (chair), Friedrich Dydak, John Ellis, Enrique Fernandez,
Helmut Haseroth, Vittorio Palladino, Ken Peach, Michel Spiro, Paolo Strolin.

\bigskip

The primary center of work is at CERN. The work of the accelerator
group was described at NuFact01 by R. Garoby~\cite{int:3}. The general
approach to a Neutrino Factory, the design of which is co-authored by
more than 90 physicists, is based upon a 2.2~GeV proton
linac~\cite{int:4}.  They have a good number of R\&D activities. These
include a large theoretical / modeling group, work on detectors
(including many universities in Europe), involvement in the scattering
experiment (MUSCAT) at Triumf (involving CLRC, Birmingham, and
Imperial College and CERN), and participation in target work at Brookhaven,
engaging in target work at CERN, RAL and Grenoble, work on low
frequency RF cavities, a major production experiment at the PS (HARP),
and involvement in the International Muon Cooling Experimental Demonstration.

\bigskip

Activities in Japan were described by S. Machida at NuFact01~\cite{int:3}.
Their general approach is a series of FFAG accelerators. No manipulation of 
longitudinal phase space and no cooling of transverse phase space is needed in 
their approach. Their R\&D program is not very extensive, but they are 
cooperating with the US on the absorber R\&D and on MUSCAT. Most 
importantly, one notes that the Japanese are in the process of constructing 
a 1 MW driver, so soon they will have a superbeam. A factory seems likely 
to follow.

\bigskip

Information is exchanged very effectively. Besides constant communication by 
means of the Web and e-mail there is a series of Neutrino Factory Workshops 
(NuFact99 -- Lyon, NuFact00 -- Monterey, NuFact01 -- Tsukuba, and NuFact02 
-- London)

\bigskip

The International Muon Cooling Experimental Demonstration effort is still at 
a very early stage. Nevertheless, there is agreement upon a process and 
procedure that will, hopefully, result in experimental activity. Appended is 
the first draft document, which outlines the proposed activities over the 
next year. Following that, groups will seek support from funding sources 
throughout the world, after which a document specifying just what each 
source will supply will need to be drawn up. By that time one should have a 
good idea of schedules, costs and expected deliverables.

\bigskip

There is the beginning of international laboratory interest. This was 
precipitated by a letter from Maiani (i.e., an initiative that started 
abroad). There have been subsequent exchanges of letters, but no agreed upon 
document has yet appeared (at present the Europeans are suggesting a 
commitment to significant support of R\&D, and the Americans are 
demurring).

\section{Towards an International Muon Cooling Experimental Demonstration}

\bigskip

\begin{center}
Alain Blondel, Rob Edgecock, Steve Geer, Helmut Haseroth, Yoshi Kuno, 
\end{center}

\begin{center}
Dan Kaplan, Michael Zisman 

\end{center}

\bigskip

\begin{center}
June 15, 2001
\end{center}

\bigskip

\subsection{Motivation}

\bigskip
Ionisation cooling of minimum ionising muons is an important ingredient in
the performance of a neutrino factory. However, it has not been demonstrated
experimentally. We seek to achieve an experimental demonstration of cooling
in a muon beam. In order to achieve this goal, we propose to continue to
explore, for the next six months or so, at least two versions of an
experiment based on existing cooling channel designs. If such an experiment
is feasible, we shall then select, on the basis of effectiveness,
simplicity, availability of components and overall cost, a design for the
proposed experiment.

\bigskip
On the basis of this conceptual design, we will then develop detailed
engineering drawings, schedule and a cost estimate. The costs and
responsibilities will be broken out by function (e.g. magnets, RF,
absorbers, diagnostics etc) and also by laboratory and region. A technical
proposal will be developed by Spring 2002, and will be used as the basis for
detailed discussions with laboratory directors and funding agencies.

\bigskip

The aim of the proposed cooling experimental demonstration is
\begin{itemize} 
\item	to show that we can design, engineer and build a section of cooling
channel capable of giving the desired performance for a neutrino factory;
\item	to place it in a beam and measure its performance, i.e.
experimentally validate our ability to simulate precisely the passage of
muons confined within a periodic lattice as they pass through liquid
hydrogen absorbers and RF cavities.
\end{itemize}
The experience gained from this experimental demonstration will provide
input to the final design of a real cooling channel.

\bigskip

The signatories to this document volunteer to organise this international
effort. It is expected that the membership of this group, referred to in
this document as the Muon Cooling Demonstration Experiment Steering
Committee (MCDESC) will evolve with time. It is proposed that the Chair of
this group should be Alain Blondel for the first year.

\bigskip

\subsection{Organisation}

\bigskip
\begin{itemize}
\item The overall organisation and coordination of the activity shall be the 
responsibility of the MCDESC.

\item The MCDESC shall assemble members of a technical team to develop the 
proposal. The members of this technical team should represent at least two 
geographical regions in each of the following aspects
\end{itemize}

\begin{enumerate}
\item Concept Development and Simulation
\item Absorbers
\item RF Cavities and Power Supplies
\item Magnets
\item Diagnostics
\item Beamlines
\end{enumerate}

\begin{itemize}

\bigskip

\item It is expected that the MCDESC will work mainly by telephone
conference and e-mail, but should meet, typically, twice each year,
preferably in association with other scheduled meetings. These meetings
should rotate around the regions. The technical team should organise its
activities as appropriate.

\end{itemize}

\bigskip

\subsection{Schedule}

The goal is to carry out a first experiment in 2004, in the expectation that
this could develop into more sophisticated tests, including possibly the
demonstration of longitudinal cooling. In order to achieve this ambitious
schedule, it will be necessary to make proposals to laboratory directors and
funding agencies in 2002.

\bigskip

{\it Therefore},
\begin{enumerate}
\item 	A short document (of order ten pages) making key technology choices
(including the choice of version of the experiment and location) should be
presented by Dec 15th 2001.

\item 	This conceptual design should be developed into a full technical
proposal by June 2002. This technical proposal would need engineering
drawings, schedules and costs, and distribution of responsibilities. This
would include the cost breakdown by component (RF, magnet, absorber,
diagnostics, beam) and by country and/or laboratory.
\end{enumerate}

\bigskip

It is the responsibility of the technical team to provide the technical
evaluations of the alternative approaches, in order for the MCDESC to be
able to make the required technology choices in the Fall of 2001.

\appendix
\chapter[Members of the Muon Collaboration]
{Members of the Neutrino Factory and Muon Collider Collaboration}
\begin{center} 
Maury~Goodman, Ahmed~Hassanein, James~H.~Norem, Claude~B.~Reed,
Dale~Smith, Lee~C.~Teng, Chun--xi~Wang\\ 
\textbf{Argonne National Laboratory, Argonne, IL  60439}\protect\\

J.~Scott~Berg, Richard~C.~Fernow, Juan~C.~Gallardo, Ramesh~Gupta, 
 Stephen~A.~Kahn, Bruce~J.~King,
Harold~G.~Kirk, David~Lissauer, Laurence~S.~Littenberg,
 William~A.~Morse,  Satoshi~Ozaki,
Robert~B.~Palmer, Zohreh~Parsa, Ralf~Prigl, Pavel~Rehak,
Thomas~Roser,
 Nick~Simos, Iuliu~Stumer, Valeri~Tcherniatine, Peter~Thieberger, Dejan~Trbojevic, Robert~Weggel,
Erich~H.~Willen, Yongxiang~Zhao\\
\textbf{Brookhaven National Laboratory, Upton, NY 11973}\protect\\

Gregory~I.~Silvestrov, Alexandr~N.~Skrinsky, Tatiana~A.~Vsevolozhskaya\\
\textbf{Budker Institute of Nuclear Physics, 630090 Novosibirsk, Russia}\protect\\

Gregory~Penn, Jonathan~S.~Wurtele$^{1}$ \\
\textbf{University of California-Berkeley, Physics Department, Berkeley, CA 94720}\protect\\

John~F.~Gunion\\
\textbf{University of California-Davis, Physics Department, CA 95616}\protect\\

David~B.~Cline, Yasuo~Fukui,  Alper~A.~Garren, Kevin~Lee, Yuriy~Pischalnikov\\
\textbf{University of California-Los Angeles, Los Angeles, CA 90095}\protect\\

Bruno~Autin, Roland~Garoby, Helmut~Haseroth, Colin~Johnson, Helge~Ravn, Edmund~J.~N.~Wilson\\
\textbf{CERN, 1211 Geneva 23, Switzerland}\protect\\

Kara~Hoffman, Kwang-Je~Kim, Mark~Oreglia, Yau~Wah\\
\textbf{The University of Chicago, Chicago, IL 60637}\protect\\

Vincent~Wu\\
\textbf{University of Cincinnati, Cincinnati, OH 45221}\protect\\

Allen~C.~Caldwell, Janet~M.~Conrad, Jocelyn~Monroe, Frank~Sciulli, Michael~H.~Shaevitz, William~J.~Willis\\
\textbf{Columbia University, Nevis Laboratory, Irvington, NY  10533}\protect\\

Hasan~Padamsee, Maury Tigner\\
\textbf{Cornell University, Newman Laboratory for Nuclear Studies, Ithaca, NY  14853}\protect\\

David~R.~Winn\\
\textbf{Fairfield University, Fairfield, CT 06430}\protect\\

Charles~M.~Ankenbrandt, Muzaffer~Atac, Valeri~I.~Balbekov,
Elizabeth~J.~Buckley-Geer, David~C.~Carey, Sam~Childress, Weiren~Chou,
Fritz~DeJongh, H.~Thomas~Diehl, Alexandr~Drozhdin, Daniel~Elvira,
David~A.~Finley, Stephen~H.~Geer, Krishnaswamy~Gounder,  D.~A.~Harris, 
Carol~Johnstone, Paul~Lebrun, Valeri~Lebedev, Joseph~D.~Lykken,
Frederick~E.~Mills, Nikolai~V.~Mokhov, Alfred~Moretti,
David~V.~Neuffer, King-Yuen~Ng, Milorad~Popovic,
Zubao~Qian, Rajendran~Raja, Panagiotis~Spentzouris, 
Ray~Stefanski, Sergei~Striganov, Alvin~V.~Tollestrup, Zafar~Usubov,
Andreas~Van~Ginneken, Steve~Vejcik\\
\textbf{Fermi National Accelerator Laboratory, P. O. Box 500, Batavia, IL 60510}\protect\\

Alain~Blondel\\
\textbf{University of Geneva, Switzerland}\protect\\

John G. Learned, Sandip Pakvasa\\
\textbf{University of Hawaii, Department of Physics, Honolulu, HI  96822}\protect\\

Massimo~Ferrario\\
\textbf{INFN-LNF, via E-Fermi 40, Frascati (Roma), Italy}\protect\\

S.~Alex~Bogacz, Swapan Chattopadhyay, Haipeng~Wang\\
\textbf{Jefferson Laboratory, 12000 Jefferson Ave., Newport News, VA 23606}\protect\\

T.~Bolton\\
\textbf{Kansas State University, Manhattan, KS 66502-2601}\protect\\

Yoshitaka~Kuno, Yoshiharu~Mori, Takeichiro~Yokoi\\
\textbf{KEK High Energy Accelerator Research Organization, 1-1 Oho, Tsukuba 305, Japan}\protect\\

Edgar~L.~Black,  Daniel~M.~Kaplan, Nickolas~Solomey, Ya\u{g}mur~Torun\\
\textbf{Illinois Institute of Technology, Physics Div., Chicago IL 60616}\protect\\

Deborah Errede, Kyoko~Makino\\
\textbf{University of Illinois, at Urbana, Urbana-Champaign, IL 61801}\protect\\
Michael~S.~Berger, Gail~G.~Hanson, Daniel~Krop,Peter~Schwandt\\
\textbf{Indiana University, Physics Department, Bloomington, IN 47405}\protect\\

Ilya~F.~Ginzburg\\
\textbf{Institute of Mathematics, Prosp.\ ac. Koptyug 4, 630090 Novosibirsk, Russia}\protect\\

Yasar~Onel\\
\textbf{University of Iowa, Physics Department, Van Allen Hall, Iowa City, IA 52242}\protect\\

Shlomo~Caspi, John~Corlett, Miguel~A.~Furman, Michael~A.~Green, C.H.~Kim
\footnote{deceased}
 Derun~Li,
Alfred~D.~McInturff, Louis~L.~Reginato,
Robert~Rimmer, Ronald~M.~Scanlan, Andrew~M.~Sessler, Brad~Shadwick,
William~C.~Turner, Simon~Yu, Michael~S.~Zisman,
Max~Zolotorev\\
\textbf{Lawrence Berkeley National Laboratory, 1 Cyclotron Rd., Berkeley, CA 94720}\protect\\

Martin~Berz, Richard York, Al Zeller\\
\textbf{Michigan State University, East Lansing, MI 48824}\protect\\

Stephen~B.~Bracker, Lucien~Cremaldi, Don~Summers\\
\textbf{University of Mississippi, Oxford, MS 38677}\protect\\

John~R.~Miller, Soren Prestemon\\
\textbf{National High Magnetic Field Laboratory, Magnet Science \& Technology, FL 32310}\protect\\

Gerald~C.~Blazey, Mary Anne Cummings, David Hedin\\
\textbf{Northern Illinois University, DeKalb, IL 60115}\protect\\

Heidi M. Schellman\\
\textbf{Northwestern University, Department of Physics and Astronomy, Evanston, IL 60208}\protect\\

Tony~A.~Gabriel, Norbert~Holtkamp, Philip~T.~Spampinato\\                  
\textbf{Oak Ridge National Laboratory, Oak Ridge, TN 37831}\protect\\

Eun-San~Kim, Moohyun~Yoon\\
\textbf{Pohang University of Science and Technology, POSTECH, Sam 31, Hyoja dong,
Pohang, Kyungbuk, 790-784, Korea}\protect\\

Changguo~Lu, Kirk~T.~McDonald, Eric~J.~Prebys\\
\textbf{Princeton University, Joseph Henry Laboratories, Princeton, NJ 08544}\protect\\

Robert~Rossmanith\\
\textbf{Research Center Karlsruhe, D-76021 Karlsruhe, Germany}\protect\\

J.~Roger~J.~Bennett\\
\textbf{Rutherford Appleton Laboratory, Chilton, Didcot, Oxon OX11 0QX, UK}\protect\\

Robert~Shrock\\
\textbf{Department of Physics and Astronomy, SUNY, Stony Brook, NY 11790}\protect\\

Odette~Benary\\
\textbf{Tel-Aviv University, Ramat-Aviv, Tel-Aviv 69978, Israel}\protect\\

Vernon~D.~Barger, Tao~Han\\
\textbf{Department of Physics, University of Wisconsin, Madison, WI 53706}\protect\\
\end{center}  

$^{1}$ also at Lawrence Berkeley National Laboratory.

\chapter[Contributors] {Participants of the Studies, Non-Member of the
Neutrino Factory and Muon Collider Collaboration }
\begin{center} 
D.~Ayres, T.~Joffe­-Minor, D.~ Krakauer, P.~Schoessow, R.~Talaga,
J.~Thron, C.~Wagner\\
\textbf{Argonne National Laboratory, Argonne, IL  60439}\protect\\

Michael~Anerella, M.~Blaskiewicz, E.B.~Blum, Joseph~M.~Brennan,
W.~Fischer, W.S.~Graves, R.~Hackenburg, Michael~Harrison,
Michael~Hemmer, Hsiao-C.~Hseuh, H.~Huang, Michael~A.~Iarocci,
J.~Keane, V.~Lodestro, D.~Lowenstein, Alfredo~Luccio, Hans~Ludewig,
Ioannis~M.~Marneris, James~Mills, Stephen~V.~Musolino, Edward~O'Brien,
Wonho~Oh, Brett~Parker, Charles~Pearson, F.~Pilat, P.~Pile,
S.~Protopopescu,  Alessandro~Ruggiero, Roman~Samulyak,
Jesse~D.~Schmalzle, Y.~Semertzidis, Mariola~Sullivan, M.J.~Tannenbaum,
J.~ Wei, W.­T.~Weng,  Shuo-Yuan~Zhang\\
\textbf{Brookhaven National Laboratory, Upton, NY 11973}\protect\\

L.~Vilchez \\
\textbf{University of California-Davis, Physics Department, CA 95616}
% ...
\protect\\

R.~Bennett, R.~Edgecock, D.~Petyt\\
\textbf{Rutherford Appleton Laboratory, Chilton, Didcot, Oxon OX11 0QX, UK}\protect\\

A.~Bodek, K.S.~McFarland\\
\textbf{University of Rochester, Rochester, NY 14627}\protect\\

G.~Apollinari, E.J.N.~Wilson\\
\textbf{Rockefeller University, New York, NY 10021}\protect\\

A.~Sery, D.~Sprehn, D.~Ritson\\
\textbf{Stanford Linear Accelerator Center, Stanford, CA 94309}\protect\\

Peter~Titus\\
\textbf{Stone and Webster Corp. (under contract to PSFC, MIT) Boston, MA }\protect\\

C.K.~Jung\\
\textbf{Department of Physics and Astronomy, SUNY, Stony Brook, NY 11790}\protect\\

W.R.~Leeson, A.~Mahmood\\
\textbf{University Texas Pan American, TX}\protect\\

T.~Patzak\\
\textbf{Tufts University, Medford, MA 02155}\protect\\

R.V.~Kowalewski\\
\textbf{University of Victoria, BC Canada}\protect\\ 
\end{center}


%
\vfill
\eject

\begin{thebibliography}{100}
\bibitem{study2}Much of the text quoted here appears in the Study II document
\cite{EPP:studyii}. 
We would like to thank Robert Shrock for helpful comments.
\end{thebibliography}
\end{document}